\documentclass[12pt]{interact}

\usepackage[numbers,sort&compress]{natbib}
\bibpunct[, ]{[}{]}{,}{n}{,}{,}

\usepackage{amsmath, amssymb, amsfonts}
\usepackage{graphicx}
\usepackage{bm}
\usepackage{physics}
\usepackage{hyperref}
\usepackage{setspace}
\usepackage{booktabs}
\usepackage{multirow}
\usepackage{caption}
\usepackage[right]{lineno}

\renewcommand{\vec}{\mathbf}
\title{\textbf{Ordering, correlation functions and phase transitions in molecular systems}}

\author{Yashwant Singh \\
    \small Department of Physics, Institute of Science, Banaras Hindu University, Varanasi-221005, India \\
    \small Email: singh.yas44@gmail.com
}

\begin{document}
\maketitle
\begin{abstract}
Although the classical density functional theory (DFT) of inhomogeneous fluids was formulated more than four decades ago, its application to broken symmetry phases of molecular systems remained a challenge. Approximate free energy functionals proposed in the past failed to give accurate description of relative stability of phases, phase transitions, and of properties arising due to broken symmetry. In a DFT pair correlation functions (PCFs) play a fundamental role. While in the case of homogeneous fluids, PCFs are routinely determined using experimental, theoretical or simulation methods, determination of PCFs of broken symmetry phases remained a problem. Breaking of symmetry at the transition point gives rise a new contribution to correlation functions which may differ significantly from that of the coexisting higher symmetry phase. We review methods which have been developed in the last few years to calculate PCFs of broken symmetry phases and their inclusion in the expressions of the grand potential and the intrinsic free-energy. This leads to formulation of an exact DFT. We describe application of the theory to freezing of variety of fluids into ordered phases and transition from an ordered phase of higher symmetry to a phase of lower symmetry. Comparison of results found from different versions of DFT and simulations reveal their accuracy. A brief description of basics of statistical mechanics is included to make the article self-contained.
\end{abstract}


\tableofcontents
\newpage
\section{Introduction}

We are quite familiar with the fact that when an isotropic liquid composed of spherical molecules is slowly cooled, it transforms at its freezing point into a crystalline solid. On freezing, molecules get localized on a regular array of points called lattice and develop a long-range order (LRO) in their distribution. On the other hand, in a liquid, molecular distribution has only some resemblance of  order over a length of a few molecular diameters; such an order is called short-range order (SRO) which defines  characteristic features of fluids (liquids and dense gases). If molecules of a system have a large  asymmetry  in shape ( e.g. rod-shaped, disc-shaped, bent-shaped etc.), the system may exhibit orientational order or disorder in addition to translational order and disorder. As a consequence, a number of phases may appear in between isotropic fluid and a fully ordered crystal. Their number depends on the asymmetry in the molecular shape and in the intermolecular interactions \cite{Chaikin1995,deGennes1993,Chandrasekhar1992,Singh2005}. These phases are called liquid crystals. Liquid crystals are systems in which a \textit{liquid-like} feature exists at least in one direction of space and in which some degree of long-range ordering in the orientation of molecules or aggregates formed by the association of molecules is present. Phases are characterized by the average positions and orientations of molecules and by the intermolecular spatial and orientational correlations.

An isotropic fluid with a structure that is invariant under arbitrary translations and rotations has the highest possible symmetry and will be referred to as an isotropic phase or fluid phase. On the other hand, crystalline solids with frozen structure that are invariant only with respect to certain discrete lattice translations and point group operations have a much lower symmetry compared to that of the isotropic phase. The transition from the fluid to the crystal breaks the symmetry of the isotropic phase, and the crystal is often referred to as the broken symmetry phase. The liquid-crystals have intermediate symmetries  between the isotropic phase and the crystal phase, and are classified on the basis of broken symmetries. The symmetry of a system is a set of geometrical transformations which leave the molecular distribution in the system invariant.

An abrupt change of one phase of a macroscopic system into another caused by variation of parameters of the state, like temperature T, pressure P, external fields like the magnetic field H, etc. is called \textit{phase transition}. Depending on whether phases can coexist in an equilibrium contact or their distinction vanishes as the transition point is approached, two different types of phase transitions occur: the first one is called \textit{first order} while the latter is called \textit{continuous or second-order phase transition}. Phases are distinguished by a physical quantity called \textit{order parameter}, which characterizes the order that has emerged due to broken symmetry and is generally defined in such a way that it is zero in the higher symmetry phase and \textit{non-zero} in the lower symmetry phase. A phase may need more than one order parameter for its description. Features such as order parameters of various broken symmetries, anisotropic scaling behaviors, coupled order parameters, defect mediated transitions, space of low dimensionality, multiply re-entrant topologies etc. are observed in molecular systems  \cite{Chaikin1995,deGennes1993,Chandrasekhar1992,Singh2005}. Understanding them in terms of microscopic(molecular) variables has been the subject of investigations for last several  decades.\\

The theory of molecular systems has been developing using: (i) a phenomenological approach based on the Landau free energy functional\cite{Chaikin1995,deGennes1993} (ii) the Kosterlitz–Thouless –Halperin–Nelson–Young (KTHNY) \cite{Kosterlitz1973,Halperin1978,Young1979} theory of defect-mediated melting and (iii) a microscopic theory based on density functional formulation \cite{Evans1979,Singh1991}. Application of the Landau theory to a wide range of liquid crystalline systems by de Gennes \cite{deGennes1993} led to an enormous growth of theoretical activity, specially the application of the renormalization group techniques to the analysis of the Landau-de Gennes free energies. However, phenomenological theories do not address the basic questions of how molecular parameters influence macroscopic behavior (except in a very general way of molecular symmetry), nor they can predict the numerical values of transition temperature, elastic moduli and the like. While the theory of KTHNY has been quite successful in two dimensions, particularly in predicting two-stage melting of a two-dimensional solid, its success in the case of three dimensions remains limited \cite{Brock1989}.

Density functional theory, which was initially developed to describe inhomogeneous fluids \cite{Evans1979} , provides an overarching framework to investigate phase transitions (first order) and properties of ordered phases \cite{Singh1991}. In 1979, Ramakrishnan and Yussouff (RY) \cite{Ramakrishnan1979} proposed a theory to describe the freezing of fluids into crystalline solids. The key idea employed in the RY theory is that the grand thermodynamic potential of a system can be projected onto an order parameter space. The expression for the potential was derived using a functional Taylor expansion in powers of one--particle density about the isotropic fluid \cite{Haymet1981}; the derivatives being the fluid phase direct correlation functions. It was assumed that the correlation functions of the ordered phase at the transition point can be replaced by that of the isotropic phase at the same temperature and chemical potential. The theory has, however, been found to have limited success in describing the phenomenon of freezing and relative stability of ordered phases in many cases \cite{Barrat1987,deKuijper1990,Wong1999}. In an alternative approach, known as weighted-density approximation (WDA), the free-energy functional of an inhomogeneous system is constructed by replacing the actual one-particle density at a given point by that of a coarse-grained density ~\cite{Tarazona1985,Curtin1985,Denton1989}. The difficulty in implementing the theory lies in making an appropriate choice of weight-function for the coarse--graining (see Section 5). 

In a recently developed version of the theory \cite{Mishra2006,Bharadwaj2013} exact expressions for free-energy functional and grand thermodynamic potential are derived. These expressions involve correlation functions of ordered phases. Methods for calculation of these correlation functions are developed and the theory is applied to investigate phase transitions and properties of broken symmetry phases. Our aim in this article is to provide a comprehensive review of the theory with brief but self-contained background formalism.

The article is organized as follows. In Section 2 we give a summary of statistical mechanics of inhomogeneous systems based on the formalism of grand canonical ensemble. The Hamiltonian  of the system includes a term representing the interaction of molecules with some spatially and orientationally varying external field. The effect of the term is to break translational and rotational symmetry of the system. The formulation allows one to derive important relations between correlation functions and  for thermodynamic potentials. In Section 3 expressions for the grand thermodynamic potential and intrinsic free
energy are derived and expressed in terms of direct correlation functions. All quantities in these expressions are functional of one-particle density. The condition for locating phase transition point is discussed. Order parameters which measure the degree and nature of ordering in a broken symmetry phase are introduced in a general way in Section 4. In Section 5, we introduce density functional theory of freezing and make use of it to specify approximations involved in the two approximate theories mentioned above. Breaking of symmetry at the transition point brings forth a new contribution to correlation function which may differ significantly from that of the coexisting fluid. In Section 6 we show how this symmetry breaking contribution to direct pair correlation function is included in the expression of free-energy functional and grand potential. In the remaining part of the article we describe methods for calculating pair correlation function of broken symmetry phases and phase transitions. In Section 7, calculation of pair correlation functions in nematic liquid crystal and isotropic-nematic transition are described. In Sections 8 and 9 calculations of pair correlation functions in 3- and 2-dimensional crystals are presented. The fluid-solid and solid-solid phase transitions are described, respectively, in Sections 10 and 11. Wherever possible, results found from the theory are compared with computer simulation results. The article ends with a summary of key features and prospects of the theory in Section 12.

\section{Inhomogeneous systems: Correlation functions}
We consider a classical many-body system contained in volume $V$, at temperature $T$ and derive expressions for molecular distribution functions which are of fundamental importance in the equilibrium theory of systems. The probability density that depends on momenta, positions, orientations, etc. of molecules are expressed in an ensemble in terms of a Hamiltonian. We limit our presentation to rigid linear molecules which have only positional and orientational degrees of freedom. For a system of $N$ molecules, the Hamiltonian is,
\[
H(\vec{x}^{\,N}, \vec{p}^{\,N}, \vec{J}^{\,N}) =
\sum_{i=1}^{N}
\left(
\frac{|\vec{p}_i|^2}{2m}
+
\frac{|\vec{J}_i|^2}{2I}
\right)
+ U_N(\vec{x}^{\,N})
+
\sum_{i=1}^{N} u^{(e)}(\vec{x}_i),
\tag{2.1}
\]
where vectors $\vec{p}_i$ and $\vec{J}_i$ denote, respectively, translational momentum and angular momentum perpendicular to
the molecular symmetry axis of molecule $i$; $m$ and $I$ are, respectively, molecular mass and moment of inertia. The vector $\vec{x}_i$ denotes both the position $\vec{r}_i$ of the centre of mass and orientation $\vec{\Omega}_i$ of molecule $i$. For a linear molecule, $\vec\Omega$ is described by Euler angles $\theta,\phi$. In Eq.~(2.1) $u^{(e)}(\vec{x}_i)$ is the potential energy of a molecule at $\vec{x}_i$ due to external forces and $U_N(\vec{x}^{\,N})$ is the total potential energy   due to interactions between molecules of the system in a configuration $\vec{x}^{\,N} = \vec{x}_1, \vec{x}_2, \vec{x}_3, \ldots, \vec{x}_N$. It is common to assume $U(\vec{x}^{\,N})$ as a sum of pairs,
\begin{equation}
U_N\!\left(\vec{x}^{\,N}\right)
=
\sum_{i<j}^{N}
u\!\left(\vec{x}_i,\vec{x}_j\right),
\label{eq:pairwise}
\tag{2.2}
\end{equation}
where $u(\vec{x}_i,\vec{x}_j)$ is the intermolecular potential energy of a pair of molecules $i$ and $j$. In the grand canonical ensemble, the equilibrium phase-space probability density $f_0$ is a function of the number of molecules $N$ as well as phase variables
$\vec{x}^{\,N}$, $\vec{p}^{\,N}$, $\vec{J}^{\,N}$ and expressed as ~\cite{Hansen2006,GrayGubbins1984}
\begin{equation}
f_0\!\left(\vec{x}^{\,N},\vec{p}^{\,N},\vec{J}^{\,N};N\right)
=
\frac{1}{\Xi}
\exp\!\left[-\beta\left(H-\mu N\right)\right],
\label{eq:gc_dist}
\tag{2.3}
\end{equation}
where $\mu$ is the chemical potential, and
$\beta = (k_B T)^{-1}$ ($k_B$ is the Boltzmann constant).
The grand partition function $\Xi$ is expressed as
\begin{equation}
\Xi(\mu,V,T) = \sum_{N\ge0} e^{\beta \mu N} Q_N(V,T),
\label{eq:grand_pf}
\tag{2.4}
\end{equation}
where $Q_N = Q_N^{(t)} Q_N^{(r)} Q_N^{(c)}$ are canonical partition
functions and are written as ~\cite{GrayGubbins1984}
\begin{align}
Q_N^{(t)} &=
\frac{1}{h^{3N}}
\int d\vec{p}^{\,N}
\exp\!\left[
-\beta \sum_{i=1}^{N} \frac{p_i^{2}}{2m}
\right]
=
\left(\frac{2\pi m}{\beta h^2}\right)^{\frac{3N}{2}}
= \Lambda_t^{-3N},
\tag{2.5a}
\label{eq:Qt}
\\[6pt]
Q_N^{(r)} &=
\frac{(4\pi)^N}{h^{2N}}
\int d\vec{J}^{\,N}
\exp\!\left[
-\beta \sum_{i=1}^{N} \frac{J_i^{2}}{2I}
\right]
=
\left(\frac{8\pi^2 I}{\beta h^2}\right)^N
= \Lambda_r^{-N},
\tag{2.5b}
\label{eq:Qr}
\\[6pt]
Q_N^{(c)} &=
\frac{1}{(4\pi)^N\,N!}
\int d\vec{x}^{\,N}
\exp\!\left[
-\beta U_N(\vec{x}^{\,N})
-\beta \sum_{i=1}^{N} u^e(\vec{x}_i)
\right].
\tag{2.5c}
\end{align}
where $h$ is the Planck constant. The quantity $\Lambda_t = (\frac{\beta h^2}{2\pi m})^{\frac{1}{2}}$ is the thermal de Broglie wavelength of the
molecule, and $\Lambda_r^{-1}$, where $\Lambda_r = \frac{\beta h^2}{8\pi^2 I}$, is the rotational partition function for a linear molecule. The quantity $Q_N^{(c)}$ is the configurational canonical partition function. In Eq.~(2.5c) volume element $d\vec{x}$ for linear molecules is $d\vec{r}\, d{\vec \Omega}$., where $d{\Omega} = \sin\theta \, d\theta \, d\phi$. Combining above relations one gets following expression for the grand partition function,
\[
\Xi\,(\mu, V, T) = \exp \left[ W(\mu, V, T)\right]
\]
\[
= \sum_{N\ge0} \,\,  \frac{1}{N!}
\int d\vec{x}^{\,N}
\left( \prod_{i=1}^{N} z(\vec{x}_i) \right)
\exp(-\beta U_N),
\tag{2.6}
\]
where
\begin{equation}
z(\vec{x}_i)
=
\frac{1}{\Lambda^3}
\exp\!\left(\beta\mu - \beta u^{(e)}(\vec{x})\right)
=
\frac{1}{\Lambda^3}
\exp\!\left(\psi(\vec{x})\right),
\tag{2.7}
\label{eq:z_def}
\end{equation}
with  $\Lambda^3 = \Lambda_t^3 \Lambda_r $. The quantity $\psi(\vec{x}) = \beta (\mu - u^{(e)}(\vec{x}))$ is the reduced intrinsic chemical potential, and $-W = -\ln \Xi$ is the grand thermodynamic potential; both $\psi$ and $W$ are expressed in units of $k_B T$.
\subsection{Particle densities and distribution functions}
The equilibrium phase-space probability density $f_0$ defined by Eq (2.3) can be factorized  into the kinetic and potential terms ~\cite{Hansen2006}. This factorization is used to define $m$–particle density.
\[
\rho^{(m)}(\vec{x}^{\,m})
= \frac{1}{\Xi}
\sum_{N \ge m} \frac{1}{(N-m)!}
\int d\vec{x}^{\,N-m}
\left( \prod_{i=1}^{N} z(\vec{x}_i) \right)
\exp(-\beta U_N),
\tag{2.8}
\]
where $\rho^{(m)}(\vec{x}^{\,m})\, d\vec{x}^{\,m}$ gives probability of
finding $m$ molecules of the system within volume element
$d\vec{x}^{\,m} = d\vec{r}^{\,m} d\vec{\Omega}^{\,m}$ irrespective of positions and orientations of remaining $N-m$ molecules and independent of all angular and translational momenta.\\
From Eq.~(2.8) one gets for one–particle density,
\[
\rho^{(1)}(\vec{x})
= \frac{1}{\Xi}
\sum_{N \ge 1} \frac{1}{(N-1)!}
\int d\vec{x}^{\,N-1}
\left( \prod_{i=1}^{N} z(\vec{x}_i) \right)
\exp(-\beta U_N)
\]
\[
= \left\langle \sum_{i=1}^{N} \delta(\vec{x}-\vec{x}_i) \right\rangle
\]
\[
= \left\langle \sum_{i=1}^{N}
\delta(\vec{r}-\vec{r}_i)\,
\delta(\vec{\Omega}-\vec{\Omega}_i) \right\rangle
\tag{2.9}
\]
\[
= \langle \hat{\rho}(\vec{r},\vec{\Omega}) \rangle
= \rho(\vec{x}),
\]
where angular bracket indicates ensemble average and $\hat\rho(\vec{x})$ is the density operator. 
The quantity $\rho(\vec{x})$ gives probability of finding a
molecule at position $\vec{r}$ with orientation $\vec{\Omega}$ and
can be a measure of inhomogeneity at the length of
molecular diameter. 
Similarly two–particle density is
\[
\rho^{(2)}(\vec{x},\vec{x'})
= \frac{1}{\Xi}
\sum_{N \ge 2} \frac{1}{(N-2)!}
\int d\vec{x}^{\,N-2}
\left( \prod_{i=1}^{N} z(\vec{x}_i) \right)
\exp(-\beta U_N)
\]
\[
= \left\langle
\sum_{i=1}^{N}\sideset{}{'}\sum_{j=1}^{N}
\delta(\vec{x}-\vec{x}_i)\,
\delta(\vec{x}'-\vec{x}_j)
\right\rangle 
\]
\[
= \left\langle
\sum_{i=1}^{N} \sideset{}{'}\sum_{j=1}^{N}
\delta(\vec{r}-\vec{r}_i)\,
\delta(\vec{\Omega}-\vec{\Omega}_i)\,
\delta(\vec{r}'-\vec{r}_j)\,
\delta(\vec{\Omega}'-\vec{\Omega}_j)
\right\rangle .
\tag{2.10}
\]
Here prime on summation indicates that $i \neq j$.

\subsection{Grand Potential as a generating functional for density--density correlation functions}

From Eqs. (2.6) and (2.7) it is obvious that $\Xi$ and $W$
depend on values taken by $\psi(\vec{x})$ in the whole volume $V$,
they are functional of $\psi(\vec{x})$. To emphasize this fact,
they are sometimes written as
\[
\Xi = \Xi[\psi], \qquad W = W[\psi]
\tag{2.11}
\]
When one takes functional derivative of $W$ with respect to
$\psi(\vec{x})$, one gets one–particle density $\rho(\vec{x})$
defined by Eq.~(2.9),
\[
\frac{\delta W[\psi]}{\delta \psi(\vec{x}_i)}
= z(\vec{x}_i)\,\frac{1}{\Xi}\,
\frac{\delta \Xi[\psi]}{\delta z(\vec{x}_i)}
\tag{2.12}
\]
\[
= \frac{1}{\Xi}
\sum_{N \ge 1} \frac{z(\vec{x}_i)}{(N-1)!}
\int d\vec{x}^{\,N-1}
\left( \prod_{j \ne i}^{N} z(\vec{x}_j) \right)
\exp(-\beta U_N)
\]
\[
= \langle \hat{\rho}(\vec{x}_i) \rangle
= \rho(\vec{x}_i)
\tag{2.13}
\]
Since the functional derivative determines the change in $W$ resulting from a change in $\psi$ at a particular value of variable $\vec{x}_i$, $\rho(\vec{x}_i)$ is the value of one–particle density at the point $\vec{x}_i$ in the system and as it measures the inhomogeneity at the molecular scale it may be sharply peaked. On the other hand, in a homogeneous isotropic system, $\rho(\vec{x})$ is independent of variable $\vec{x}$ and is simply the number density $\rho = \langle N \rangle / V$, where $\langle N \rangle$ is the average number of molecules in the system. The second derivative yields
\[
\frac{\delta^2 W[\psi]}
{\delta \psi(\vec{x}_i)\, \delta \psi(\vec{x}_j)}
= \frac{\delta \rho(\vec{x}_i)}{\delta \psi(\vec{x}_j)}
= \frac{\delta}{\delta \psi(\vec{x}_j)}
\left[
z(\vec{x}_i)\,\frac{1}{\Xi}\,
\frac{\delta \Xi[\psi]}{\delta z(\vec{x}_i)}
\right]
\tag{2.14}
\]
\[
= z(\vec{x}_i)\, z(\vec{x}_j)
\left[
\frac{1}{\Xi}\,
\frac{\delta^2 \Xi[\psi]}
{\delta z(\vec{x}_i)\, \delta z(\vec{x}_j)}
- \frac{1}{\Xi^2}\,
\frac{\delta \Xi[\psi]}{\delta z(\vec{x}_i)}\,
\frac{\delta \Xi[\psi]}{\delta z(\vec{x}_j)}
\right]
\]
\[
= \langle \hat{\rho}(\vec{x}_i)\, \hat{\rho}(\vec{x}_j) \rangle
- \langle \hat{\rho}(\vec{x}_i) \rangle
  \langle \hat{\rho}(\vec{x}_j) \rangle
\tag{2.15}
\label{rhoxixj}
\]
\[
= \langle \delta \hat{\rho}(\vec{x}_i)\,
\delta \hat{\rho}(\vec{x}_j) \rangle
= H^{(2)}(\vec{x}_i,\vec{x}_j).
\tag{2.16}
\]
In Eq.~(2.15)
$\langle \hat\rho(\vec{x}_i)\,\hat\rho(\vec{x}_j) \rangle$ gives probability of
finding simultaneously two molecules in volume elements
$d\vec{x}_i$ and $d\vec{x}_j$ and can be written as
\begin{align*}
n(\vec{x}_1,\vec{x}_2)
&=
\langle \hat\rho(\vec{x}_1)\,\hat\rho(\vec{x}_2) \rangle
\\[6pt]
&=
\left\langle
\sum_i \sum_j
\delta(\vec{x}_1 - \vec{x}_i)\,
\delta(\vec{x}_2 - \vec{x}_j)
\right\rangle
\tag{2.17}
\\[6pt]
&=
\left\langle
\sum_i \sum_{j\ne i}
\delta(\vec{x}_1 - \vec{x}_i)\,
\delta(\vec{x}_2 - \vec{x}_j)
\right\rangle
+
\rho(\vec{x}_1)\,
\delta(\vec{x}_1 - \vec{x}_2)
\\[6pt]
&=
\rho^{(2)}(\vec{x}_1,\vec{x}_2)
+
\rho(\vec{x}_1)\,
\delta(\vec{x}_1 - \vec{x}_2).
\tag{2.18}
\label{eq:rho2_def}
\end{align*}
The quantity $\rho^{(2)}(\vec{x}_1,\vec{x}_2)$ gives the probability of finding simultaneously two distinct molecules in volume elements $d\vec{x}_1$ and $d\vec{x}_2$. Two particle distribution function $g^{(2)}(\vec{x}_1,\vec{x}_2)$ is  defined as ~\cite{Hansen2006}
\begin{equation}
g^{(2)}(\vec{x}_1,\vec{x}_2)
=
\frac{\rho^{(2)}(\vec{x}_1,\vec{x}_2)}
{\rho(\vec{x}_1)\,\rho(\vec{x}_2)}.
\tag{2.19a}
\end{equation}
It is expected that $g^{(2)}(\vec{x}_1,\vec{x}_2)$ decays to $1$ as spatial separation between two molecules tends to infinity. This leads to the definition of the total pair correlation function, $h^{(2)}(\vec{x}_1,\vec{x}_2)$,
\begin{equation*}
h^{(2)}(\vec{x}_1,\vec{x}_2)
= g^{(2)}(\vec{x}_1,\vec{x}_2) - 1
\tag{2.19b}
\end{equation*}
and from Eqs. (2.16)-(2.19),
\begin{equation}
H^{(2)}(\vec{x}_1,\vec{x}_2)
=
\rho(\vec{x}_1)\,\rho(\vec{x}_2)
\left[
1 + h^{(2)}(\vec{x}_1,\vec{x}_2)
\right]
+
\rho(\vec{x}_1)\,
\delta(\vec{x}_1 - \vec{x}_2).
\tag{2.20}
\end{equation}\\
One can continue taking derivatives of $W[\psi]$ with respect to $\psi$ to derive expressions for $H^{(n)}$, $\rho^{(n)}$ and $g^{(n)}$ for $n$--particles,where $n$ can be any number. We note that $W[\psi]$, which has $\psi(\vec{x})$ as its natural variable, acts as a generating functional for density-density correlations. One can use these relations to derive expressions for the thermodynamic potentials. However, working with $\psi(\vec{x})$ as a field variable is found to be inconvenient and is often avoided. Instead, one prefers to work with generating functional where field variable is $\rho(\vec{x})$.

\subsection{Intrinsic free--energy and direct correlation functions}
The intrinsic free energy, $A$ which is functional
of $\rho(\vec{x})$ and is free from $\psi(\vec{x})$ is defined as  
\[
A[\rho] = -W([\psi]) + \int d\vec{x}\, \psi(\vec{x})\, \rho(\vec{x})
\tag{2.21}
\]
The relation (2.21) is the Legendre transformation which
relate functional $A[\rho]$ with $W[\psi]$. The Hohenberg–Kohn–Mermin ~\cite{HohenbergKohn1964,Mermin1965,Ebner1976} theorem states that at fixed values of $\mu, V$ and $T$ the minimum of the functional $W([\psi],[\rho])$ occurs when $\rho(\vec{x})$ is the equilibrium one–particle
density, i.e.
\[
\left.
\frac{\delta W([\psi],[\rho])}{\delta \rho(\vec{x})}
\right|_{\mu,T,V}
= 0 \qquad (\text{min})
\tag{2.22}
\]
The  quantity $\rho(\vec{x})$ at the minimum of $W$ is the equilibrium
density and  $W[\psi] = min \, \rho(x) $
$ W([\psi],[\rho])$ is
the equilibrium value of the grand potential.
Using Eq.(2.22) we find from Eqs.(2.21) and (2.13) that at equilibrium 
\[
\left.
\frac{\delta A[\rho]}{\delta \rho(\vec{x})}
\right|_{\mu,T,V}
= -
\left.
\frac{\delta W([\psi],[\rho])}{\delta \rho(\vec{x})}
\right|_{\mu,T,V}
+ \psi(\vec{x})
\tag{2.23a}
\]
\[
= \psi(\vec{x})
\]
and
\[
\left.
\frac{\delta A[\rho]}{\delta \psi(\vec{x})}
\right|_{\mu,T,V}
= -
\frac{\delta W([\psi])}{\delta \psi(\vec{x})}
+ \rho(\vec{x})
= 0
\tag{2.23b}
\]
The Helmholtz free energy  $F$ is related with the intrinsic free energy as,
\[
\beta F = A +\beta \int d\vec{x} \, u^{(e)}(x) \, \rho(\vec x)
\tag{2.24}
\]
For evaluation of the intrinsic free-energy, one can use the factorization and evaluate ideal and excess parts separately. The ideal part is evaluated exactly as its partition function $\Xi (z, V, T)$ reduces to 
\[
\Xi_{\mathrm{id}}
= \sum_{N=0}^{\infty} \frac{1}{N!}
\left[ \int d\vec{x}\, z(\vec{x}) \right]
= \exp \left[ \int d\vec{x}\, z(\vec{x}) \right]
\tag{2.25a}
\]
and from Eq~(2.6) 
\[
W_{\mathrm{id}}
= \int d\vec{x}\, \frac{1}{\Lambda^3}\, \exp \psi(\vec{x}) .
\tag{2.25b}
\]
The functional derivative of (2.25) gives
one–particle density
\[
\frac{\delta W_{\mathrm{id}}}{\delta \psi(\vec{x})}
= \rho(\vec{x})
= \frac{1}{\Lambda^3} \exp \psi(\vec{x}) .
\tag{2.26}
\]
Eq~(2.26) provides a relation between the intrinsic chemical potential and one--particle density for an ideal gas,

\[
\psi(\vec{x}) = \ln \big( \rho(\vec{x}) \Lambda^3 \big).
\tag{2.27}
\]
Substituting Eq. (2.27) into Eq. (2.23) and performing functional integration, one finds the following expression for the ideal part of the intrinsic free energy
\[
A_{\mathrm{id}}[\rho]
= \int d\vec{x}\, \rho(\vec{x})
\left[ \ln \big( \rho(\vec{x}) \Lambda^3 \big) - 1 \right] 
\tag{2.28}
\]
\\
The  total intrinsic free energy $A$ of the system  is a sum of ideal part
$A_{id}$ and excess part $\Delta A$, arising due to intermolecular
interactions,
\begin{equation}
A = A_{id} + \Delta A   \,\,.
\tag{2.29}
\end{equation}
We now use relations (2.23),(2.28) and (2.29) to write
\[
\psi(\vec{x})
= \frac{\delta (A_{\mathrm{id}} + \Delta A)}{\delta \rho(\vec{x})}
= \ln \big( \rho(\vec{x}) \Lambda^3 \big) - c^{(1)}(\vec{x}) .
\tag{2.30a}
\]
or
\[
\rho(\vec{x})
= \frac{1}{\Lambda^3}
\exp \big[ \psi(\vec{x}) + c^{(1)}(\vec{x}) \big] 
\tag{2.30b}
\]
where
\[
\frac{\delta \Delta A[\rho]}{\delta \rho(\vec{x})}
= -\, c^{(1)}(\vec{x},[\rho]) 
\tag{2.31}
\]
Eq.(2.31) defines the one-particles direct correlation function $c^{(1)}(x)$. Eq. (2.30b) suggests that $-k_BTc^{(1)}(\vec x)$ is a \textit{system induced effective potential} acting at $\vec x$.\\

Two-particle direct correlation function is defined as the second functional derivative of $\Delta A$. Thus,
\begin{equation}
\frac{\delta^2 \Delta A[\rho]}{\delta \rho(\vec{x}_1)\,\delta \rho(\vec{x}_2)}
= -\frac{\delta c^{(1)}(\vec{x}_1,[\rho])}{\delta \rho(\vec{x}_2)}
= -c^{(2)}(\vec{x}_1,\vec{x}_2,[\rho]) \tag{2.32}
\end{equation}
From Eqs.~(2.30) and (2.32) one gets for second derivative of $A$,
\begin{equation*}
\frac{\delta \psi(\vec{x}_1)}{\delta \rho(\vec{x}_2)}
= \frac{\delta^2 (A_{\mathrm{id}}+\Delta A[\rho])}{\delta \rho(\vec{x}_1)\,\delta \rho(\vec{x}_2)}
= \frac{\delta^2 A[\rho]}{\delta \rho(\vec{x}_1)\,\delta \rho(\vec{x}_2)}
\end{equation*}
\begin{equation}
= \frac{\delta(\vec{x}_1-\vec{x}_2)}{\rho(\vec{x}_1)}
- c^{(2)}(\vec{x}_1,\vec{x}_2)
= K^{(2)}(\vec{x}_1,\vec{x}_2) 
\tag{2.33}
\end{equation}
Similarly, the $n$th-order derivatives of $A[\rho]$ leads to
\begin{equation*}
\frac{\delta^n A[\rho]}{\prod_{i=1}^{N}\delta \rho(\vec{x}_i)}
= \frac{(-1)^n (n-2)!}{[\rho(\vec{x}_1)]^{\,n-1}}
\,\delta(\vec{x}_1-\vec{x}_2)\cdots\delta(\vec{x}_{n-1}-\vec{x}_n)
- c^{(n)}(\vec{x}_1,\vec{x}_2,\ldots,\vec{x}_n)
\end{equation*}
\begin{equation}
= K^{(n)}(\vec{x}_1,\vec{x}_2,\ldots,\vec{x}_n) \tag{2.34}
\end{equation}
Note that in Eqs.~(2.33) and (2.34) first term arises due to ideal gas part
$A_{\mathrm{id}}$ and second part due to $\Delta A$. Thus $\Delta A$, acts as a generating functional for direct correlation functions;
\[
\frac{\delta^n \Delta A[\rho]}{\prod_{i=1}^n \delta \rho(\vec{x}_i)}
=\frac{\delta^{n-2} c^{(n)}(\vec{x}_1,\vec{x}_2)}{\prod_{i=3}^n \delta \rho(\vec{x}_i)}
= -\,c^{(n)}(\vec{x}_1,\ldots,\vec{x}_n;[\rho])
\tag{2.35}
\]
The excess free energy $\Delta A$ can also be expressed in terms of
two-particle density. From relations (2.6) and (2.21) one gets
\[
\frac{\delta W}{\delta \beta u(\vec{x_1},\vec{x_2})}
=-\frac{\delta \Delta A[\rho]}{\delta \beta u(\vec{x}_1,\vec{x}_2)}
= -\frac{1}{2}\,\rho^{(2)}(\vec{x}_1,\vec{x}_2;[\rho,u])
\]
\[
= -\frac{1}{2}\,\rho(\vec{x}_1)\,\rho(\vec{x}_2)\,
g(\vec{x}_1,\vec{x}_2;[\rho,u]).
\tag{2.36}
\]

\subsection{Orstein--Zernike Equations}
Using the identity
\begin{equation*}
\int d\vec{x}_3 \,
\frac{\delta \psi(\vec{x}_1)}{\delta \rho(\vec{x}_3)}
\frac{\delta \rho(\vec{x}_3)}{\delta \psi(\vec{x}_2)}
= \delta(\vec{x}_1-\vec{x}_2),
\end{equation*}
and Eqs.~(2.16) and (2.33) one gets
\begin{equation}
\int d\vec{x}_3 \,
K^{(2)}(\vec{x}_1,\vec{x}_3)\,
H^{(2)}(\vec{x}_3,\vec{x}_2)
= \delta(\vec{x_1}-\vec{x_2})
\tag{2.37}
\end{equation}
which leads to
\begin{equation}
h^{(2)}(\vec{x}_1,\vec{x}_2)
= c^{(2)}(\vec{x}_1,\vec{x}_2)
+ \int d\vec{x}_3 \,
c^{(2)}(\vec{x}_1,\vec{x}_3)\,\rho(\vec{x_3})
h^{(2)}(\vec{x}_3,\vec{x}_2),
\tag{2.38}
\end{equation}
This is the \textit{Ornstein--Zernike} (OZ) equation relating total and direct-pair correlation functions and is one of the most important relations in the theory of molecular systems. Note that the OZ equation is exact, valid in both ordered and disordered phases of the system.\\
The above procedure can be extended to show that $H^{(n)}$ and $K^{(n)}$ satisfy the generalized OZ equations involving $n$-order correlation functions ~\cite{Stell1976}. For example for $n=3$,
\begin{equation}
H^{(3)}(\vec{x}_1,\vec{x}_2,\vec{x}_3)
= \int d\vec{x}_4\, d\vec{x}_5\, d\vec{x}_6\;
K^{(3)}(\vec{x}_4,\vec{x}_5,\vec{x}_6)\,
H^{(2)}(\vec{x}_1,\vec{x}_4)\,
H^{(2)}(\vec{x}_2,\vec{x}_5)\,
H^{(2)}(\vec{x}_3,\vec{x}_6)
\tag{2.39}
\end{equation}
These generalized OZ equations are exact, valid in both ordered and
disordered phases. They are useful in formulation of a wide variety of problems connected with many variables and have counterparts in Euclidean field theories like Schwinger and Green’s functions. However, their use in determining correlation functions has so far been limited only to pair function of isotropic phase. Since in an isotropic phase $\rho(\vec{x})$ is independent of  variable $\vec{x}$, Eq.~(2.38) reduces to 
\begin{equation}
h^{(2)}(\vec{x}_1,\vec{x}_2)
= c^{(2)}(\vec{x}_1,\vec{x}_2)
+ \rho \int d\vec{x}_3\,
h^{(2)}(\vec{x}_1,\vec{x}_3)\,
c^{(2)}(\vec{x}_2,\vec{x}_3)
\tag{2.40}
\end{equation}
Since Eq.~(2.40) involves two unknowns, $h$ and $c$, one needs one more equation
involving these functions  to determine values of functions $h$ and $c$. This equation is often referred to as \textit{closure relation} and are derived in Appendix A.\\

Using the fact that space is homogeneous and isotropic,one can derive important relations relating one--body density to pair correlation functions. Because of the symmetry the Hamiltonian and correlation functions remain invariant if molecules and the external field are simultaneously spatially displaced or rotated by a fixed amount. This leads to the following relations:
\begin{equation}
\nabla_{r_1}\!\left[\beta u^{e}(\vec{x}_1)+\ln\rho(\vec{x}_1)\right]
= \int d\vec{x}_2 \,
c^{(2)}(\vec{x}_1,\vec{x}_2)\,
\nabla_{r_2}\rho(\vec{x}_2),
\tag{2.41a}
\end{equation}
and
\begin{equation}
\nabla_{\Omega_1}\!\left[\beta u^{e}(\vec{x}_1)+\ln\rho(\vec{x}_1)\right]
= \int d\vec{x}_2 \,
c^{(2)}(\vec{x}_1,\vec{x}_2)\,
\nabla_{\Omega_2}\rho(\vec{x}_2).
\tag{2.41b}
\end{equation}
In above equations, $\nabla_r$ and $\nabla_\Omega$ are, respectively, radial
and angular gradients. In writing above relations it was assumed that when spatial displacement was given, orientations were kept fixed, and when molecules were given rotation positions were kept fixed. Eqs.~(2.41) yield integro-differential equations for single-particle density distribution and are known as Lovett equations~\cite{Lovett1976,Wertheim1976}. The first order expression for the Yvon--Born--Green (YBG) hierarchy ~\cite{BornGreen1949} also provides another way of relating one--body density to two--body distribution;
\begin{equation}
\nabla_{1} ln\rho(\vec{x}_1)+\beta \nabla_{1} u^{e}(\vec{x}_1) = -\beta \int d\vec{x}_2 \,
\nabla_{1}u(\vec{x}_1,\vec{x}_2)\,
\rho(\vec{x}_2)g(\vec{x}_1,\vec{x}_2),
\tag{2.42}
\end{equation}
where $\nabla_{1}$ gradient operator acting at molecule at $\vec{x}_1$
\bigskip
\noindent

\section{Expressions for grand potential and intrinsic free-energy in terms of direct pair correlation function}
\par
Exact expression for excess free energy $\Delta A$ is found from Eqs.~(2.31) and (2.32) by performing functional integrations. The
integration takes the systems from some initial density $\rho_0(\vec{x})$ to the final density
$\rho(\vec{x})$ along a path defined by a parameter $\lambda$ such that
$\rho(\vec{x},\lambda)=\rho_0(\vec{x})+\lambda\,\Delta\rho(\vec{x})$ where
$\Delta\rho(\vec{x})=\rho(\vec{x})-\rho_0(\vec{x})$. The result depends only on the initial and final densities, not on the chosen path~\cite{Ebner1976}.
\par
The integration of Eq.~(2.31) yields ~\cite{Hansen2006}
\begin{align}
\Delta A[\rho(\vec{x})]-\Delta A[\rho_0(\vec{x})]
&= -\int_0^1 d\lambda \int d\vec{x}_1
\frac{\partial \rho(\vec{x}_1,\lambda)}{\partial \lambda}
\,c^{(1)}(\vec{x},\lambda) \nonumber \\
&= -\int_0^1 d\lambda \int d\vec{x}_1
\Delta\rho(\vec{x}_1)\,c^{(1)}(\vec{x},\lambda),
\tag{3.1}
\end{align}
Similarly, one gets from the integration of Eq.~(2.32),
\begin{equation}
c^{(1)}(\vec{x}_1,\lambda)-c^{(1)}(\vec{x}_1)
= \int_0^{\lambda} d\lambda' \int d\vec{x}_2
\Delta\rho(\vec{x}_2)\,
c^{(2)}(\vec{x}_1,\vec{x}_2,\lambda'),
\tag{3.2}
\end{equation}
Substituting Eq.~(3.2) into Eq.~(3.1) one gets
\begin{align}
\Delta A[\rho(\vec{x})]
&= \Delta A[\rho_0(\vec{x})]
- \int d\vec{x}_1 \Delta\rho(\vec{x}_1)\,
c^{(1)}(\vec{x}_1,[\rho_0(\vec{x}_1)]) \nonumber \\
&\quad - \int_0^1 d\lambda \int_0^{\lambda} d\lambda'
\int d\vec{x}_1 d\vec{x}_2
\Delta\rho(\vec{x}_1)\Delta\rho(\vec{x}_2)\,
c^{(2)}(\vec{x}_1,\vec{x}_2,\lambda') \nonumber \\
&= \Delta A[\rho_0(\vec{x})]
- \int d\vec{x}_1 \Delta\rho(\vec{x}_1)\,
c^{(1)}(\vec{x}_1,[\rho_0(\vec{x}_1)]) \nonumber \\
&\quad - \frac{1}{2}
\int d\vec{x}_1 d\vec{x}_2
\Delta\rho(\vec{x}_1)\Delta\rho(\vec{x}_2)\,
\bar{c}(\vec{x}_1,\vec{x}_2),
\tag{3.3}
\end{align}
where
\begin{align}
\bar{c}(\vec{x}_1,\vec{x}_2)
&= 2\int_0^1 d\lambda \int_0^{\lambda} d\lambda'
\,c^{(2)}(\vec{x}_1,\vec{x}_2,\lambda') \nonumber \\
&= 2\int_0^1 d\lambda\,(1-\lambda)\,
c^{(2)}\!\left(\vec{x}_1,\vec{x}_2;
\rho_0(\vec{x})+\lambda\Delta\rho(\vec{x})\right).
\tag{3.4}
\end{align}
For notational convenience, we use only one integration sign unless otherwise mentioned to
indicate the possible multiple integrations over all those variables whose volume element appear.\\
Combining ideal gas part with Eq.~(3.3) for $\Delta A$, the intrinsic energy $A$ becomes 
\begin{align}
A[\rho(\vec{x})]
&= \int d\vec{x}\, \rho(\vec{x})
\left[ \ln \left( \rho(x)\,\Lambda^3 \right) - 1 \right]
+ \Delta A[\rho_0(\vec{x})]
- \int d\vec{x}_1 \Delta\rho(\vec{x}_1)\,
c^{(1)}\!\left(\vec{x}_1,[\rho_0(\vec{x}_1)]\right)
\nonumber \\
&\quad - \frac{1}{2}
\int d\vec{x}_1 d\vec{x}_2
\Delta\rho(\vec{x}_1)\Delta\rho(\vec{x}_2)\,
\bar{c}(\vec{x}_1,\vec{x}_2),
\tag{3.5}
\end{align}
The grand potential in the absence of the external potential $u^{(e)}(\vec{x})$ is defined as (see Eq.~(2.21))
\begin{equation}
-\widetilde{W}
= A[\rho] - \beta\mu \int d\vec{x}\,\rho(\vec{x}).
\tag{3.6}
\end{equation}
Using Eq.~(3.5) for $A[\rho]$, we get
\begin{align}
-\widetilde{W}
&= \int d\vec{x}\, \rho(\vec{x})
\left[ \ln \left( \rho(x)\,\Lambda^3 \right) - 1 \right]
- \beta\mu \int d\vec{x}\, \rho(\vec{x})
+ \Delta A[\rho_0(\vec{x})]
\nonumber \\
&\quad - \int d\vec{x}_1 \Delta\rho(\vec{x}_1)\,
c^{(1)}\!\left(\vec{x}_1,[\rho_0(\vec{x}_1)]\right)
- \frac{1}{2}
\int d\vec{x}_1 d\vec{x}_2
\Delta\rho(\vec{x}_1)\Delta\rho(\vec{x}_2)\,
\bar{c}(\vec{x}_1,\vec{x}_2),
\tag{3.7}
\end{align}
The initial choice of density $\rho_0(\vec{x})$ depends on the problem under consideration. If one takes $\rho_0(\vec{x})$ equal to zero, then from
Eq.~(3.3),
\begin{equation}
\Delta A[\rho(\vec{x})]
= -\frac{1}{2}
\int d\vec{x}_1 d\vec{x}_2
\rho(\vec{x}_1)\rho(\vec{x}_2)\,
\bar{c}(\vec{x}_1,\vec{x}_2),
\tag{3.8}
\end{equation}
and the total intrinsic energy:
\begin{equation}
A[\rho]
= \int d\vec{x}\, \rho(\vec{x})
 \left[\ln \left( \rho(x)\,\Lambda^3 \right) - 1 \right]
- \frac{1}{2}
\int d\vec{x}_1 d\vec{x}_2
\rho(\vec{x}_1)\rho(\vec{x}_2)\,
\bar{c}(\vec{x}_1,\vec{x}_2),
\tag{3.9}
\end{equation}
where
\begin{equation}
\bar{c}(\vec{x}_1,\vec{x}_2)
= 2 \int_0^1 d\lambda\,(1-\lambda)\,
c^{(2)}\!\left(\vec{x}_1,\vec{x}_2;\lambda\rho(\vec{x})\right).
\tag{3.10}
\end{equation}
The grand potential $\tilde{W}$ becomes
\begin{equation}
\begin{aligned}
-\tilde{W} = & \int d\vec{x} \ \rho(\vec{x}) [ \ln(\rho(\vec{x})\Lambda^3) - 1 ] - \beta \mu \int \rho(\vec{x}) d\vec{x} \\
& - \frac{1}{2} \int d\vec{x}_1 d\vec{x}_2 \ \rho(\vec{x}_1) \rho(\vec{x}_2) \ \bar{c}(\vec{x}_1, \vec{x}_2)
\end{aligned} \tag{3.11}
\end{equation}

Above equations of intrinsic free energy and grand potential are exact, and can be used to investigate properties including phase transitions of different phases of a molecular system. However, these equations involve one-particle density and direct pair correlation function (DPCF). To evaluate value of $\bar{c}(\vec{x}_1, \vec{x}_2)$ from Eq (3.10) one needs to know values of DPCF for all values of $\rho(\vec{x})$ along the path of integration which covers both disordered and ordered phases. Before we discuss DPCF of ordered phases we identify order parameters characterizing the ordering in the systems.

\section{One--particle density and order parameters}
In experimental studies of molecular systems, one uses techniques such as X-ray and neutron scattering, NMR, IR and Raman spectroscopy, textures and miscibility criteria to distinguish one phase from the other ~\cite{Chaikin1995,deGennes1993,Chandrasekhar1992}. Phases are distinguished on the basis of ordering in distribution of molecules in the system. It is natural, therefore, to introduce in a general way the order parameters as expansion coefficients of one-particle density in a suitable basis set. From Eq.(2.9), we have
\begin{equation}
\rho(\mathbf{x}) \equiv \rho(\mathbf{r},\mathbf{\Omega})
= N \left\langle \delta(\mathbf{r}-\mathbf{r}\,')\,
\delta(\mathbf{\Omega}-\mathbf{\Omega}\,') \right\rangle ,
\tag{4.1}
\end{equation}
where ensemble average is over the primed variables.\\ Using the Fourier integral representation of the spatial Dirac function,
\begin{equation}
\delta(\mathbf{r}) = \frac{1}{V}\sum_{{G}} \exp(i\mathbf{G}\cdot\mathbf{r}),
\tag{4.2}
\end{equation}
where $\mathbf{G}$ are the reciprocal lattice vectors of a positionally ordered structure,
and for the angular Dirac function,
\begin{equation}
\delta(\mathbf{\Omega}-\mathbf{\Omega}\,') =
\sum_{lmn} (2l+1)\,
D^{\,l}_{mn}(\mathbf{\Omega})\,
D^{\,l*}_{mn}(\mathbf{\Omega}\,'),
\tag{4.3}
\end{equation}
where the functions $D^{\,l}_{mn}(\mathbf{\Omega})$ are the Wigner rotation matrices, which satisfy the following orthogonality condition~\cite{GrayGubbins1984} :
\begin{equation}
\int d\mathbf{\Omega}\,
D^{\,l'*}_{m'n'}(\mathbf{\Omega})\,
D^{\,l}_{mn}(\mathbf{\Omega})
=
\frac{8\pi^{2}}{2l+1}\,
\delta_{ll'}\,\delta_{mm'}\,\delta_{nn'} ,
\tag{4.4}
\end{equation}
one gets
\begin{equation}
\rho(\mathbf{r},\mathbf{\Omega})
=
\rho \sum_{\mathbf{G}} \sum_{lmn}
Q^{\,l}_{mn}(\mathbf{G})\,
\exp(i\mathbf{G}\cdot\mathbf{r})\,
D^{\,l}_{mn}(\mathbf{\Omega}),
\tag{4.5}
\end{equation}
and
\begin{equation}
Q^{\,l}_{mn}(\mathbf{G})
=
\frac{2l+1}{8\pi^{2}N}
\int d\mathbf{r}\, d\mathbf{\Omega}\,
\rho(\mathbf{r},\mathbf{\Omega})\,
\exp(-i\mathbf{G}\cdot\mathbf{r})\,
D^{\,l*}_{mn}(\mathbf{\Omega}) .
\tag{4.6}
\end{equation}
The quantity $Q^{\,l}_{mn}(\mathbf{G})$ is the order parameter and $\rho$ the mean number density of the system. Above  expressions are for a system of rigid molecules of arbitrary symmetry described by Euler angles $\theta, \phi$ and  $\psi$ and $d\mathbf{\Omega} = sin\theta d\phi d\theta d\psi$, $\int d\mathbf{\Omega} = 8\pi^2 $

From Eq. (4.6) we can identify order parameters which define positional and orientational ordered structures ~\cite{Singh1991,Zannoni1979}. For example,
\[
Q_{000}(0) = 1
\tag{4.7}
\]
corresponds to the fluid phase, and
\[
Q_{000}(G) = \mu_G
= \frac{1}{8\pi^2 N} \int d\mathbf{r} d\mathbf{\Omega} \, \rho(\mathbf{r},\mathbf{\Omega})
\exp(-i\mathbf{G}\cdot\mathbf{r})
\tag{4.8}
\]
are the positional order parameters of a monoatomic lattice. The orientational order parameters are given by 
\[
Q_{lmn}(0) = \frac{2l+1}{8\pi^2 N}
\int d\mathbf{r}\, d\Omega \,
\rho(\mathbf{r},\mathbf{\Omega}) D_{mn}^{l*}(\mathbf\Omega)
\tag{4.9}
\]
The coupling between positional and orientational ordering is measured by coupled order parameters defined as 
\[
Q_{l00}(G) = Q_{Gl} = (2l+1)\tau_{Gl} \\
= \frac{2l+1}{8\pi^2 N}
\int d\mathbf{r}\, d\mathbf\Omega \,
\rho(\mathbf{r},\mathbf{\Omega}) \exp(-i\mathbf{G}\cdot\mathbf{r})
P_l(\cos\theta)
\tag{4.10}
\]

\subsection{Orientational (nematic) ordering and order parameters}
The orientational order parameters $Q_{\ell mn}(0)$ defined by  Eq.~(4.9) can be simplified using the relation $\rho(r,\Omega) = \rho f(\Omega) $ to give,

\begin{align}
Q_{\ell mn}(0)= (2\ell+1)f_{lmn} 
&= \frac{2\ell+1}{8\pi^2}
   \int d\mathbf{\Omega}\,
   f(\mathbf\Omega)\, D^{\ell*}_{mn}(\mathbf\Omega), \nonumber 
\tag{4.11}
\end{align}
where $f(\mathbf\Omega)$ is the orientational distribution function. From Eq.~(4.5) one has

\begin{align}
f(\Omega) = \sum_{lmn} (2\ell+1)\, f_{lmn} \, D^{\ell}_{mn}(\Omega),
\tag{4.12}
\end{align}
where
\begin{align}
f_{lmn} =  < D^{\ell}_{mn}(\Omega)^*>.
\tag{4.13}
\end{align}

The $f_{lmn}$ which completely define $f(\Omega)$ are the orientational order parameters. In principle, there are $(2l + 1)^2$ order parameters for each $l$, although this number can be drastically reduced by exploiting the symmetry of the phase and of the constituent molecules. The Euler angles $\phi, \theta$ and $\chi$ in an orientationally ordered system with director $\hat{n}$ along the $z$-axis of a space-fixed coordinate frame $x, y, z$ are defined with help of three orthogonal unit vectors $\vec{a}, \vec{b}$ and $\vec{c}$ attached to a molecule in such a way that vector $\vec{c}$ is along the maximum symmetry (unique) molecular axis ~\cite{GrayGubbins1984}. Angle between the $y$-axis and a normal to $z$-$c$ plane is $\phi$, $\theta$ is the angle between the $z$-axis and the unit vector $\vec{c}$. Angle $\phi$ describes a rotation around the $z$-axis and angle $\theta$ measures deviation of the unique molecular axis from $z$-axis. The angle $\chi$ describes the rotation around the molecular unique $c$-axis.

In a uniaxial phase the orientational distribution function must be invariant under rotation about the axis of alignment of molecules denoted by unit vector $\hat{n}$ called director. If the $z$-axis of the space-fixed coordinate frame is along $\mathbf{n}$, then $m$ must be zero in $\langle D_{mn}^l(\Omega)^* \rangle$. In addition if the system has mirror plane perpendicular to the director ($D_{\infty h}$ symmetry) then only terms with even $l$ can appear in Eq. (4.13). We therefore have
\begin{equation}
f_{ln} = \langle D_{0n}^l(\theta, \chi)^* \rangle, \quad l \text{ even} \tag{4.14}
\end{equation}
Since
\begin{equation}
D_{0n}^l(\theta, \chi)^* = \left( \frac{4\pi}{2l+1} \right)^{1/2} Y_{ln}^*(\theta, \chi) \tag{4.15}
\end{equation}
Eq. (4.14) reduces to
\begin{equation}
f_{ln} = \left( \frac{4\pi}{2l+1} \right)^{1/2} \langle Y_{ln}^*(\theta, \chi) \rangle \tag{4.16}
\end{equation}
where $Y_{ln}^*(\theta, \chi)$ is the spherical harmonics. Now for each $l$ there are $(2l + 1)$ order parameters.

This number can further be reduced by exploiting the symmetry of the constituent molecules. The simplest case is that of cylindrically symmetric molecules. In this case rotation about the molecular symmetry axis should not modify the distribution $f(\Omega)$, which implies that $n=0$ in Eq. (4.13). In other words, for a cylindrically symmetric mesophase composed of cylindrically symmetric molecules the orientational distribution has to depend only on the angle $\theta$ between the director and the molecular symmetry axis. With these simplifications Eq.(4.12) can be rewritten as, 
\begin{equation}
f(\Omega) = f(\theta) = \sum_{l \text{ even}} (2l+1) f_l D_{00}^l(\Omega)^* = \sum_{l \text{ even}} (2l+1) f_l P_l(\cos \theta) \tag{4.17}
\end{equation}
where $P_l(cos\theta)$ is the Legendre polynomial of degree $l$ and
\begin{equation}
f_l = P_l=\langle D_{00}^l(\Omega)^* \rangle \equiv \langle P_l(\cos \theta) \rangle \tag{4.18}
\end{equation}
For each $l$ now we have only one order parameter. This is the case of a uniaxial nematic phase. In this phase, the centres of mass of molecules are distributed as in the fluid phase, but the molecular axes are statistically aligned along (arbitrary) direction defined by the director $\hat{n}$. The structure factor recorded by scattering experiments reflects the breaking of rotational symmetry in any plane perpendicular to the director but has only two-fold symmetry in any plane containing $\hat{n}$ ~\cite{Chaikin1995}.

Many mesogenic molecules do not possess cylindrical symmetry but may have two mirror planes that contain the $c$-axis and are mutually perpendicular. A prolate ellipsoid or a rectangular parallelepiped are examples of such molecular model. For this case the orientational distribution function must satisfy the following condition, ~\cite{Zannoni1979}
\begin{equation}
f(\vec\Omega) = f(\theta, \chi) = f(\theta, \pi - \chi) = f(\theta, \pi + \chi)
\tag{4.19}
\end{equation}
which implies that for each $l$ the number of order parameters is reduced from $(2l+1)$ to $(1 + l/2)$. For $l = 2$, the order parameters are
\begin{equation}
f_{20} = {P}_2 = \langle P_2(\cos\theta) \rangle = \langle\frac{1}{2}(3\cos^2\theta - 1)\rangle
\tag{4.20}
\end{equation}

\begin{equation}
f_{2,2} = f_{-2,2} = \eta = \left(\frac{3}{8}\right)^{1/2} \langle \sin^2\theta \cos 2\chi \rangle
\tag{4.21}
\end{equation}
For $l = 4$, the number of order parameters are three:
\begin{equation}
f_{40} = {P}_4 = \langle P_4(\cos\theta) \rangle
\tag{4.22}
\end{equation}

\begin{equation}
f_{4,2} = f_{4,-2} = \left(\frac{5}{32}\right)^{1/2} \langle \sin^2\theta (\cos^2\theta - 1)\cos 2\chi \rangle
\tag{2.23}
\end{equation}
and
\begin{equation}
f_{4,4} = f_{4,-4} = \left(\frac{35}{128}\right)^{1/2} \langle \sin^4\theta \cos 4\chi \rangle
\tag{4.24}
\end{equation}
While the order parameters ${P}_l$ measure the degree of alignment of the unique molecular axis along the $z$-axis, the other order parameters measure the difference in the degree of alignment of $\vec a$ and $\vec b$ molecular axes along the $z$-axis. This is obvious from the fact that the projection of molecular axes $\vec a$, $\vec b$, and $\vec c$ on the director $\mathbf{n}$, i.e.
\begin{equation}
c_z = \cos\theta, \quad a_z = -\sin\theta \cos\chi, \quad b_z = \sin\theta \sin\chi
\tag{4.25}
\end{equation}
For example the order parameter $\eta$ can be written as
\begin{equation}
\eta = \left(\frac{3}{8}\right)^{1/2} \left\langle a_z^2 - b_z^2 \right\rangle
\tag{4.26}
\end{equation}
which gives the difference in the alignment of the molecular $\vec a$ and $\vec b$ axes along the director.

In another example, let us consider a mesophase having symmetry of three mutually orthogonal mirror planes with inversion symmetry through their intersection. This is an example of a biaxial ordering ~\cite{LuckhurstSluckin2015}. If the constituent molecules also have the same symmetry as that of the mesophase, one needs four order parameters at $l = 2$ level to describe the orientational ordering. They are

\begin{equation*}
\bar{P}_2 = \langle D^{2}_{00}(\vec \Omega)^* \rangle = \langle P_2(\cos\theta) \rangle
\end{equation*}

\begin{equation*}
\eta = \langle D^{2}_{02}(\vec \Omega)^* \rangle = \left(\frac{3}{8}\right)^{1/2} \langle \sin^2\theta \cos 2\chi \rangle
\end{equation*}

\begin{equation*}
b = \langle D^{2}_{20}(\vec \Omega)^* \rangle = \left(\frac{3}{8}\right)^{1/2} \langle \sin^2\theta \cos 2\phi \rangle
\end{equation*}

\begin{equation}
\zeta = \langle D^{2}_{22}(\vec \Omega)^* \rangle = \frac{1}{4} \left\langle (1 + \cos\theta)^2 
\left( \cos 2\phi \cos 2\chi - \sin 2\phi \sin 2\chi \right) \right\rangle
\tag{4.27}
\end{equation}

While the first two order parameters, i.e., ${P}_2$ and $\eta$, have the same physical meaning as in the case of a uniaxial phase, the other two order parameters $b$ and $\zeta$ are the measure of the biaxial ordering in the system.

Note that a priori we have no knowledge of the convergence of the series of Eq.~(4.12) or even in the simplest case of Eq.~(4.17). One may, however, argue that since the number of zeros in the Legendre polynomials increases with their rank, one may expect that the contributions to decrease with $l$ if the distribution function is peaked at $\theta = 0$.\\
In an alternative approach one can use direction cosines $i_{\alpha}$ $(i = a,b,c$ and $\alpha = x,y,z)$ of the space-fixed and molecular-fixed coordinate frames. Using these direction cosines one defines what is called the {\it ordering matrix}.

\begin{equation}
S_{ij}^{\alpha \beta} = \frac{1}{2} \left\langle (3 i_{\alpha} j_{\beta} - \delta_{ij} \delta_{\alpha \beta}) \right\rangle
\tag{4.28}
\end{equation}
where $\delta_{\alpha \beta}$ and $\delta_{ij}$ are Kronecker delta functions. $S_{ij}^{\alpha \beta}$ is a symmetric tensor in $ij$ and $\alpha\beta$ and is traceless with respect to either pair, i.e.,

\begin{equation}
S_{ij}^{\alpha \alpha} = 0, \quad S_{ii}^{\alpha \beta} = 0
\tag{4.29}
\end{equation}

\subsection{Positional ordering and order parameters}

A crystal, in which particles are localized on lattice points, is a system of extreme inhomogeneities, where the value of density $\rho(\mathbf{r})$ shows a several orders of magnitude difference between its value at lattice sites and in the interstitial regions. This fact leads one to parametrize $\rho(\mathbf{r})$ by a collection of overlapping Gaussian profiles centred over lattice sites $\{\mathbf{R}_m\}$ ~\cite{Tarazona1984}.

In a $d$-dimensional system,

\begin{equation}
\rho(\mathbf{r}) = \sum_i \left( \frac{\alpha}{\pi} \right)^{d/2}
\exp \left[ -\alpha (\mathbf{r}-\mathbf{R}_i)^2 \right]
\tag{4.30}
\end{equation}
where $\alpha$ is a variational parameter which characterizes the width of the Gaussian; the square root of $\alpha$ is inversely proportional to the width of the peak. It thus measures the inhomogeneity; the value $\alpha = 0$ corresponds to the uniform fluid (infinitely broad Gaussians create uniformity), and an increasing value of $\alpha$ corresponds to increasing localization of particles about their respective lattice sites.

When $\rho(\mathbf{r},\mathbf{\Omega})$ in Eq.~(4.8) is replaced by $\rho(\mathbf{r})$
given by Eq.~(4.30) one gets

\begin{equation}
\mu_G = \exp\!\left(-G^2/4\alpha\right)
\tag{4.31}
\end{equation}
The Fourier transform of relation (4.13) gives
\begin{equation}
\rho(\mathbf{r}) = \rho + \sum_{G \neq 0} \rho_{G} \exp(i\mathbf{G}\cdot\mathbf{r})
\tag{4.32}
\end{equation}
where $\rho_G = \rho \mu_G$, is amplitude of a density
wave of wavelength $2\pi/|\mathbf{G}|$. For isotropic fluid
$\rho_G = 0$ for $G \neq 0$.

The order parameters $\mu_G$ is determined experimentally
from the scattering experiments. The structure
factor for a monatomic crystal has a form (see
derivation in Appendix C) given as

\begin{equation}
S(\mathbf{q}) = (2\pi)^3 \sum_G \rho |\mu_G|^2 \delta(\mathbf{q}-\mathbf{G})
+ \frac{1}{N} H^{(2)}(\mathbf{q})
\tag{4.33}
\end{equation}

The intensity of the Bragg peak at $\mathbf{G}$ is proportional
to the square of $\mu_G$. The second term in Eq.~(4.33) arises due to the correlation between density--density fluctuations and represents non-Bragg angles scattered background of the scattering function. Though, in general, we expect Bragg peaks at every RLV, thermal motions of particles in the crystal lead not only to a finite width to the Bragg peaks but also to decay in intensity of Bragg peaks at large values of $G$. However, the number of RLVs and the corresponding $\mu_G$ that survive are large and need to be considered in the theory. A complete description of the fluid–crystal transition, therefore requires minimization of $\Delta W$ with respect to $\mu_G$ for all possible RLVs.\\

There may be other ordered phases where positional ordering is either in 1- or 2-dimensions rather than in all the three directions. For instance, one may have systems in which centre of mass of molecules are randomly stacked along one direction forming liquid tubes. The
tubes themselves are arranged in a planer lattice. This is a case of a 2-dimensionally ordered system in 3-dimensions and is called \textit{columnar phase} ~\cite{Chandrasekhar1992} formed by disks--shaped  mesogenic molecules. The other example which corresponds to a 1-dimensional order in 3-dimensions is that of a system in which a set of 2-dimensional liquid layers are stacked on each other with a well defined interlayer spacing. The molecules in each layer of the system have liquid like motion and have almost no correlation of position between different layers. A system which has this kind of positional order and also have long range order in the  orientation of molecules is called \textit{smectic
liquid crystal} ~\cite{Chaikin1995,deGennes1993,Chandrasekhar1992,Singh2005}.

\subsection{Smectic ordering and order parameters}
For the smectic phase with positional order in one
dimension only (e.g. SmA or SmC) Eq. (4.5) reduces to
\[
\rho(\mathbf{r}, \mathbf\Omega) = \rho \sum_{G_z} \sum_{lmn}
Q_{lmn}(G_z)\,
\exp(-i G_z z)\, D^{\,l}_{mn}({\Omega}),
\tag{4.34}
\]
where \( G_z = \mathbf{G}\cdot\hat{z} \) is parallel to the layer normal. Again, Eq.~(4.34) can be simplified by using the symmetry of the phase and of constituent molecules. Thus, for a
phase having symmetry \( D_{\infty h} \wedge T(z) \) (SmA) and composed of cylindrically symmetric molecules, \( \rho(\mathbf{r},\mathbf{\Omega}) \) has to depend only on angle \( \theta \) between the director and molecular axis. Accordingly, we have
\[
\rho(\mathbf{r}, \mathbf{\Omega}) =
\rho \sum_{l}\,' \sum_{q}
Q_{ql}\, \cos(2\pi q z/d)\, P_l(\cos\theta)
\tag{4.35}
\]
where \( G_z = 2\pi qz/d \), \( d \) being the average interlayer
spacing. The prime on the sum indicates the condition \( l \) is even. Again \( Q_{q0} \) represents positional, \( Q_{0l} = (2l+1) P_l \) the orientational order parameters.
The \( Q_{ql} \) is the mixed (coupled) order parameters for SmA phase.

The one--dimensional positional ordering in the smectic phase can be expressed as
\[
\rho_s(\mathbf{r}) = \rho + \sum_{m\ne0}\mu_m \exp(im\mathbf{q}_0\cdot{\vec r})
\tag{4.36}
\]
where \( \mathbf{q}_0 =\frac{2\pi}{d} \hat{e}_z \). Here \(\hat{e}_z \) is a unit vector along z--axis and \(d\) is interlayer spacing. Since harmonics with \(m > 1\) of the density modulation are not experimentally visible in a wide class of smectics, one can write Eq.~(4.36) as,
\[
\rho_s(\mathbf{r}) = \rho + 2\mu_q \cos(\mathbf{q}_0\cdot{z})
\tag{4.37}
\]
The fourier transform  of Eq.~(4.37) leads to only two Bragg peaks away from \(q_z=0\) in the scattering function.
\[
I({q}) = |\mu_q|^2 (2\pi)^3
\left[
\delta(\mathbf{q}-q_0\hat{e}_{z}) +
\delta(\mathbf{q}+q_0\hat{e}_{z})
\right]
\tag{4.38}
\]
where $\hat{e}_z$ is a unit vector along $z$-axis.
The Sm C phase differs from Sm A by a tilt $\theta$ of the director $\hat{n}$ with respect to the normal vector $\hat{e}_z$ to the smectic layer. The remaining non-chiral smectic phases are more ordered than Sm A; they are 3-dimensional realizations of bond orientational ordered structure which existence is indicated by the occurrence of a six--fold modulation in the intensity of the diffused ring at \(q=\frac{2\pi}{d}\) in the scattering function ~\cite{Chaikin1995,deGennes1993}.

\section{Density functional theory of freezing: Approximate free-energy functionals}
As stated in Section 2 (see Eq.~(2.27)), the functional $W([\psi],[\rho])$
at fixed values of $\mu$, $V$ and $T$ is minimum when $\rho(\vec{x})$ is the equilibrium one-particle density and $W[\psi] = \min_{\rho(\vec{x})} W([\psi],[\rho])$ is the grand potential of the system. This variational principle ~\cite{HohenbergKohn1964,Mermin1965,Ebner1976} forms the basis of density functional theory of inhomogeneous systems ~\cite{Evans1979,Evans1991} and of phase transition in molecular systems ~\cite{Singh1991,Hansen2006}. 

Expression for the grand potential in absence of external potential $u^{(e)}(\vec{x})$ given by Eq.~(3.7) can be used to define $-\Delta \tilde{W} = W_h - W_l$, where subscripts `h' and `l' refer to phases of higher and lower symmetries, respectively.

\begin{equation}
\begin{aligned}
-\Delta \tilde{W} =
& \int d\vec{x} \left[
\rho_l(\vec{x}) \ln \frac{\rho_l(\vec{x})}{\rho_h(\vec{x})}
- \Delta \rho(\vec{x})
\right] \\
& - \frac{1}{2} \int d\vec{x}_1 d\vec{x}_2 \,
\Delta \rho(\vec{x}_1)\Delta \rho(\vec{x}_2)\,
\bar{c}(\vec{x}_1,\vec{x}_2)
\end{aligned}
\tag{5.1}
\end{equation}

where $\Delta \rho(\vec{x}) = \rho_l(\vec{x}) - \rho_h(\vec{x})$, and
(see Eq.~(3.10))
\begin{equation}
\bar{c}(\vec{x}_1,\vec{x}_2)
= 2 \int d\lambda (1-\lambda)\,
c^{(2)}(\vec{x}_1,\vec{x}_2;
\rho_h(\vec{x}) + \lambda \Delta \rho(\vec{x}))
\tag{5.2}
\end{equation}

In writing Eq.~(5.1) use has been made of the relation
(see Eq.~(2.30b)),
\begin{equation}
\ln \left(\rho_h(\vec{x}) \Lambda^3\right)
- c^{(1)}(\vec{x},[\rho_h(\vec{x})]) = \beta \mu
\tag{5.3}
\end{equation}
and the fact that at the coexistence point $\mu_l=\mu_h=\mu$.
The condition that at fixed values of $\mu$, $\rho_h(\vec{x})$,
$T$ and $V$ the equilibrium density $\rho_l(\vec{x})$ satisfies
\begin{equation}
\left.
\frac{\delta \Delta \tilde{W}}{\delta \rho(\vec{x})}
\right|_{\rho(\vec{x})=\rho_l(\vec{x})}
= 0
\tag{5.4}
\end{equation}

leads to
\begin{equation}
\ln \frac{\rho_l(\vec{x})}{\rho_h(\vec{x})}
= \int d\vec{x}_2 \,
\Delta \rho(\vec{x}_2)\,
\underline{c}(\vec{x},\vec{x}_2)
\tag{5.5}
\end{equation}

where
\begin{equation}
\underline{c}(\vec{x}_1,\vec{x}_2)
= \int_0^1 d\lambda\,
c^{(2)}(\vec{x}_1,\vec{x}_2;
\rho_h(\vec{x}) + \lambda \Delta \rho(\vec{x}))
\tag{5.6}
\end{equation}
Eqs.~(5.1) and (5.5) are the basic equations of the density functional theory of phase transition. In order to locate the transition, one may attempt to find solutions of these equations which have symmetries of the coexisting phases. In practice, however, one does not attempt to find bifurcation points; instead, an attempt is made to find an ordered structure of lowest free energy. although the choice of a particular structure is not \emph{a priori} needed, it simplifies the computational work considerably. The minimization of $\Delta \tilde{W}$ is carried out in terms of order parameters of an ordered phase. However, any approximation introduced in choosing the order parameters or in the value of the quantity $\bar{c}$ may lead to $\Delta \tilde{W}$ not being a convex functional of $\rho(\vec{x})$. As a consequence, there may exist solutions which are extremal points of Eq.~(5.1) but not local minima. It is therefore necessary to introduce suitable constraints to identify the true minimum.

The other point which needs to be emphasized is that the DPCF that
appears in the theory has a length scale comparable to that of the
intermolecular interactions. Therefore, long-wavelength
density--density fluctuations, if present in the system, as they are
near the critical point of a continuous transition, are not explicitly included in the theory. In view of this, the theory is applicable only to those situations where long-wavelength fluctuations are insignificant. We also note that the quantity $\bar{c}(\vec{x}_1,\vec{x}_2)$, which couples the density fluctuations $\Delta \rho(\vec{x})$ of the emerging structure at $\vec{x}_1$ and $\vec{x}_2$, is crucial. Only when its value attains a certain level does the transformation take place from a higher-symmetry to a lower-symmetry phase; otherwise, the system remains in the high-symmetry phase. Therefore, it is essential to use accurate values of $\underline{c}(\vec{x}_1,\vec{x}_2)$ and $\bar c(\vec{x}_1,\vec{x}_2)$ in the calculations.

\subsection{Theory of freezing of Ramakrishnan and Yussouff (RY)}

Consider a case in which the high-symmetry phase is a
homogeneous, isotropic fluid. For this case, Eqs.~(5.1) and (5.2)
can be rewritten as

\begin{equation}
\begin{aligned}
-\Delta \tilde{W} =
& \int d\vec{x} \left[
\rho(\vec{x}) \ln \frac{\rho(\vec{x})}{\rho_0}
- \Delta \rho(\vec{x})
\right] \\
& - \frac{1}{2} \int d\vec{x}_1 d\vec{x}_2 \,
\Delta \rho(\vec{x}_1)\Delta \rho(\vec{x}_2)\,
\bar{c}(\vec{x}_1,\vec{x}_2)
\end{aligned}
\tag{5.7}
\end{equation}

\begin{equation}
\bar{c}(\vec{x}_1,\vec{x}_2)
= 2 \int_0^1 d\lambda (1-\lambda)\,
c^{(2)}(\vec{x}_1,\vec{x}_2;
\rho_0 + \lambda (\rho(\vec{x})-\rho_0))
\tag{5.8}
\end{equation}

where $\Delta \rho(\vec{x})=\rho(\vec{x})-\rho_0$.where $\rho_0$ and $\rho(\vec{x})$ are the densities of the fluid and the ordered phase.

The integration over $\lambda$ in Eq.~(5.8) covers the region of density space that corresponds to the ordered phase and the weight factor $(1-\lambda)$ ensures that as $\lambda$ increases, the contribution to $\bar{c}$ from the ordered region decreases. This may prompt one to conclude that, since the contribution to $\bar{c}$ from the DPCF of the
ordered phase is small, it may be reasonable to replace $\bar{c}$ in Eq.~(5.7) by $c^{(2)}(\vec{x}_1,\vec{x}_2;\rho_0)$. In such a case,
Eqs.~(5.7) and (5.5) become

\begin{equation}
\begin{aligned}
-\Delta \tilde{W} =
& \int d\vec{x} \left[
\rho(\vec{x}) \ln \frac{\rho(\vec{x})}{\rho_0}
- \Delta \rho(\vec{x})
\right] \\
& - \frac{1}{2} \int d\vec{x}_1 d\vec{x}_2 \,
\Delta \rho(\vec{x}_1)\Delta \rho(\vec{x}_2)\,
c^{(2)}(\vec{x}_1,\vec{x}_2;\rho_0)
\end{aligned}
\tag{5.9}
\end{equation}

\begin{equation}
\ln \frac{\rho(\vec{x})}{\rho_0}
= \int d\vec{x}_2 \,
\Delta \rho(\vec{x}_2)\,
c^{(2)}(\vec{x},\vec{x}_2;\rho_0)
\tag{5.10}
\end{equation}

In Eqs.~(5.9) and (5.10), $c^{(2)}$ is the DPCF of the coexisting
fluid of density $\rho_0$ and chemical potential equal to that of
the ordered phase.

The above equations, (5.9) and (5.10), were first proposed by Ramakrishnan and Yussouff ~\cite{Ramakrishnan1979} to describe the freezing of atomic systems (see also Ref.~\cite{Haymet1981}) and are referred to as equations of the RY theory. These equations are also known as equations of the second-order density functional theory (SODFT). The reason for calling the theory second order is that the functional Taylor expansion of $\Delta {A}[\rho]$ in powers of $\big(\rho(\vec{x}) - \rho_0\big)$, upto quadratic order, and adding the ideal gas part leads to Eq.~(5.9).

Since its proposal, the RY theory has been used to investigate the freezing of a variety of pure fluids and mixtures ~\cite{Singh1991,Hansen2006,Lowen1994,Ram2014}. It has also been used to calculate properties of ordered phases such as elasticity, defects, interfaces, etc.\ (see references
listed in Ref.~\cite{Singh1991}). The theory works reasonably in case of hard-spheres, but results start becoming less accurate as the potential becomes softer. The theory predicts freezing of solids into face-centred-cubic (fcc) structure irrespective of softness of the potential whereas computer simulation results show that for softer potentials, the stable solid has a body-centred-cubic (bcc) structure. Though the RY theory was an important step in explaining the freezing of fluids, it failed to give accurate description of the freezing transition for a large class of intermolecular potentials ~\cite{Barrat1987,deKuijper1990,Wong1999}.

\subsection{Theory based on weighted density approximation (WDA)}

The theory originated due to efforts ~\cite{Tarazona1985,Curtin1985,Denton1989} directed to find nonperturbative free-energy functionals. In WDA a coarse-grained density $\bar{\rho}(\vec{x})$ is defined as a weighted average of $\rho(\vec{x})$ over a volume comparable with the volume of a particle.

\begin{equation}
\bar{\rho}(\vec{x}) =
\int d\vec{x}' \, w(\vec{x},\vec{x}') \rho(\vec{x}')
\tag{5.11}
\end{equation}
where $w(\vec{x})$ is some suitable weight function normalized such that
\begin{equation}
\int d\vec{x}\, w(\vec{x}) = 1
\tag{5.12}
\end{equation}
The excess part of the free energy $\Delta A$ is expressed as,
\begin{equation}
\Delta A^{\text{WDA}} =
\int d\vec{x} \, \rho(\vec{x}) \,
\Delta a_0(\bar{\rho}(\vec{x}))
\tag{5.13}
\end{equation}
where $\Delta a_0$ denotes the excess free energy per particle of a uniform fluid.
The choice of the weight function is made by requiring $\Delta A^{\text{WDA}}$ to satisfy the known relation
for DPCF of the uniform fluid
\begin{equation}
c^{(2)}(\vec{x}_1,\vec{x}_2;\rho_0)
= \lim_{\rho(\vec{x}) \to \rho_0}
\left(
\frac{\delta^2 \Delta A^{\text{WDA}}}
{\delta \rho(\vec{x}_1)\delta \rho(\vec{x}_2)}
\right)
\tag{5.14}
\end{equation}
Using Eqs.~(5.13) and (5.14) one gets a differential equation, the solution of which leads to
the weight function $w(\vec{x},\bar{\rho}(\vec{x}))$. Since such an approach demands huge
computational efforts, Denton and Ashcroft ~\cite{Denton1989} proposed a ``modified'' version, known as
modified WDA or MWDA in which one replaces Eq.~(5.13) by
\begin{equation}
\Delta A^{\text{MWDA}} =
N \Delta a_0(\bar{\rho})
\tag{5.15}
\end{equation}
where
\begin{equation}
\bar{\rho} =
\frac{1}{N}
\int d\vec{x} \rho(\vec{x})
\int d\vec{x}' \rho(\vec{x}')
\bar{w}(\vec{x},\vec{x}';\bar{\rho})
\tag{5.16}
\end{equation}
Requiring that the weight factor $\bar{w}$ satisfy Eq.~(5.14) exactly, one finds [18],
\begin{equation}
\begin{aligned}
\bar{w}(\vec{x},\vec{x}';\rho_0)
= -\left[2\Delta a_0'(\rho_0)\right]^{-1}
\Big[
c^{(2)}(\vec{x},\vec{x}';\rho_0)
+ \frac{1}{V}\rho_0 \Delta a_0''(\rho_0)
\Big]
\end{aligned}
\tag{5.17}
\end{equation}
where V is the volume of system. In the MWDA the inherent nonlocal functional dependence in the excess free energy is entirely subsumed in the effective density.

Khein and Ashcroft~~\cite{KheinAshcroft1999} showed that WDA and MWDA are simplified versions of a more general approach, they called it generalized density functional theory (GDFT). For a system of hard spheres, a theory that supersedes all versions of WDA is the {\it fundamental measure theory} (FMT). It was developed by Rosenfeld ~\cite{Rosenfeld1989,Rosenfeld1997} and modified by Tarazona~\cite{Tarazona2000}. In the FMT, the excess free-energy density is taken to be a function of not just one density but of several weighted densities, defined by weight-functions that emphasize the geometrical characteristics of the particles. Since its introduction the FMT has been applied to many problems of hard-sphere systems~~\cite{Rosenfeld1997,Roth2010}. Recently it has been used to calculate the DPCF of hard-sphere crystals~~\cite{Lin2021}.

\section{Effect of symmetry breaking on correlation functions and on free--energy functional}

Breaking of symmetry changes the molecular distribution and adds new contributions to correlation functions. For example, when orientational symmetry of an isotropic fluid is broken, system becomes orientationally ordered (nematic phase) and molecular orientational distribution lose their rotational invariance. When this symmetry is imposed on molecular distribution (see Section 7) one finds two qualitatively different contributions: one that preserves the symmetry of the fluid and the one that breaks it ~\cite{Holovko1999,PhuongSchmid2003}. This is true for all broken symmetry phases. One can therefore write correlation functions of an ordered phase as a sum of two qualitatively different contributions ~\cite{Mishra2006,SinghSingh2009}.\\
\begin{equation}
h(\vec{x}_1,\vec{x}_2)
=
h^{(0)}(\vec{x}_1,\vec{x}_2,[\rho_h])
+
h^{(b)}(\vec{x}_1,\vec{x}_2;[\rho_l]),
\tag{6.1}
\end{equation}

\begin{equation}
c(\vec{x}_1,\vec{x}_2)
=
c^{(0)}(\vec{x}_1,\vec{x}_2,[\rho_h])
+
c^{(b)}(\vec{x}_1,\vec{x}_2;[\rho_l]),
\tag{6.2}
\end{equation}
and
\begin{equation}
\rho_l(\vec{x})
=
\rho_h(\vec x)
+
\Delta\rho(\vec{x}),
\tag{6.3}
\end{equation}
where superscripts $0$ and $b$, refer, respectively, to symmetry conserving and symmetry breaking contributions and subscripts $h$ and $l$, as before, represent phases of higher and lower symmetries, respectively. Hereafter the superscript (2) from $h$ and $c$ are dropped for notational simplification.

In order to avoid confusion, we prefer to continue our discussion with the higher symmetry phase being isotropic and homogeneous, and the lower symmetry phase, an ordered phase. A case in which both phases which coexist at the phase boundary are broken symmetry phases will be discussed in Section 11. In a case where higher symmetry phase is a fluid, the symmetry conserving components $h^{(0)}$ and $c^{(0)}$ correspond to a homogeneous system and depend only on the magnitude of molecular separation and are function of average density $\rho$. The broken symmetry components $h^{(b)}$ and $c^{(b)}$ are, however, invariant only under a discrete set of symmetry operations of the ordered phase and are functional of $\rho(\vec{x})$.

Above relations can be used to split the inhomogeneous OZ equation (2.38) into two parts; one that contains $h^{(0)}$, $c^{(0)}$ and $\rho$, whereas the other that contains $h^{(b)}$, $c^{(b)}$ and $\rho^{(b)}(\vec x)=\rho(\vec x)-\rho $ along with
$h^{(0)}$, $c^{(0)}$ and $\rho$:
\begin{equation}
h^{(0)}(\vec{x}_1,\vec{x}_2)
=
c^{(0)}(\vec{x}_1,\vec{x}_2)
+
\rho
\int d\vec{x}_3\,
h^{(0)}(\vec{x}_1,\vec{x}_3)\,
c^{(0)}(\vec{x}_3,\vec{x}_2)
\tag{6.4}
\end{equation}
and
\begin{align}
h^{(b)}(\vec{x}_1,\vec{x}_2)
&=
c^{(b)}(\vec{x}_1,\vec{x}_2)
+
\int d\vec{x}_3\,
h^{(0)}(\vec{x}_1,\vec{x}_3)\,
\rho^{(b)}(\vec{x}_3)\,
c^{(0)}(\vec{x}_3,\vec{x}_2)
\nonumber\\
&\quad
+
\int d\vec{x}_3\,
\rho(\vec{x}_3)
\Big[
h^{(0)}(\vec{x}_1,\vec{x}_3)\,
c^{(b)}(\vec{x}_3,\vec{x}_2)
+
h^{(b)}(\vec{x}_1,\vec{x}_3)\,
c^{(0)}(\vec{x}_3,\vec{x}_2)
\nonumber\\
&\qquad\qquad
+
h^{(b)}(\vec{x}_1,\vec{x}_3)\,
c^{(b)}(\vec{x}_3,\vec{x}_2)
\Big]
\tag{6.5}
\end{align}
Eq. (6.4) is the well known OZ equation of isotropic fluids and is solved along with a closure relation to determine values of $h^{(0)}$, $c^{(0)}$ and their derivatives
with respect to density $\rho$. Eq. (6.5) is the OZ equation for the symmetry broken part of correlation functions. However, in order to make use of it to find values of $h^{(b)}$ and $c^{(b)}$ for a given value of $\rho(\vec{x})$ one needs one more relation (closure relation) that connects $h^{(b)}$, $c^{(b)}$ and intermolecular interactions. Alternatively, if values of one of them is known, then (6.5) can be used to determine values of the other function. 

In view of the above discussions, one can rewrite excess
free energy as ~\cite{Mishra2006,SinghSingh2009}
\begin{equation}
\Delta A[\rho] = \Delta A_{0}(\rho) + \Delta A_{b}[\rho]
\tag{6.6}
\end{equation}
where
\begin{equation}
\frac{\delta^2 \Delta A_{0}(\rho)}
{\delta \rho(\vec{x}_1)\,\delta \rho(\vec{x}_2)}
=
-\,c^{(0)}(\vec{x}_1,\vec{x}_2;\rho)
\tag{6.7}
\end{equation}
and
\begin{equation}
\frac{\delta^2 \Delta A_{b}[\rho]}
{\delta \rho(\vec{x}_1)\,\delta \rho(\vec{x}_2)}
=
-\,c^{(b)}(\vec{x}_1,\vec{x}_2;[\rho])
\tag{6.8}
\end{equation}
Since $c^{(0)}$ is a function of density, the functional integration
of Eq.~(6.7) in the density space is done taking an
isotropic fluid of density $\rho_0$ as a reference. This leads
to
\begin{equation}
\Delta A_{0}(\rho)
=
\Delta A_{0}(\rho_0)
-
\frac{1}{2}
\int d\vec{x}_1\, d\vec{x}_2\,
\Delta \rho(\vec{x}_1)\,
\Delta \rho(\vec{x}_2)\,
\bar{c}^{(0)}(\vec{x}_1,\vec{x}_2)
\tag{6.9}
\end{equation}
where
\[
\Delta \rho(\vec{x}) = \rho(\vec{x}) - \rho_0
\]
and
\begin{equation}
\bar{c}^{(0)}(\vec{x}_1,\vec{x}_2)
=2 \int_0^1 d\lambda\,(1-\lambda)\,
c^{(0)}(\vec{x}_1,\vec{x}_2;\rho_0+\lambda(\rho-\rho_0))
\tag{6.10}
\end{equation}
The functional integration of (6.8) in which $c^{(b)}$ depends
on order parameters in addition to $\rho$, the path of
integration is characterized by two parameters $\lambda$ and $\xi$.
These parameters vary from 0 to 1. The parameter $\lambda$
raises the density from zero to $\rho$ and it varies from 0
to 1 whereas the parameter $\xi$ raises the order parameters
from zero to their final values. The integration gives
\begin{equation}
\Delta A_{b}[\rho]
=
-\frac{1}{2}
\int d\vec{x}_1\, d\vec{x}_2\,
\rho^{(b)}(\vec{x}_1)\,
\rho^{(b)}(\vec{x}_2)\,
\bar{c}^{(b)}(\vec{x}_1,\vec{x}_2)
\tag{6.11}
\end{equation}
where
\begin{equation}
\bar{c}^{(b)}(\vec{x}_1,\vec{x}_2)
=
4
\int_0^1 d\lambda\,(1-\lambda)
\int_0^1 d\xi\,(1-\xi)\,
c^{(b)}(\vec{x}_1,\vec{x}_2;\lambda \rho ;\xi \rho^{(b)}),
\tag{6.12}
\end{equation}
While integrating over $\lambda$ the order parameters are kept fixed and while integrating over $\xi$ the density is kept fixed. The results do not depend on the order of integration ~\cite{Mishra2006,SinghSingh2009}. The free energy $A$ and the grand potential $W$ of an ordered phase can now be  written as a sum of three terms, ideal gas component, symmetry conserving  and symmetry broken components.
\begin{equation}
A[\rho]
=
A_{\mathrm{id}}[\rho]
+
\Delta A_{0}
+
\Delta A_{b}
\tag{6.13}
\end{equation}
\begin{equation}
\Delta W
=
\Delta W_{\mathrm{id}}
+
\Delta W_{0}
+
\Delta W_{b}
\tag{6.14}
\end{equation}

The free energy of an ordered phase expressed in terms of one-particle density and DPCF is
\begin{align*}
    A &= \int d\vec{x} \rho(\vec{x}) \left[ \ln \left(\rho(\vec{x})\Lambda^3\right) - 1 \right] \\
    &\quad -\frac{1}{2} \int d\vec{x}_1 d\vec{x}_2 \rho(\vec{x}_1) \rho(\vec{x}_2) \bar{c}^{(0)}(\vec{x}_1, \vec{x}_2) \\
    &\quad -\frac{1}{2} \int d\vec{x}_1 d\vec{x}_2 \rho^{(b)}(\vec{x}_1) \rho^{(b)}(\vec{x}_2) \bar{c}^{(b)}(\vec{x}_1, \vec{x}_2) \tag{6.15}
\end{align*}

In writing Eq (6.15) we set the reference density $\rho_0 = 0$. The expression of $\Delta \tilde{W}$ is,
\begin{align*}
    -\Delta \tilde{W} &= \int d\vec{x} \left[ \rho(\vec{x}) \ln \frac{\rho(\vec{x})}{\rho_0} - \Delta\rho(\vec{x}) \right] \\
    &\quad -\frac{1}{2} \int d\vec{x}_1 d\vec{x}_2 \Delta\rho(\vec{x}_1) \Delta\rho(\vec{x}_2) \bar{c}^{(0)}(\vec{x}_1, \vec{x}_2) \\
    &\quad -\frac{1}{2} \int d\vec{x}_1 d\vec{x}_2 \rho^{(b)}(\vec{x}_1) \rho^{(b)}(\vec{x}_2) \bar{c}^{(b)}(\vec{x}_1, \vec{x}_2) \tag{6.16}
\end{align*}

Minimization of $\Delta \tilde{W}$ leads to
\begin{align*}
    \ln \frac{\rho(\vec{x})}{\rho_0} &= \lambda' + \int d\vec{x}_2 \Delta\rho(\vec{x}_2) \underline{c}^{(0)}(\vec{x}_1, \vec{x}_2) \\
    &\quad + \int d\vec{x}_2 \rho^{(b)}(\vec{x}_2) \underline{c}^{(b)}(\vec{x}_1, \vec{x}_2) \tag{6.17}
\end{align*}

where
\begin{equation*}
    \underline{c}^{(0)}(\vec{x}_1, \vec{x}_2) = \int_0^1 d\lambda c^{(0)}(\vec{x}_1, \vec{x}_2; \rho_0 + \lambda(\rho - \rho_0)) \tag{6.18}
\end{equation*}
and
\begin{equation*}
    \underline{c}^{(b)}(\vec{x}_1, \vec{x}_2) = \int_0^1 d\lambda \int_0^1 d\xi c^{(b)}(\vec{x}_1, \vec{x}_2; \lambda\rho, \xi\rho^{(b)}) \tag{6.19}
\end{equation*}

Eq. (6.17) is often referred to as order parameter equation. The parameter $\lambda'$ is included to ensure that $\rho(x)$ satisfies the required condition discussed in Section 5.

Eqs (6.15) -- (6.18) are basic equations of density functional theory of molecular systems. To distinguish it from other versions of DFT we call it exact DFT (EDFT). It is exact in the sense that no approximation has been introduced in the derivation of the above equations. However, the accuracy of the results depends on the accuracy of DPCF and on the evaluation of values of $\bar{c}^{(b)}$ and $\underline{c}^{(b)}$.

\section{Nematic phase: correlation functions and isotropic--nematic transition}

In a nematic phase, as stated in Section.$4$, molecules remain positionally disordered but align preferentially along one (arbitrary) direction. At the isotropic-nematic transition, the isotropy of space is spontaneously broken, and as a consequence, the correlations in the distribution of molecules lose their  rotational invariance except about an axis along the director $\hat{n}$. In the simplest case of axially symmetric molecules, the rotational group $SO(3)$ or $O(3)$ of the isotropic phase is replaced by one of the uniaxial symmetry group $D_{\infty h}$ or $D_{\infty}$.The phase which has  $R^3  \Lambda    D_{\infty h}$, (denoting the semi-direct product of the translational group $R^3$ and the rotational group $D_{\infty h}$) is known as the uniaxial nematic phase. The correlation functions of the phase can be expanded in terms of basis set of spherical harmonics.

An expansion of pair functions in a general basis set of rotation matrices
$D^{\ell}_{mn}$ is given (see Appendix B)
by
\begin{equation}
A(r, \Omega_1, \Omega_2)
=
\sum_{\ell_1,\ell_2,\ell} \hspace{0.2cm}
\sum_{m_1,m_2,m}  \hspace{0.2cm}
\sum_{n_1,n_2}
A_{\ell_1 \ell_2 \ell ,m_1 m_2 m}^{n_1 n_2}(r)
D^{\ell_1 *}_{m_1 n_1}(\Omega_1)
D^{\ell_2 *}_{m_2 n_2}(\Omega_2)
Y_{\ell m}^{*}(\hat{r}),
\tag{7.1}
\end{equation}

which, for the case of axially symmetric molecules, reduces to
\begin{equation}
A(r, \Omega_1, \Omega_2)
=
\sum_{\ell_1,\ell_2,\ell} \hspace{0.2cm}
\sum_{m_1,m_2,m}
A_{\ell_1 \ell_2 \ell m_1 m_2 m}(r)
Y_{\ell_1 m_1}(\Omega_1)
Y_{\ell_2 m_2}(\Omega_2)
Y_{\ell m}^{*}(\hat{\mathbf{r}}).
\tag{7.2}
\end{equation}

Here, $A$ stands for pair functions $h$, $c$, or $g$.
The expansion in Eq.(7.1) does not assume any symmetry other than translational
invariance and can be applied to both isotropic and  nematic phases.
In the nematic phase, if we choose the director $\hat{\mathbf{n}}$ along the
$z$-axis of the Cartesian coordinate system (diector frame) then in terms of the Euler angles, any  rotation $\phi'$ about this axis must leave
$A(\mathbf{r}, \Omega_1, \Omega_2)$ unchanged. Thus, using the relation ~\cite{GrayGubbins1984}
\begin{equation}
Y_{\ell m}(\Omega)
=
\sum_{m'}
D^{\ell}_{m' m}(\phi',0,0)\,
Y_{\ell m'}(\Omega'),
\tag{7.3}
\end{equation}
where
\begin{equation*}
D^{\ell}_{m' m}(\phi',0,0)
= e^{- i m \phi'}\,\delta_{m' m},
\end{equation*}

one obtains

\begin{align*}
A(r, \mathbf{\Omega_1}, \mathbf{\Omega_2})
&=\sum
A_{\ell_1 \ell_2 \ell m_1 m_2 m}(r)\,e^{-im_1\phi'}Y_{l_1m_1}(\Omega_1')e^{-im_2\phi'}Y_{l_2m_2}(\Omega_2') e^{im\phi'} Y_{lm}^{*}(\hat r')\\
&= \sum 
e^{-i(m_1+m_2-m)\phi'}A_{\ell_1 \ell_2 \ell m_1 m_2 m}(r)
Y_{\ell_1 m_1}(\Omega_1')
Y_{\ell_2 m_2}(\Omega_2')
Y_{\ell m}^{*}(\hat r')
\tag{7.4}
\end{align*}
The invariance implies $m_1 + m_2 - m = 0$.\\

Thus, for a uniaxial nematic phase of  axially symmetric molecules in director frame one can use following expansion for one-particle and two-particle functions: (see Eqs (4.14)-(4.18))

\begin{equation}
\rho(\mathbf{r}, \mathbf{\Omega})
=
\rho\, f(\mathbf{\Omega})
=
\frac{\rho}{\sqrt{4\pi}}
\sum_{\ell \,\text{even}}
f_{\ell}\, Y_{\ell 0}(\Omega),
\tag{7.5}
\end{equation}

where
\[
f_{\ell} = \sqrt{2\ell + 1}\, P_{\ell},
\qquad
f_{0} = 1.
\]

and
\begin{equation}
A(r, \mathbf{\Omega}_1, \mathbf{\Omega}_2)
=
\sum_{\ell_1,\ell_2,\ell}
\sum_{m_1,m_2,m}
A_{\ell_1 \ell_2 \ell m_1 m_2 m}(r)\,
Y_{\ell_1 m_1}(\Omega_1)\,
Y_{\ell_2 m_2}(\Omega_2)\,
Y_{\ell m}^{*}(\hat{r}),
\tag{7.6}
\end{equation}

Note the difference in the definition of $f(\Omega)$ and $f_l$ given above from those given by Eqs. (4.17) and (4.18). The expression of $f(\Omega)$ given here involves a factor of $\frac{1}{4\pi}$ and $f_l$ is $(2l+1)^{1/2}$ times $P_l$. In the case of an isotropic phase (see Appendix B),
six indices harmonic coefficients appearing in (7.6)
reduce to
\begin{equation}
A_{\ell_1 \ell_2 \ell m_1 m_2 m}(r)
=
A_{\ell_1 \ell_2 \ell}(r)\,
C_{g}(\ell_1 \ell_2 \ell; m_1 m_2 m),
\tag{7.7}
\end{equation}
where $C_{g}$ is the Clebsch--Gordan (CG) coefficient.
The absence of the CG coefficient in Eq. (7.6) removes the
limitation $|\ell_1 - \ell_2| \leq \ell \leq \ell_1 + \ell_2$ on  values of the index $\ell$. So the coefficients such as $A_{200000}(r)$ and $A_{020000}(r)$ are nonzero in nematic whereas they do not survive in the isotropic case. These additional harmonic coefficients which survive are considerably higher in comparison to that of symmetry conserving numbers and are due to broken symmetry. For example, with $\ell_{\max} = 2, 4, 6$ the number of coefficients that survive in nematic phase are 38, 351, and 1842 respectively, out of which
respectively, 5, 14, and 23 are the symmetry conserving
and the rest are broken symmetry contributions. By computer simulation of a system of ellipsoids, Phuong and Schmid ~\cite{PhuongSchmid2003} evaluated the effect of breaking of rotational symmetry on pair correlation functions and found that in a nematic phase there are two qualitatively different contributions, one that preserves  rotational invariance and the other that  breaks it and vanishes in the isotropic
phase. The symmetry-preserving part of the correlation
function passes smoothly without any abrupt change through the transition.

The uniaxial nematic is the simplest broken symmetry phase of molecular systems and has been studied since long ~\cite{deGennes1993,Chandrasekhar1992}. Originally proposed  theories of Onsager~\cite{Onsager1949} and Maier-Saupe~\cite{Maier1958} were based on different premises; in the Onsager theory nematic ordering appears due to a large asymmetry in molecular shape, whereas in the Maier-Saupe theory the anisotropy in the attraction part of intermolecular interaction is the cause of orientational ordering. These two different models are still being used to investigate the nematic phase, as they capture essential features of the phase.

\subsection{Spherical potential with anisotropic attraction}

We consider a pair potential which has a spherical core and anisotropic
attraction and was used by Holovko and Sokolovska~\cite{Holovko1999}
to calculate pair correlation functions of the nematic phase.
The potential is written as
\begin{equation}
u(\mathbf{r}_1, \mathbf{r}_2)
=
u_{\mathrm{hs}}(r)
+
u_0(r)
+
u_2(r, \mathbf{\Omega}_1, \mathbf{\Omega}_2),
\tag{7.8}
\end{equation}
where $u_{\mathrm{hs}}$ is the hard-sphere potential,
\[
u_{\mathrm{hs}}(r)
=
\begin{cases}
\infty, & r < \sigma \\
0, & r > \sigma    \hspace{0.5cm}  ,
\end{cases}
\]
the long-range attraction has an isotropic part;
\begin{equation}
u_0(r)
=
- a_0 (z_0 \sigma)^2
\frac{\exp(-z_0 r)}{r / \sigma},
\tag{7.9}
\end{equation}
and an anisotropic part;
\begin{equation}
u_2(r, \mathbf{\Omega}_1, \mathbf{\Omega}_2)
=
- a_2 (z_2 \sigma)^2
\frac{\exp(-z_2 r)}{r / \sigma}
P_2(\cos \Omega_{12}),
\tag{7.10}
\end{equation}
where $P_2(\cos \Omega_{12})$ is the second-order Legendre
polynomial of relative molecular orientations. The distance of closest approach of two molecules
coincides with the diameter of hard spheres $\sigma$,
and is independent of the intermolecular separation vector
$\mathbf{r}$. Because of this, the harmonic
coefficients that survive in the expansion
of correlation functions have $l=m=0$ and $m_1=-m_2$. In spite of these simplifications, the model is able to bring forth the effect of broken symmetry on correlation functions.


The fact that the harmonic coefficients that survive in the expansion of pair correlation functions have to satisfy $l=m = 0 \quad \text{and} \quad m_1 = -m_2$, allows notational simplifications from six indices to three. Thus, for a system interacting via the pair potential of Eq. (7.8), the pair functions can be written as ~\cite{Holovko1999}

~\cite{Maier1958}

\begin{align}
A(r,\Omega_1,\Omega_2)
&= A_{000}(r)
+ A_{200}(r)\left[
Y_{20}(\Omega_1) + Y_{20}(\Omega_2)
\right] \nonumber \\
&\quad + \sum_{m=-2}^{2} A_{22m}(r)\,
Y_{2m}(\Omega_1)\, Y_{2m}^{*}(\Omega_2)
\tag{7.11}
\end{align}
 
It is helpful to use the Hankel transform to write the
harmonic coefficients in the Fourier space ~\cite{GrayGubbins1984}. The inhomogeneous OZ equation ~(2.40) in $k$-space reduces to ~\cite{Mishra2007}

\begin{align}
\gamma_{\ell_1 \ell_2 0 m \underline{m} 0}(k)
&= \gamma_{\ell_1 \ell_2 m}(k) \nonumber \\
&= \frac{\rho}{4\pi}
\sum_{\ell'_3,\ell''_3}
c_{\ell_1 \ell'_3 m}(k)\,
h_{\ell''_3 \ell_2 m}(k)
\sum_{L} P_{L} \sqrt{2L+1}\,
\Gamma^{\ell'_3 \ell''_3 L}_{\underline{m}\, m\, 0}
\tag{7.12}
\end{align}
where $\underline{m} = -m$,  $ \gamma_{\ell_1 \ell_2 m}(k)
=h_{\ell_1 \ell_2 m}(k)
- c_{\ell_1 \ell_2 m}(k)$\\
and
\begin{equation*}
\Gamma^{\ell_1 \ell_2 \ell}_{m_1 m_2 m}
=
\int d\mathbf{\Omega}\,
Y_{\ell m}^{*}(\Omega)\,
Y_{\ell_1 m_1}(\Omega_1)\,
Y_{\ell_2 m_2}(\Omega)\\
\end{equation*}
or
\begin{equation}
\Gamma^{\ell_1 \ell_2 \ell}_{m_1 m_2 m}
=\left(
\frac{(2\ell_1+1)(2\ell_2+1)}{4\pi(2\ell+1)}
\right)^{1/2}
C(\ell_1 \ell_2 \ell; 0\,0\,0)\,
C(\ell_1 \ell_2 \ell; m_1\,m_2\,\underline{m}).
\tag{7.13}
\end{equation}\\
Since there is no summation over $m$ on the right-hand side of Eq.~(7.12), the OZ equation for harmonics with different values of $m$ decouples.\\

\subsubsection{Mean spherical approximation (MSA) and analytical solution for pair correlation function in the nematic phase}

In the mean spherical approximation (MSA)~\cite{Hansen2006},
\begin{align}
h(\mathbf{x}_1, \mathbf{x}_2) &= -1,
\qquad |\mathbf{r}_1 - \mathbf{r}_2| < \sigma,
\tag{7.14} \\
c(\mathbf{x}_1, \mathbf{x}_2) &= -\beta
u_a(r, \mathbf{\Omega}_1, \mathbf{\Omega}_2),
\qquad |\mathbf{r}_1 - \mathbf{r}_2| > \sigma,
\tag{7.15}
\end{align}
where $u_a$, given by Eqs.~(7.9) and (7.10), is the attractive part of the potential (7.8).\\
Condition (7.14) is exact for the potential model of
Eq.~(7.8), since \(g(r, \mathbf{\Omega}_1, \mathbf{\Omega}_2)
\) \\ \( =1 + h(r, \mathbf{\Omega}_1, \mathbf{\Omega}_2)
=0 \quad \text{for } r < \sigma.
\)
Although it is only for large intermolecular separation
$r$ that $c(\mathbf{r},\mathbf{\Omega}_1, \mathbf{\Omega}_2)$ is asymptotic to $ - \beta u(r, \mathbf{\Omega}_1, \mathbf{\Omega}_2)$, in the MSA it is assumed that $c = - \beta u $ for all $ r \; (r > \sigma)$. In spite of this, the MSA is found to give good results for many cases ~\cite{Hansen2006}. 

Using Eq.~(7.11) and Eq.~(7.14),
one gets for $r < \sigma$, ~\cite{Mishra2007}
\begin{equation}
c_{000}(r)
=
- \gamma_{000}(r)
-
(4\pi)^{3/2},
\tag{7.16}
\end{equation}
and
\begin{equation}
c_{\ell_1 \ell_2 m}(r)
=
- \gamma_{\ell_1 \ell_2 m}(r).
\tag{7.17}
\end{equation}

For $r > \sigma$, from Eqs.~(7.15) and (7.11), one gets
\begin{equation}
c_{iim}(r)
=
(-1)^m
\frac{(4\pi)^{3/2}}{2i + 1}
\,
\beta a_i (z_i \sigma)^2
\frac{\exp(-z_i r)}{r/\sigma},
\tag{7.18}
\end{equation}
where $i = 0, 2$ and $m = 0, 1, 2$.

Holovko and Sokolovska ~\cite{Holovko1999} used analytical approach to solve the OZ equation~(7.12) with the MSA relations (7.16)--(7.18) and the Lovett equation (2.41b) which in the present case can be written as
\begin{equation}
1
=
\rho
\left\langle
|Y_{21}(\Omega)|^{2}
\right\rangle
\int d\mathbf{r}\,
c_{221}(r),
\tag{7.17}
\end{equation}
where
\[
\left\langle
|Y_{21}(\Omega)|^{2}
\right\rangle
=
\int d\Omega\,
f(\Omega)\,
Y_{21}(\Omega)\,
Y_{21}^{*}(\Omega),
\]
self-consistently.

\subsubsection{Numerical solution of the OZ equation with MSA and Percus--Yevick relations for pair correlation functions in the nematic phase}

Mishra \textit{et al.} ~\cite{Mishra2007} used the numerical procedure to solve the OZ equation~(7.12) with the MSA relations and also with the PY relation (see Appendix B), which is written as

\begin{equation}
c(\mathbf{r}, \mathbf{\Omega}_1, \mathbf{\Omega}_2)
= 
M(\mathbf{x_1},\mathbf{x_2})
\left[
g(r,\Omega_1, \Omega_2)
-
c(r,\Omega_1,\Omega_2)
\right],
\end{equation}
where
\[
M(\mathbf{x_1},\mathbf{x_2})
=
\exp\!\left[
- \beta u(\mathbf{x_1},\mathbf{x_2})
\right]
-
1
\]
is the Mayer function. The expansion in spherical harmonics with the constraint
$\ell = m = 0$ leads to
\begin{align}
c_{\ell_1 \ell_2 m}(r)
&=
\frac{1}{\sqrt{4\pi}}
\sum_{\ell_1',\ell_2',m'} \sum_{\ell_1'',\ell_2'',m''}
M_{\ell_1' \ell_2' m'}(r)
\nonumber\\
&\qquad \times
\Big[
\gamma_{\ell_1'' \ell_2'' m''}(r)
+
(4\pi)^{3/2}
\delta_{\ell_1'' 0}
\delta_{\ell_2'' 0}
\delta_{m'' 0}
\Big]
\Gamma^{\ell_1' \ell_1'' \ell_1}_{m' \,m'' m} \Gamma^{\ell_2' \ell_2'' \ell_2}_{\underline{m}' \,\underline{m}'' \underline{m}},
\tag{7.18}
\end{align}

In numerical procedures involving iterations, the
following care has to be taken ~\cite{PhuongSchmid2003,Mishra2007}. Since some harmonic coefficients of $h$ decay slowly and may extend to large values of $r^* = \left( \frac{r}{\sigma} \right)$, one has to choose the range of $r^*$ large enough to ensure proper convergence. The other important issue is related to the pronounced
long-range tails in coefficients $h_{\ell_1 \ell_2 m}(r)$
with $m = \pm 1$ (see Fig.7.2), which exhibit a
$1/r^{*}$ behavior at large distances. It is therefore advisable, as suggested in
Refs.~\cite{PhuongSchmid2003}, to use power laws such as $a + \frac{b}{r^{*}}$ to fit the data points of these harmonic coefficients beyond a cutoff $r > r_0$, shift them by $a$ and extrapolate  to infinity before taking  the Hankel transform in each iteration. This as observed by Mishra and Singh ~\cite{Mishra2006} removes the finite-size effects on the tail.\\

\begin{figure}
    \centering
  \includegraphics[width=0.85\linewidth]{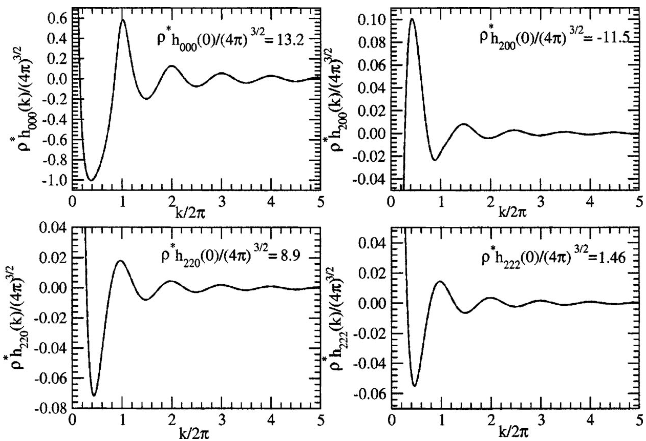}
    \caption*{\textbf{Fig.7.1}
Comparison of some of the harmonics of the total pair correlation function in the Fourier space for the nematic phase ($\beta a_2 = 1$, $\beta a_0 = 0.1$, $z_0 {\sigma} = z_2 {\sigma} = 1$, $\eta = 0.315$, $P_2 = 0.63$, $P_4 = 0.27$) obtained by the analytical
solution ~\cite{Roth2010} (dashed line) with the results found using the numerical method (full line) for the MSA. The two curves are indistinguishable at the scale of the figure. Reproduced with permission from Ref ~\cite{Mishra2007}
}
\end{figure}

In Fig.~7.1 we report  and compare results of the Fourier transform of some of the harmonic coefficients $h_{l_1l_2m}(r^*)$ obtained analytically ~\cite{Holovko1999} with the results found using the numerical method in Ref.~ ~\cite{Mishra2007} for the MSA. Both results are for $\eta = 0.315$, $P_2 = 0.63$, and $P_4 = 0.27$. These values of the order parameters correspond to a minimum of the intrinsic  free energy defined by Eq.~(6.15). Note that in case of the MSA the order parameter $P_4$ does not appear explicitly, it appears indirectly as shown in Ref.~\cite{Mishra2007}. Both curves shown in the figure overlap indicating an excellent agreement between the two results.
In Fig.~7.2 the harmonic coefficients $c_{221}(r^*)$ and $h_{221}(r^*)$ are compared; these harmonic coefficients appear in the nematic elastic constants ~\cite{Singh1991}. The decay of $h_{221}(r^*)$ as $1/r^{*}$at  large distances is clearly seen.

\begin{figure}
    \includegraphics[width=0.85\linewidth]{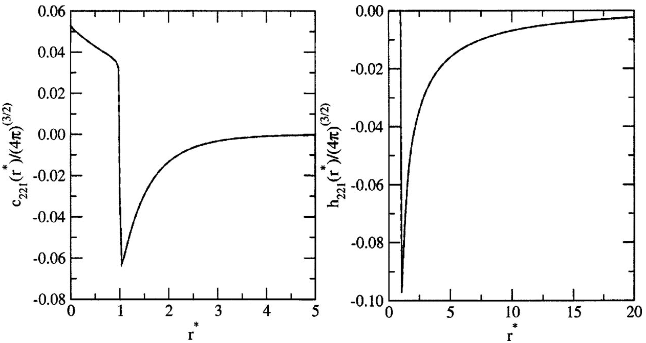}
    \caption*{\textbf{Fig.7.2}
Comparison of the harmonic coefficients $c_{221}(r^{*})$ and $h_{221}(r^{*})$ obtained by the analytical
solution (Ref.~\cite{Roth2010}) (dashed line) with the results found using the numerical method (full line) for the MSA.
Parameters are the same as in Fig.~7.1. A $1/r^{*}$ tail in the harmonic coefficient $h_{221}(r^{*})$ is seen. Reproduced with permission from Ref ~\cite{Mishra2007}.}
    \label{fig:placeholder}
\end{figure}

The nematic structure factor is defined as
\begin{equation}
S(k) = 1 + \rho^* \int d\Omega_1 \, d\Omega_2 \,
f(\Omega_1) \, h(k, \Omega_1, \Omega_2) \, f(\Omega_2).
\tag{7.19} 
\end{equation} 
where $\rho^*=\rho\sigma^3$. Using harmonic expansions (7.5) and (7.6), and restrictions $l=m=0$, one gets
\begin{equation}
S(k) = 1 + \frac{\rho^*}{(4\pi)^{3/2}}
\left[
h_{000}(k)
+ 2\sqrt{5} \, P_2 \, h_{200}(k)
+ 5 P_2^2 \, h_{220}(k)
\right].
\tag{7.20}
\end{equation}

\begin{figure}
    \centering
    \includegraphics[width=0.7\linewidth]{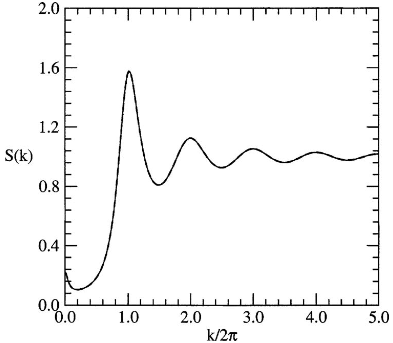}
  \caption*{\textbf{Fig.7.3}
Comparison of the structure factor curves for the nematic phase ($\beta a_2 = 1$, $\beta a_0 = 0.1$, $z_0{\sigma} = z_2{\sigma} = 1$, $\eta = 0.315$, $P_2 = 0.63$, $P_4 = 0.27$). The dashed line is obtained with the analytical solution (Ref.~\cite{Roth2010}) while the full line is obtained by our numerical method for the MSA. The two curves overlap at all values of $k$. Reproduced with permission from Ref ~\cite{Mishra2007}.
}

    \label{fig:placeholder}
\end{figure}
\begin{figure}
    \centering
    \includegraphics[width=0.9\linewidth]{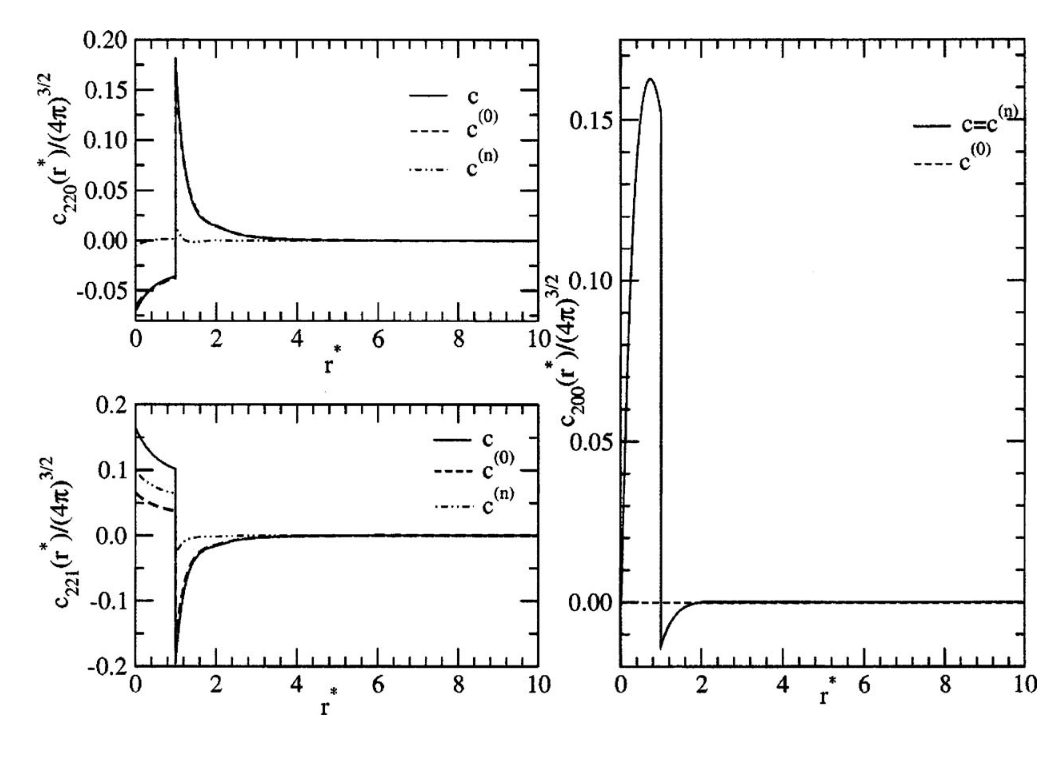}
    \caption*{ \textbf{Fig.7.4} Plot of the harmonic coefficients of the direct pair correlation
functions $c_{220}(r^{*})$, $c_{221}(r^{*})$, and $c_{200}(r^{*})$
obtained by solving the PY integral equation theory ($\beta a_{2}=1$, $\beta a_{0}=0.1$, $z_{0}{\sigma}=z_{2}{\sigma}=1$, $\eta=0.30$). While the contribution of the symmetry breaking part in
$c_{220}(r^{*})$ is small in $c_{221}(r^{*})$ it is comparable inside the core. The harmonic coefficient $c_{200}(r^{*})$ arises due to symmetry breaking only. The symmetry conserving contribution is shown by dotted line and the broken symmetry part by dot--dashed line and the total by full line. Reproduced with permission from Ref.~\cite{Mishra2007}.
}
    \label{fig:placeholder}
\end{figure}

In Fig.~7.3, $S(k)$ is plotted as a function of $k$. A small peak at $k \simeq 0$ is attributed to the appearance of an additional effective attraction due to parallel alignment of molecules. Note that in Eq.~(7.20) the harmonic $h_{200}(k)$, which arises due to broken symmetry, compensates the divergence of $h_{220}(k \to 0)$, so that $S(k)$ at $k \to 0$ behaves similarly to the isotropic case ~\cite{Holovko1999}.

\begin{figure}
    \centering
    \includegraphics[width=0.7\linewidth]{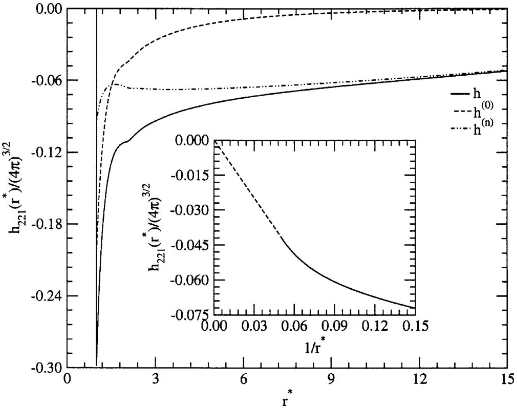}
    \caption*{\textbf{Fig.7.5} Harmonic coefficient $h_{221}(r^{*})$ in the director frame.
Details are the same as in Fig.~7.4.
Inset shows the plot of $h_{221}(r^{*})$ with respect to $1/r^{*}$; the dashed line shows the extrapolated part. The origin of the tail is due to orientational symmetry breaking. Reproduced with permission from Ref.~\cite{Mishra2007}.
}
    \label{fig:placeholder}
\end{figure}

The PY theory results found by Mishra \textit{et al.} ~\cite{Mishra2007} for the model potential given in Eq.~(7.8) are shown in Figs.~7.4 and 7.5 for $\beta a_2 = 1$, $\eta = 0.30$, $P_2 = 0.69$, and $P_4 = 0.32$. These values of the order parameters have been found from the minimization of the free-energy functional. While the harmonic coefficients $c_{220}(r^*)$ and $c_{221}(r^*)$, shown in Fig.~7.4, appear for both the isotropic and nematic phases, the harmonic coefficient $c_{200}(r^*)$ appears only in the nematic phase and vanishes in the isotropic phase. The harmonic coefficient $h_{221}(r^*)$, plotted in Fig.~7.5, where its $1/r^*$ behavior is shown in the inset. The structure factor obtained from the PY theory is given by

\begin{equation}
S(k) = 1 + \frac{\rho^*}{(4\pi)^{3/2}}
\sum_{l_1,l_2}
\sqrt{(2l_1+1)(2l_2+1)} \,
P_{l_1} P_{l_2} \,
h_{l_1 l_2 0}(k).
\tag{7.21}
\end{equation}

which reduces to Eq.~(7.20) when $\ell_1$ and $\ell_2$ are restricted to
$0$ and $2$. However, values of harmonic coefficients
$h_{\ell_1 \ell_2 0}(k)$ for same values of $\ell_1$ and $\ell_2$ will be different.

\subsubsection{Isotropic--nematic (I--N) transition}

One can adopt a procedure in which the OZ equation, closure relation and the order parameter equation (6.17) are solved together. The final output of the calculation is a set of self-consistent values of order parameters and PCFs. In the calculation one starts with assumed values of $P_l$ and $\Delta\rho^*$ and solves OZ equation and closure relation to find values of structural parameters $\bar{c}^{(0)}$, $\bar{c}^{(b)}$, $\underline{c}^{(0)}$ and $\underline{c}^{(b)}$ which are then substituted in the order parameter equations (Eqs (7.22) -- (7.25) given below) to find new set of order parameters. This cycle continues till the self-consistency is achieved. The final output gives values of order parameters and PCFs at the chosen values of density and temperature. In the plane of the order parameters the minimum value of the grand potential $W$ decides the equilibrium phase.

Using the relation $\rho(\vec{x}) = \rho f(\vec{\Omega})$, where $f(\vec{\Omega})$ satisfies the condition $\int d\hat{\Omega} \, f(\hat{\Omega}) = 1$ ,
and $\rho = \rho_0 (1 + \Delta \rho^*)$, one gets the following relations, from Eq.~(6.17), 

\begin{equation}
f(\vec{\Omega}) =
\frac{\exp(\text{sum})}
{\int_{-1}^{1} d(\cos\theta)\, \exp(\text{sum})}
\tag{7.22}
\end{equation}

and

\begin{equation}
1 + \Delta \rho^* =
\frac{1}{2} \int_{-1}^{1} d(\cos\theta)\, \exp(\text{sum})
\tag{7.23}
\end{equation}

where
\begin{equation}
\text{sum} =
\int d\vec{x}_2 \, \Delta \rho(\vec{x}_2) \,
\underline{c}^{(0)}(\vec{x}_1,\vec{x}_2)
+
\int d\vec{x}_2 \, \rho^{(0)}(\vec{x}_2) \,
\underline{c}^{(b)}(\vec{x}_1,\vec{x}_2)
\tag{7.24}
\end{equation}

The order parameters $P_\ell$ are found from the relation
\begin{equation}
P_\ell =
\int d\vec{\Omega} \, P_\ell(\cos\theta)\, f(\vec{\Omega}),
\tag{7.25}
\end{equation}
where $f(\Omega)$ is given by Eqs (7.21) and  (7.24).

To locate the isotropic--nematic transition point and associated transition parameters one has to substitute self-consistent values of order and structural parameters in the expression of $\Delta \tilde{W}$ given in Eq. (6.16). Phase coexistence occurs at the values of $\rho$ and $T$ at which $-\Delta \tilde{W} = 0$ and derivatives of $\Delta{W}$ with respect to order parameters are zero. In Fig.~(7.6) we show phase diagram in plane of density $\rho$ and pressure $P^* = \beta P / \rho$ obtained from the MSA and PY results for $\beta a_2 = 1$. The pressure is obtained using the compressibility relation ~\cite{Hansen2006}.

In the case of the isotropic phase,
\begin{equation}
\frac{\beta P}{\rho}
= 1 - \frac{1}{\rho} \int_0^\rho d\rho' \, \rho' \, \hat{c}_{000}(0,\rho')
\tag{7.26}
\end{equation}

For the nematic phase the relation is found to be ~\cite{Mishra2007}
\begin{equation}
\frac{\beta P}{\rho}
= \frac{1}{\rho} \int_0^\rho 
\frac{d\rho'}{1 + \phi(\rho')}
\tag{7.27}
\end{equation}

where
\begin{equation}
\phi(\rho') =
\frac{\rho'}{(4\pi)^{1/2}}
\sum_{\ell_1,\ell_2}
\sqrt{(2\ell_1+1)(2\ell_2+1)} \,
P_{\ell_1} P_{\ell_2}
\int d{r} r^2 \, h_{\ell_1 \ell_2 0}({r}, \rho')
\tag{7.28}
\end{equation}

The plateau corresponds to the change in the density at the transition.
The difference in the results of the two theories are due to the differences in their values of DPCFs.

\bigskip

\begin{figure}[h]
    \centering
   \includegraphics[width=0.7\linewidth]{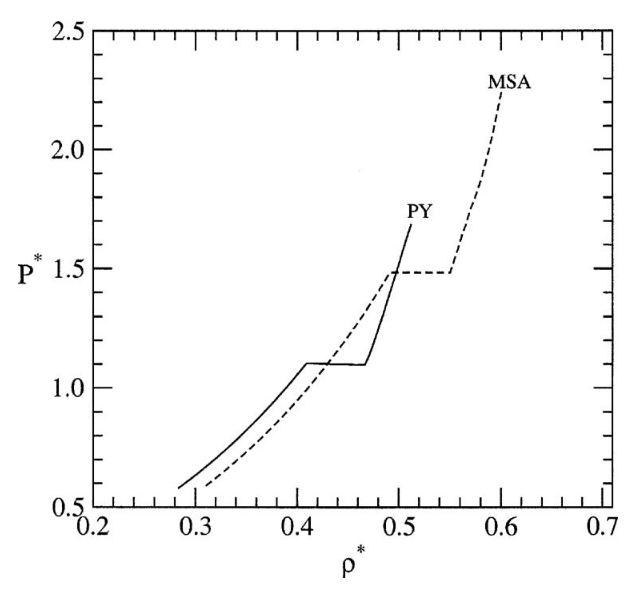}
    \caption*{\textbf{Fig.7.6} $P^*$--$\rho^*$ isotherms for the MSA and PY closures for $\beta a_2 = 1$. 
    Other potential parameters are the same as in Fig.~7.1. The plateau 
    corresponds to the change in density at the isotropic--nematic transition. 
    Reproduced with permission from Ref.~\cite{Mishra2007}.}
    \label{fig:pstar_rho_isotherm}
\end{figure}

\noindent
\subsection{Nematic phase of a system of molecules of broken spherical symmetry}

One of the potential models that mimics the asymmetry in molecular shape and have anisotropic attraction between molecules is the Gay--Berne (GB) pair potential ~\cite{Gay1981}. In the GB potential, molecules are viewed as rigid units with axial symmetry. Each individual molecule $i$ is represented by a centre-of-mass position $\mathbf{r}_i$ and an orientational unit vector $\hat{\mathbf{e}}_i$, which is in the direction of the main symmetry axis of the molecule. The GB interaction energy between a pair of molecules $(i,j)$ is given by 

\begin{equation}
u(\mathbf{r}_{ij}, \hat{\mathbf{e}}_i, \hat{\mathbf{e}}_j)
=
4\epsilon(\hat{\mathbf{e}}_i, \hat{\mathbf{e}}_j, \hat{\mathbf{r}}_{ij})
\left( R^{-15} - R^{-6} \right)
\tag{7.25}
\end{equation}
where
\begin{equation}
R =
\frac{r_{ij} - \sigma(\hat{\mathbf{e}}_i, \hat{\mathbf{e}}_j, \hat{\mathbf{r}}_{ij}) + \sigma_0}
{\sigma_0}
\tag{7.27}
\end{equation}
Here $\sigma_0$ is a constant defining the molecular diameter, $r_{ij}$ is the distance between the centres of mass of molecules $i$ and $j$, and 
\[
\hat{\mathbf{r}}_{ij} = \frac{\mathbf{r}_{ij}}{|\mathbf{r}_{ij}|}
\]
is a unit vector along the centre--centre vector $\mathbf{r}_{ij} = \mathbf{r}_i - \mathbf{r}_j$. $\sigma(\hat{\mathbf{e}}_i, \hat{\mathbf{e}}_j, \hat{\mathbf{r}}_{ij})$ is the distance (for a given molecular orientation) at which the intermolecular potential vanishes and is 
given by

\begin{equation}
\sigma(\hat{\mathbf{e}}_i, \hat{\mathbf{e}}_j, \hat{\mathbf{r}}_{ij})
=
\sigma_0
\left[
1 - \chi
\frac{
(\hat{\mathbf{e}}_i \cdot \hat{\mathbf{r}}_{ij})^2
+
(\hat{\mathbf{e}}_j \cdot \hat{\mathbf{r}}_{ij})^2
-
2\chi
(\hat{\mathbf{e}}_i \cdot \hat{\mathbf{r}}_{ij})
(\hat{\mathbf{e}}_j \cdot \hat{\mathbf{r}}_{ij})
(\hat{\mathbf{e}}_i \cdot \hat{\mathbf{e}}_j)
}{
1 - \chi^2 (\hat{\mathbf{e}}_i \cdot \hat{\mathbf{e}}_j)^2
}
\right]^{-1/2}
\tag{7.28}
\end{equation}

The parameter $\chi$ is a function of the ratio
\[
x_0 \equiv \frac{\sigma_e}{\sigma_s},
\]
which is defined in terms of the contact distances when the particles are end-to-end $(e)$ 
and side-by-side $(s)$:

\begin{equation}
\chi = \frac{x_0^2 - 1}{x_0^2 + 1}
\tag{7.29}
\end{equation}

The orientational dependence of the potential well depth is given by a product of two functions

\begin{equation}
\epsilon(\hat{\mathbf{e}}_i, \hat{\mathbf{e}}_j, \hat{\mathbf{r}}_{ij})
=
\epsilon_0
\epsilon^{\nu}(\hat{\mathbf{e}}_i, \hat{\mathbf{e}}_j)
\epsilon^{'\mu}(\hat{\mathbf{e}}_i, \hat{\mathbf{e}}_j, \hat{\mathbf{r}}_{ij})
\tag{7.30}
\end{equation}

where the scaling parameter $\epsilon_0$ is the well depth for the cross configuration
\[
(\hat{\mathbf{e}}_i \cdot \hat{\mathbf{r}}_{ij})
=
(\hat{\mathbf{e}}_j \cdot \hat{\mathbf{r}}_{ij})
=
(\hat{\mathbf{e}}_i \cdot \hat{\mathbf{e}}_j)
=
0.
\]
The first of these functions

\begin{equation}
\epsilon(\hat{\mathbf{e}}_i, \hat{\mathbf{e}}_j)
=
\left[
1 - \chi^2 (\hat{\mathbf{e}}_i \cdot \hat{\mathbf{e}}_j)^2
\right]^{-1/2}
\tag{7.31}
\end{equation}

favors the parallel alignment of the particle and so aids liquid crystal formations.
The second function has a form analogous to
$\sigma(\hat{\mathbf{e}}_i, \hat{\mathbf{e}}_j, \hat{\mathbf{r}}_{ij})$, i.e.

\begin{equation}
\epsilon'(\hat{\mathbf{e}}_i, \hat{\mathbf{e}}_j, \hat{\mathbf{r}}_{ij})
=
\left[
1 - \chi'
\left(
\frac{
(\hat{\mathbf{e}}_i \cdot \hat{\mathbf{r}}_{ij})^2
+
(\hat{\mathbf{e}}_j \cdot \hat{\mathbf{r}}_{ij})^2
-
2\chi'
(\hat{\mathbf{e}}_i \cdot \hat{\mathbf{r}}_{ij})
(\hat{\mathbf{e}}_j \cdot \hat{\mathbf{r}}_{ij})
(\hat{\mathbf{e}}_i \cdot \hat{\mathbf{e}}_j)
}{
1 - \chi'^2 (\hat{\mathbf{e}}_i \cdot \hat{\mathbf{e}}_j)^2
}
\right)
\right]
\tag{7.32}
\end{equation}

The parameter $\chi'$ is determined by the ratio of the well depth as

\begin{equation}
\chi' = \frac{k'^{1/\mu} - 1}{k'^{1/\mu} + 1}
\tag{7.33}
\end{equation}
Here, $k'$ is the well-depth ratio for the side-by-side and end-to-end configuration.\\

The GB model contains four parameters $(x_0, k', \mu, \nu)$ that determine the anisotropy in the repulsive and attractive forces in addition to two parameters $\sigma_0, \epsilon_0$ that scale the distance and energy, respectively. Though $x_0$ measures the anisotropy of the repulsive core, it also determines the difference in the depth of the attractive well between the side-by-side and the cross configurations. Both $x_0$ and $k'$ play an important role in stabilizing the liquid crystalline phases. The exact role of the other two parameters $\mu$ and $\nu$ are not very obvious; though they appear to affect the anisotropic attractive forces in a subtle way.\\

The phase diagram found from computer simulations for the system interacting via the GB potential of equations (7.26)--(7.33) exhibits isotropic, nematic, and Sm B phases ~\cite{DeMiguel1991,Brown1998,Bates1999} for $x_0 = 3.0$, $k' = 5.0$, $\mu = 2$ and $\nu = 1$. An island of SmA is, however, found to appear in the phase diagram at a value of $x_0$ slightly greater than $3.0$ ~\cite{Brown1998}. The range of Sm A extends to both higher and lower temperatures as $x_0$ is increased. Also, as $x_0$ is increased, the isotropic--nematic (I--N) transition is seen to move to lower density (and pressure) at a given temperature. Bates and Luckhurst ~\cite{Bates1999} have investigated the phases and phase transitions for the GB potential with $x_0 = 4.4$, $k' = 20.0$, $\mu = 1$ and $\nu = 1$ using the isothermal--isobaric Monte Carlo simulations. At low pressure, they found isotropic, Sm A, and Sm B phases but not the nematic phase. However, as the pressure is increased, the nematic phase also becomes stabilized, and a sequence of I--N--Sm A and Sm B was found.

The GB model, along with other models of broken molecular symmetry, has been studied using both approximate versions, SODFT and WDA of DFT ~\cite{Mishra2004,Singh2002}. Results found from such studies can be found in a review article by Ram ~\cite{Ram2014} and references cited therein. Here, our interest is in broken-symmetry contributions to  correlation functions and in determination of parameters of the I--N transition using EDFT.

Mishra and Singh ~\cite{Hansen2006} calculated the pair correlation functions of the nematic phase interacting via the GB potential. Since the phase is uniaxial and the constituent molecules are axially symmetric, one can use Eqs.~(7.5) and (7.6) to expand the one-particle and two-particle functions, respectively. The constraints are $m_1 +m_2 - m = 0$, which arise due to axial symmetry of the phase around the axis parallel to the director  and $l_1 + l_2 + l$ as well as each $l$ are even because of molecular symmetry. The OZ equations (6.4) and (6.5) were solved using PY closer relation which was also split in the symmetry conserving and symmetry breaking parts, as given below:
\begin{equation}
c^{(0)}(\mathbf{x}_1,\mathbf{x}_2)
=
\left[M(\vec{x}_1,\vec{x}_2)
\right]
\left[
1 + h^{(0)}(\mathbf{x}_1,\mathbf{x}_2)
- c^{(0)}(\mathbf{x}_1,\mathbf{x}_2)
\right],
\tag{7.34}
\end{equation}

and
\begin{equation}
c^{(b)}(\mathbf{x}_1,\mathbf{x}_2)
=
\left[
M(\vec{x}_1,\vec{x}_2)
\right]
\left[
h^{(b)}(\mathbf{x}_1,\mathbf{x}_2)
- c^{(b)}(\mathbf{x}_1,\mathbf{x}_2)
\right].
\tag{7.35}
\end{equation}

Mishra and Singh first solved OZ equation (6.4) and the PY relations of Eq.~(7.34) and determined values of $c^{(0)}_{l_1 l_2 l m_1 m_2 m}(r^*)$, $h^{(0)}_{l_1 l_2 l m_1 m_2 m}(r)$ for values of $l$, $l_i$ up to $l_{\max} = 8$ at reduced temperature $T^* (= k_B T/\epsilon_0) = 1.0$ and for densities $0 \le \rho^* (= \rho \sigma_0^3) \le 0.36$. To solve Eqs.~(6.5), and (7.35), they set up linear equations for $c^{(b)}_{l_1 l_2 l m_1 m_2 m}(r^*)$ and $h^{(b)}_{l_1 l_2 l m_1 m_2 m}(r^*)$, using expansions in spherical harmonics. In these equations $h^{(0)}_{l_1 l_2 l m_1 m_2 m}(r^*)$ , $c^{(0)}_{l_1 l_2 l m_1 m_2 m}(r^*)$, and  order parameter $P_l$ appear. The solution is obtained using the numerical procedure outlined above in Section.~7.1.

\begin{figure}
\centering
\includegraphics[width=\columnwidth]{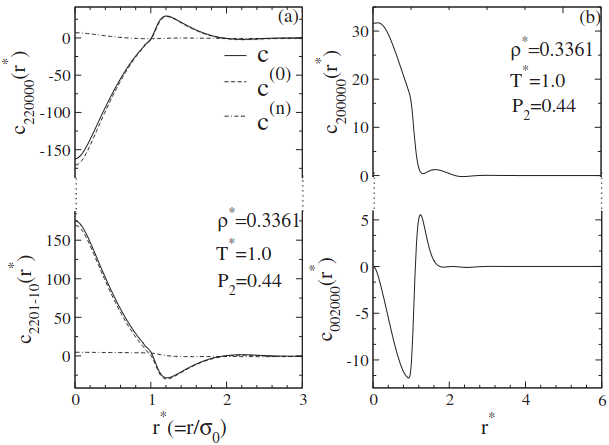}
\caption*{\textbf{Fig.7.7} 
Harmonic coefficients $c_{\ell_1l_2 lm_1m_2m}(r^*)$ in the director frame for the GB potential. In (a) we show two those coefficients which  preserve the rotational invariance while in (b) the coefficients that arise due to
symmetry breaking. Reproduced with permission from Ref.~\cite{Mishra2006}}
\end{figure}

Some harmonic coefficients of direct pair correlation function in the director frame for $T^* = 1.0$, $\rho^* = 0.3361$, and $P_2 = 0.44$ are shown in Fig. 7.7. While the coefficients $c_{220000}(r^*)$ and $c_{2201-10}(r^*)$ shown in Fig.~7.7(a) survive both in isotropic and nematic phases, the coefficients  $c_{200000}(r^*)$ and $c_{002-000}(r^*)$shown in Fig.~7.7(b) survive only in the nematic phase and vanish in the isotropic phase. The contribution arising due to symmetry breaking to coefficients $c_{220000}(r^*)$ and $c_{2201-10}(r^*)$ is found to be small compared to the symmetry conserving part.

\begin{figure}
\centering
\includegraphics[width=\columnwidth]{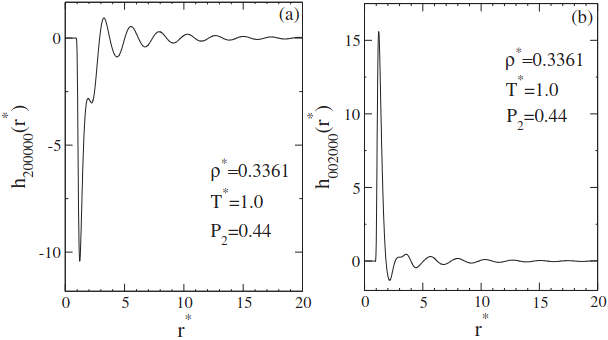}
\caption*{\textbf{Fig.7.8}
Coefficients $h_{\ell_1l_2m_1m_2 m}(r)$ which survive only in the nematic phase of a system interacting via GB potential. Reproduced with permission from ref.~\cite{Hansen2006}.}
\end{figure}

\begin{figure}
\centering
\includegraphics[width=0.7\columnwidth]{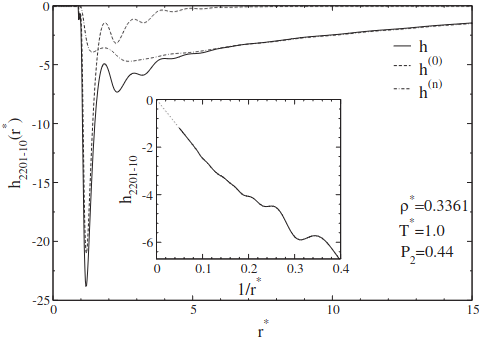}
\caption*{\textbf{Fig.7.9} Harmonic coefficient $h_{2201-10}(r^*)$ in the director frame. Details are same as in Fig. 7.7. Reproduced with permission trom Ref. ~\cite{Mishra2006}}
\end{figure}

Few selected harmonic coefficients of h are shown in Figs. 7.8 and 7.9 in director space. In Fig. 7.8 we plot the harmonic coefficients $h_{200 000}(r^*) $ and $h_{002 000}(r^*)$ which survive only in the nematic phase and note the oscillatory behavior which continues to survive for large values of $r^*$. In Fig. 7.9 we plot coefficient $h_{2201 10}(r^*)$ which is of fundamental importance as it defines nematic elastic constants ~\cite{deGennes1993} and decays as $1/r^*$ at large distance. This long--range tail behavior is attributed to the director transverse fluctuations which give rise to orientational wave excitations, i.e. the Goldstone modes. This can be seen by taking the tensor order parameter $Q_{\alpha \beta}= \frac{1}{N}\sum_{i=1}^N \frac{3}{2} \times (e_{i\alpha}e_{i\beta}-\frac{1}{3}\delta_{\alpha\beta})$, where ,$\alpha,\beta=x,y,z$ and $e_{i\alpha}$ is the $\alpha$ component of the molecular axis vector $\mathbf{e}_i$  of each molecule and $\delta_{\alpha\beta}$ the Kronecker symbol and calculating ( assuming that the director is along $z$ and the $y$ axis is perpendicular to wave vector $\mathbf{k}$) the correlation $\langle Q_{xz}(\mathbf{k})Q_{xz}(-\mathbf{k})\rangle $. The result involves coefficient  $h^{(b)}_{l_1 l_2 m_1 m_2 m}(\bar{\mathbf{k}})$ with $|{m_1}|,|{m_2}|=1$. These  coefficients which are the Fourier transform of $h^{(b)}_{l_1 l_2 m_1 m_2 m}(r^*)$ behave as $\frac{1}{\mathbf{k^2}}$ for $\mathbf{k}\xrightarrow{} 0$

One of the important features of the total pair correlation function of a nematic phase is the appearance of the $1/r$ tail in some of the harmonic coefficients of $h(\vec{x}_1,\vec{x}_2)$. This has been seen in computer simulations ~\cite{PhuongSchmid2003}, in analytical solutions of the MSA theory ~\cite{Holovko1999}, and in numerical solutions ~\cite{Mishra2006,Mishra2007} of the MSA and PY theories.

Using values of DPCFs calculated above and two order parameters $P_2$ and $P_4$ Mishra and Singh ~\cite{Mishra2006} found that at $T^* = 1$ the I--N transition takes place with $\rho_0^* = 0.317$, $\Delta\rho_0^* = 0.026$, $P_2 = 0.644$, and $P_4 = 0.332$. These values are in very good agreement with computer simulation values, $\rho_0^* = 0.317$, $P_2 = 0.66$, and $P_4 = 0.29$ ~\cite{Longa2001}.

\section{Pair correlation functions of crystalline solids}

Pair correlation functions (PCFs) of fluids are determined using scattering experiments, integral equation theory (IET) and computer simulations ~\cite{Hansen2006}. However, these methods suffer from severe limitations when applied to crystals. In the structure factor of a crystal measured by scattering experiments, contribution due to PCFs is buried in the non-Bragg angles diffusively scattered background (see Eq (4.33) and Appendix C). On the theory side, solving the inhomogeneous OZ equation (2.38) is difficult for reasons stated in Sections 2 and 6.

In a method developed by Singh and coworkers ~\cite{Bharadwaj2013,SinghSingh2009,Jaiswal2014,Singh2011,Bharadwaj2017,Jaiswal2013} use is made of Eqs (6.1)-(6.5) to separate correlation functions and the OZ equation (2.38) into two parts; one which corresponds to a homogeneous system (symmetry conserving part) and the other in which broken symmetry part appears. For evaluation of symmetry conserving components of PCFs and their derivatives with respect to density $\rho$ integral equation theory (IET) of homogeneous fluids (see Appendix A) is used. For broken symmetry component of direct pair correlation function (DPCF) , a series expressed in ascending powers of order parameters is used ~\cite{Bharadwaj2013,SinghSingh2009,Singh2011}. This series involves $m$-particles (where $m \ge 3$) direct correlation functions. Using known values of DPCF, the OZ equation (6.5) is solved to give total pair correlation function ~\cite{Jaiswal2014}. In this section we implement the approach summarized above to calculate DPCF of a 3-dimensional ideal crystal and in the next section to calculate both the DPCF and the total pair correlation function of a 2-dimensional crystal.

The pair potentials used in the calculation are inverse power potentials (IPPs), $u(r) = \epsilon(\sigma/r)^n$, where $\epsilon, \sigma$ and $n$ are potential parameters and $r$ is the molecular separation. The parameter $n$ measures the softness of the potential. The IPPs have a simple scaling property according to which density and temperature can be combined to give a single parameter defined in 3-dimensions as,
\begin{equation}
    \gamma = \rho \sigma^3 (\beta\epsilon)^{3/n} = \rho^* T^{*-3/n} \tag{8.1}
\end{equation}

Using the scaling relation the potential is written as:
\begin{equation}
    \beta u(r) = \left(\frac{4\pi}{3} \gamma\right)^{n/3} \frac{1}{r^n} \tag{8.2}
\end{equation}
where $r$ is measured in the unit of $a_0 = (3/4\pi\rho)^{1/3}$.

The reason for preferring IPPs is that the range of the potential can be varied by changing the value of $n$ and the fact that the equations of state and phase coexistence boundaries of these potentials have been extensively investigated by computer simulations \cite{Alder1968,Prestipino2005,Agrawal1995,Davidchack2005,Tan2011} for several values of n giving "exact" results for comparison. The more repulsive ($n \ge 7$) systems have been found to freeze into a face-centred cubic (fcc) structure while the soft repulsion ($n < 7$) freeze into a body-centered cubic (bcc) structure. The fluid-bcc-fcc triple point is found at $1/n \simeq 0.15$. The atomic arrangements in the two cubic structures differ substantially; the fcc is close packed in real space and the density inhomogeneity is much sharper than the bcc, which is open structure in real space but close packed in Fourier space. However, in spite of this difference in the atomic arrangement, the two structures have a small difference in free energy (or chemical potential) at the fluid-solid transition and, therefore, a correct prediction of the relative stability of the two cubic structures is a stringent test of any theory.

\subsection{Determination of symmetry conserving components $h^{(0)}(r)$, $c^{(0)}(r)$ and their derivatives with respect to density}

The IET of homogeneous systems (fluids), described in Appendix A is used to calculate $h^{(0)}$, $c^{(0)}$ and their derivatives with respect to $\rho$. In the calculation the OZ equation (6.4) and the Roger and Young (RY) ~\cite{Rogers1984} closure relation given by Eq (A.13) (in Appendix A) are used. For the IPPs potential the RY closure relation gives very accurate values of PCFs and their derivatives with respect to density ~\cite{Bharadwaj2013}. The differentiation of the OZ equation (6.4) and the RY closure relation (A-13) with respect to $\rho$ yields the following relations~\cite{Bharadwaj2013}:

\begin{align}
    \frac{\partial h^{(0)}(r)}{\partial \rho} &= \frac{\partial c^{(0)}(r)}{\partial \rho} + \int d\vec{r}' c^{(0)}(r') h^{(0)}(|\vec{r}' - \vec{r}|) \nonumber \\
    &\quad + \rho \int d\vec{r}' \frac{\partial c^{(0)}(r')}{\partial \rho} h^{(0)}(|\vec{r}' - \vec{r}|) \nonumber \\
    &\quad + \rho \int d\vec{r}' c^{(0)}(r') \frac{\partial h^{(0)}(|\vec{r}' - \vec{r}|)}{\partial \rho} \tag{8.3}
\end{align}

and

\begin{equation}
    \frac{\partial h^{(0)}(r)}{\partial \rho} = \exp[-\beta u(r)] \exp[\chi(r)f(r)] \frac{\partial \chi(r)}{\partial \rho} \tag{8.4}
\end{equation}

\begin{align}
\frac{\partial^2 h^{(0)}(r)}{\partial \rho^2}
&= \frac{\partial^2 c^{(0)}(r)}{\partial \rho^2}
+ 2 \int d\vec{r}' \Bigg[
\frac{\partial c^{(0)}(r')}{\partial \rho}
\, h^{(0)}(|\vec{r}' - \vec{r}|)
+ c^{(0)}(r')
\, \frac{\partial h^{(0)}(|\vec{r}' - \vec{r}|)}{\partial \rho}
\Bigg] \nonumber \\
&\quad + \rho \int d\vec{r}' \Bigg[
2 \frac{\partial c^{(0)}(r')}{\partial \rho}
\, \frac{\partial h^{(0)}(|\vec{r}' - \vec{r}|)}{\partial \rho}
+ c^{(0)}(r')
\, \frac{\partial^2 h^{(0)}(|\vec{r}' - \vec{r}|)}{\partial \rho^2}
\nonumber \\
&\qquad\qquad\quad
+ \frac{\partial^2 c^{(0)}(r')}{\partial \rho^2}
\, h^{(0)}(|\vec{r}' - \vec{r}|)
\Bigg]
\tag{8.5}
\end{align}

and

\begin{equation}
    \frac{\partial^2 h^{(0)}(r)}{\partial \rho^2} = \exp[-\beta u(r)] \exp[\chi(r)f(r)] \left[ \frac{\partial^2 \chi(r)}{\partial \rho^2} + \left( \frac{\partial \chi(r)}{\partial \rho} \right)^2 f(r) \right] \tag{8.6}
\end{equation}

\begin{figure}
\centering
\includegraphics[width=\columnwidth]{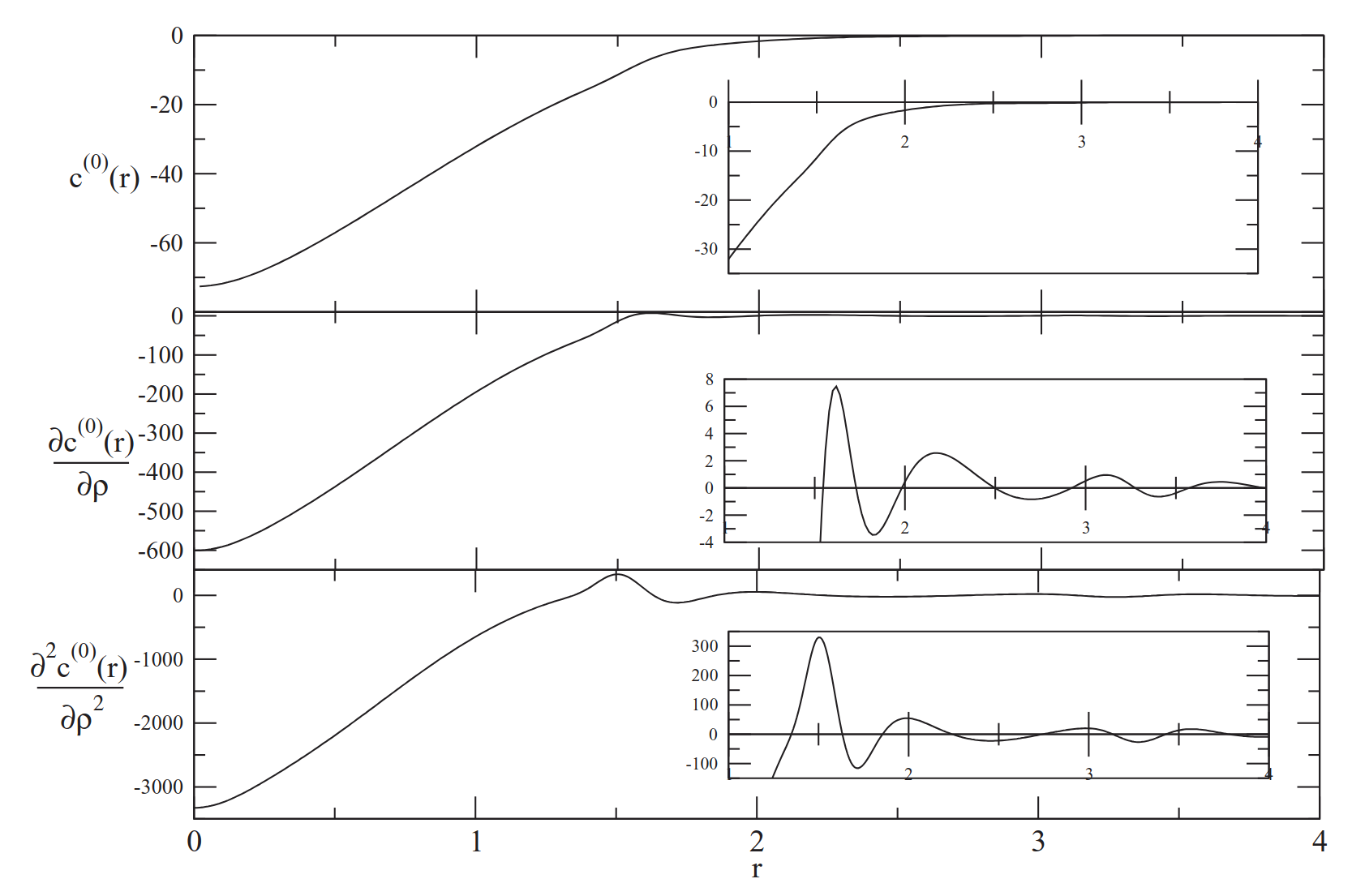}
\caption*{\textbf{Fig.8.1} 
Plots of $c^{(0)}(r)$, $\dfrac{\partial c^{(0)}(r)}{\partial \rho}$, and $\dfrac{\partial^2 c^{(0)}(r)}{\partial \rho^2}$ as functions of $r$ for thr IPP $n=6$ and $\gamma = 2.30$, which is close to the freezing point. The distance $r$ is expressed in units of $a_0 = \left( \dfrac{3}{4\pi\rho} \right)^{1/3}$. Insets show magnified views of the respective quantities for $r \geq 1$. Reproduced with permission from Ref.~\cite{Bharadwaj2013}}
\end{figure}

The solution of the closed set of coupled equations (65),(A.13) and (8.3)--(8.6) gives the values of $h^{(0)}(r)$, $c^{(0)}(r)$, $\dfrac{\partial h^{(0)}(r)}{\partial \rho}$, $\dfrac{\partial c^{(0)}(r)}{\partial \rho}$, $\dfrac{\partial^2 h^{(0)}(r)}{\partial \rho^2}$, and $\dfrac{\partial^2 c^{(0)}(r)}{\partial \rho^2}$ as functions of $r$ for a given potential $u(r)$.

In Fig.~8.1, the functions $c^{(0)}(r)$, $\dfrac{\partial c^{(0)}(r)}{\partial \rho}$, and 
$\dfrac{\partial^2 c^{(0)}(r)}{\partial \rho^2}$ are plotted for the inverse power potential (IPP) with $n=6$ and $\gamma = 2.30$, which is close to the freezing point (see Sec.~10).

\subsection{Determination of three- and four-particle direct correlation functions of a homogeneous systems}

The $m$-particle direct correlation function $c_m^{(0)}$ is related with the ${m-2}^{th}$  order functional derivative of pair function with respect to density $\rho(\vec{r})$ (Eq. 2.35):
\begin{equation}
    \frac{\delta^{m-2} c(\vec{r}_1, \vec{r}_2)}{\delta\rho(\vec{r}_3) \dots \delta\rho(\vec{r}_m)} = c_m(\vec{r}_1, \vec{r}_2, \dots, \vec{r}_m) \tag{8.7}
\end{equation}

In the limit of the system being homogeneous, Eq. (8.7) yields the following relations:
\begin{equation}
    \frac{\partial c^{(0)}(r, \rho)}{\partial \rho} = \int d\vec{r}_3 c_3^{(0)}(\vec{r}_1, \vec{r}_2, \vec{r}_3; \rho) \tag{8.8}
\end{equation}

\begin{equation}
    \frac{\partial^2 c^{(0)}(r, \rho)}{\partial \rho^2} = \int d\vec{r}_3 \frac{\partial c_3^{(0)}(\vec{r}_1, \vec{r}_2, \vec{r}_3; \rho)}{\partial \rho} = \int d\vec{r}_3 d\vec{r}_4 c_4^{(0)}(\vec{r}_1, \vec{r}_2, \vec{r}_3, \vec{r}_4; \rho) \tag{8.9}
\end{equation}

and so on, where $c_m^{(0)}$ is the $m$-particle direct correlation function of the fluid of density $\rho$. Note that these equations are exact and connect the derivatives of $c^{(0)}(r)$ with respect to density $\rho$ with higher-order direct correlation functions. Using known values of the derivatives, these equations can be solved to find values of $c_m^{(0)}$. However, in view of the nature of these equations, the solution is not straightforward; it requires some sort of simplification. Barrat et al.~\cite{Barrat1987} determined $c_3^{(0)}$ from Eq. (8.8) by first writing it as a product of an arbitrary function $t(r)$,

\begin{equation}
    c_3^{(0)}(\vec{r}_1, \vec{r}_2, \vec{r}_3) = t(r_{12}) t(r_{13}) t(r_{23}) \tag{8.10a}
\end{equation}

which allowed separation of variables $r_{12}$, $r_{13}$ and $r_{23}$ and then determining the values of $t(r)$ from Eq. (8.8). Method of Ref.~\cite{Barrat1987} was used and extended by Bharadwaj et al. ~\cite{Bharadwaj2013} to calculate $c_3^{(0)}$ and $c_4^{(0)}$. The description given below is based on the one given in Ref.~\cite{Bharadwaj2013}.

Eq.(8.10a) can be rewritten using a diagram as 

\begin{equation}
c^{(0)}_{3}(\mathbf{r}_1,\mathbf{r}_2,\mathbf{r}_3)
\equiv
\raisebox{-0.4cm}{\includegraphics[width=1.8cm]{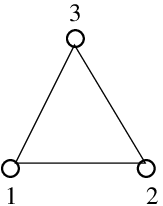}}
\tag{8.10b}
\end{equation}
where a line linking particles $i$ and $j$ denotes a $t(r)$ function and each circle (representing a particle) carries weight unity. Similarly the relation (8.8) is written as 
\begin{equation}
\frac{\partial c^{(0)}(r,\rho_0)}{\partial \rho}
=
\raisebox{-0.35cm}{\includegraphics[width=1.8cm]{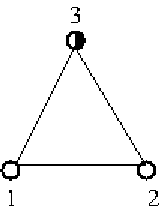}}
\tag{8.11}
\end{equation}
where the half-black circle represents the particle over which
integration is performed over its all configurations and all circles carry weight unity. Using known values of $\frac{\partial c^{(0)}(r,\rho_0)}{\partial \rho}$, Eq.(8.11) is solved to find values of $t(r)$. Values of $t(r)$ as a function of $r$ are shown in Fig.8.1 for a system interacting via IPPs for $n = 6, 4$ and $\gamma = 2.30,5.60$ respectively. It is seen that the dependence of function $t(r)$  on $\gamma$ is similar to that of $c^{(0)}(r)$ but that magnitude is about 10 times smaller.\\
Taking the derivative of both sides of Eq.~(8.10a) with respect to $\rho$ one gets
\begin{equation}
\frac{\partial c^{(0)}_3(\vec r_1,\vec r_2,\vec r_3)}{\partial \rho}
=
\frac{\partial t(r_{12})}{\partial \rho} t(r_{13})t(r_{23})
+ t(r_{12})\frac{\partial t(r_{13})}{\partial \rho}t(r_{23})
+ t(r_{12})t(r_{13})\frac{\partial t(r_{23})}{\partial \rho}.
\tag{8.12}
\end{equation}
Substitution of this into Eq.~(8.9) leads to

\begin{equation}
\frac{\partial^2 c^{(0)}(r,\rho)}{\partial \rho^2}
=
\int d\vec r' \left[
\frac{\partial t(r)}{\partial \rho_0}t(r')t(|\vec r' - \vec r|)
+
t(r)\frac{\partial t(r')}{\partial \rho}t(|\vec r' - \vec r|)
+
t(r)t(r')\frac{\partial t(|\vec r' - \vec r|)}{\partial \rho}
\right].
\tag{8.13}
\end{equation}
where $r_{12} = r$, $r_{13} = r'$, and $r_{23} = |\mathbf{r}' - \mathbf{r}|$. As values of $t(r)$ are known, Eq.~(8.13) is used to find values of $\frac{\partial t(r)}{\partial \rho}$. Guided by the relation (8.11) we write $\frac{\partial t(r)}{\partial \rho}$ as 

\begin{align}
\frac{\partial t(r)}{\partial \rho}
&=
s(r)\int d\mathbf{r}' \,
s(r')\, s\!\left(|\mathbf{r}'-\mathbf{r}|\right)
\tag{8.14a} \\
&\equiv
\raisebox{-0.4cm}{\includegraphics[width=1.8cm]{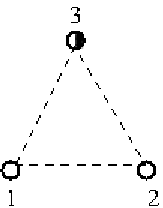}}
\tag{8.14b}
\end{align}
where a dashed line connecting the particles $i$ and $j$ is $s(r)$ function. Using the already determined values of $\partial t(r)/\partial \rho$ at a given value of $\rho$, one determines values of  $s(r)$ in the same way as for values of $t(r)$ from the known values of $\partial c^{(0)}(r)/\partial \rho_0$. In Fig.~8.3, values of $s(r)$ for $n = 6, 4$ and $\gamma = 2.30, 5.60$  as a function of $r$ are plotted for the IPP fluid. Unlike $t(r)$, the function $s(r)$ shows many oscillations before it decays to zero.
\begin{figure}
\centering
\includegraphics[width=11cm]{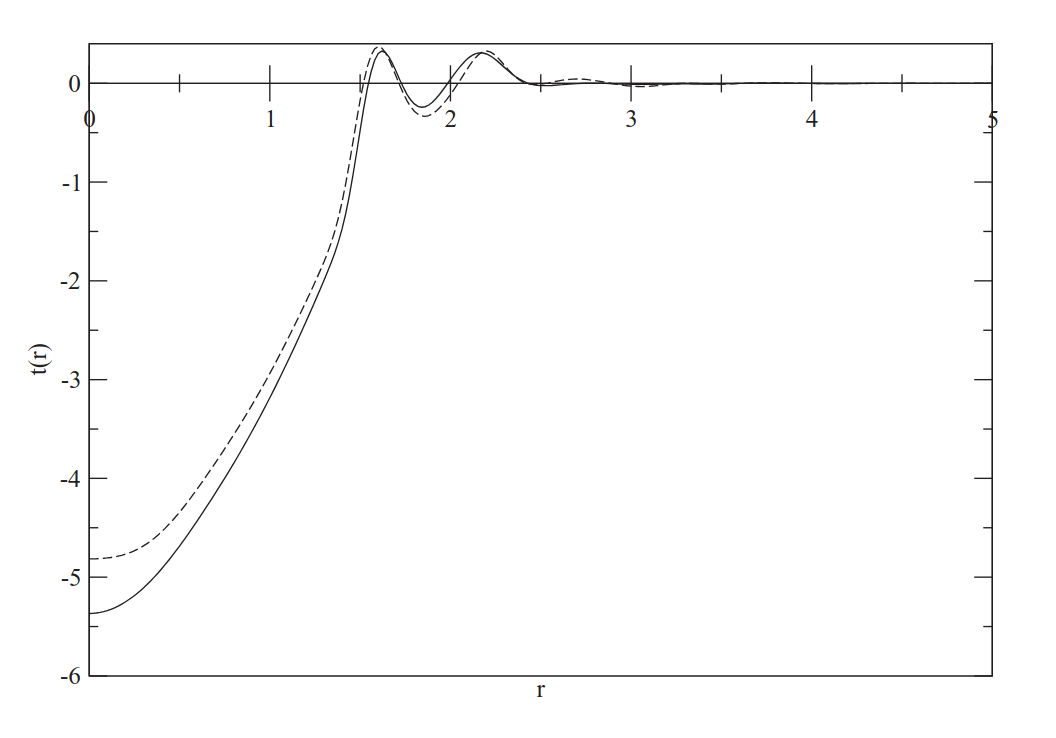}
\caption*{\textbf{Fig. 8.2} Plot of $t(r)$ vs $r$ for a system interacting via IPPs $n = 6$ at $\gamma = 2.32$ and $n=4$ at $\gamma = 5.6$. the distance $r$ is in unit of $a_0=\left(\frac{3}{4\pi\rho}\right)^{1/3}$. The dashed line represents values for $n=4$ and full line for $n=6$ . Reproduced with permission from Ref.~\cite{Bharadwaj2013}}
\end{figure}
\begin{figure}
\centering
\includegraphics[width=11cm]{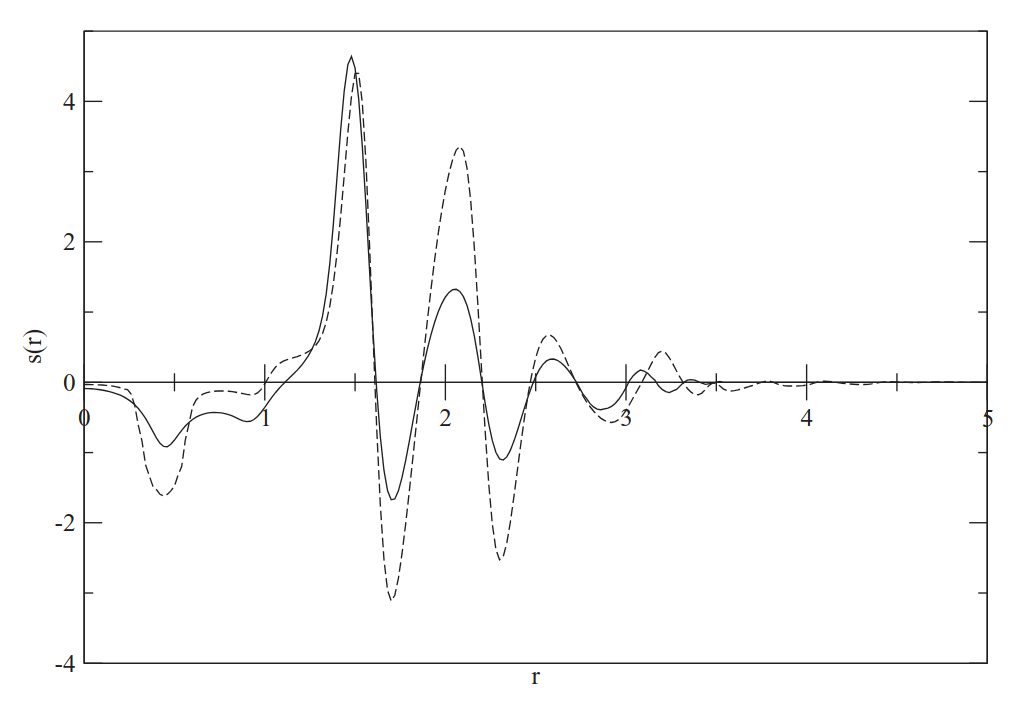}
\caption*{\textbf{Fig. 8.3} Plot of $s(r)$ vs $r$ for $n = 6$ at $\gamma = 2.32$ and $n = 4$ at $\gamma = 5.60$ for IPP fluid. The distance $r$ is in unit of $a_0 = (\frac{3}{4\pi \rho})^{1/3}$. The dashed curve represents values for $n=4,\gamma_l=5.60$ and full curve for $n=6,\gamma_l=2.32$. Reproduced with permission from Ref.~\cite{Bharadwaj2013}}
\end{figure}\\
From Eqs.(8.9),(8.12)and(8.14) we have
\begin{equation}
c^{(0)}_{4}(\mathbf{r}_1,\mathbf{r}_2,\mathbf{r}_3,\mathbf{r}_4)
\equiv
\raisebox{-0.4cm}{\includegraphics[width=1.8cm]{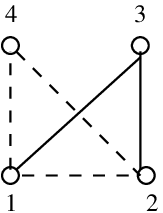}}+\raisebox{-0.4cm}{\includegraphics[width=1.8cm]{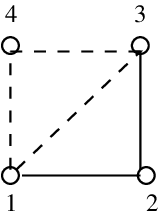}}+\raisebox{-0.4cm}{\includegraphics[width=1.8cm]{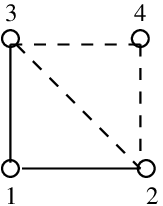}},
\tag{8.15}
\end{equation}
where a dashed line represents the $s(r)$ bond and a full line the $t(r)$ bond.

Each diagram of Eq. (8.15) has two circles connected by three bonds, two $s$ bonds (dashed line) and one $t$ bond (full line), where one of the remaining circles is connected by two $t$ bonds and the other by two $s$ bonds. By permuting circles in the last two diagrams one can convert one diagram into another.

Using notation, $r = |\vec{r}_2 - \vec{r}_1|$, $r' = |\vec{r}_3 - \vec{r}_1|$, $r'' = |\vec{r}_4 - \vec{r}_1|$, $|\vec{r}' - \vec{r}| = |\vec{r}_3 - \vec{r}_2|$, $|\vec{r}'' - \vec{r}| = |\vec{r}_4 - \vec{r}_2|$ and $|\vec{r}'' - \vec{r}'| = |\vec{r}_2 - \vec{r}_3|$, Eqs. (8.16) and (8.17) can be written as:

\begin{equation}
    c_3^{(0)}(\vec{r}_1, \vec{r}_2, \vec{r}_3) = t(r) t(r') t(|\vec{r}' - \vec{r}|) \tag{8.16}
\end{equation}
and
\begin{align}
    c_4^{(0)}(\vec{r}_1, \vec{r}_2, \vec{r}_3, \vec{r}_4) &= s(r) t(r') s(r'') t(|\vec{r}' - \vec{r}|) s(|\vec{r}'' - \vec{r}|) \nonumber \\
    &\quad + t(r) s(r') s(r'') t(|\vec{r}' - \vec{r}|) s(|\vec{r}'' - \vec{r}'|) \nonumber \\
    &\quad + t(r) t(r') s(|\vec{r}' - \vec{r}|) s(|\vec{r}'' - \vec{r}|) s(|\vec{r}'' - \vec{r}'|) \tag{8.17}
\end{align}
This change in variables shows that $c_3^{(0)}$ depends on two variables $r$ and $r'$ while $c_4^{(0)}$ depends on three variables $r, r'$ and $r''$.\\
Functions $t$ and $s$ which involve two variables can be expanded in spherical harmonics ~\cite{Bharadwaj2013}, leading to:

\begin{equation}
    t(|\vec{r}' - \vec{r}|) = \frac{2}{\pi} \sum_{lm} A_l(r, r') Y_{lm}(\hat{r}) Y_{lm}^*(\hat{r}') \tag{8.18}
\end{equation}
where
\begin{equation}
    A_l(r, r') = \int_0^\infty dq q^2 t(q) j_l(qr) j_l(qr'), \tag{8.19}
\end{equation}
$j_l(x)$ is the spherical Bessel function and $Y_{lm}(\hat{r})$ the spherical harmonics. Substitution of these expansions for $t$ and $s$ functions in Eqs. (8.16) and (8.17) yields expressions for $c_3^{(0)}$ and $c_4^{(0)}$ which can be used to calculate their values from the known values of $t(r)$ and $s(r)$. In Ref.~\cite{Bharadwaj2013} values of $c_3^{(0)}$ and $c_4^{(0)}$ calculated for systems interacting via IPPs of Eq. (8.2) are given.

\subsection{Broken symmetry component of PCFs of a cystal}

In a crystal the broken symmetry contributions
$h^{(b)}$ and $c^{(b)}$ are invariant only under a discrete set of
translations corresponding to lattice vectors $\mathbf{R}_i$

\begin{equation*}
h^{(b)}(\mathbf{r}_1,\mathbf{r}_2)
=
h^{(b)}(\mathbf{r}_1+\mathbf{R}_i,
\mathbf{r}_2+\mathbf{R}_i),
\end{equation*}
\begin{equation*}
    \tag{8.20}
\end{equation*}
\begin{equation*}
c^{(b)}(\mathbf{r}_1,\mathbf{r}_2)
=
c^{(b)}(\mathbf{r}_1+\mathbf{R}_i,
\mathbf{r}_2+\mathbf{R}_i).
\end{equation*}
If one chooses a center of mass variable
$
\mathbf{r}_c=\frac{1}{2}(\mathbf{r}_1+\mathbf{r}_2)
$
and a distance variable
$\mathbf{r}=\mathbf{r}_2-\mathbf{r}_1 ,
$
then $h^{(b)}$ and $c^{(b)}$ can be expressed as a periodic
function of the center of mass variable and a continuous function
of the difference variable $\mathbf{r}$ ~\cite{McCarley1997};

\begin{equation*}
h^{(b)}(\mathbf{r}_1,\mathbf{r}_2)
=
\sum_{\mathbf{G}}
\exp(i\mathbf{G}\cdot\mathbf{r}_c)\,
h^{(\mathbf{G})}(\mathbf{r}),
\end{equation*}
\begin{equation*}
    \tag{8.21}
\end{equation*}
\begin{equation*}
c^{(b)}(\mathbf{r}_1,\mathbf{r}_2)
=
\sum_{\mathbf{G}}
\exp(i\mathbf{G}\cdot\mathbf{r}_c)\,
c^{(\mathbf{G})}(\mathbf{r}),
\end{equation*}
where $\mathbf{G}$ are reciprocal lattice vectors (RLVs). Since $h^{(b)}$ and $c^{(b)}$ are real and symmetric with respect to
interchange of $\mathbf r_1$ and $\mathbf r_2$,
$h^{(\mathbf G)}(\mathbf r)=h^{(-\mathbf G)}(\mathbf r)$ and
$h^{(\mathbf G)}(-\mathbf r)=h^{(\mathbf G)}(\mathbf r)$.
Similar relations hold for $c^{(\mathbf G)}(\mathbf r)$.
Eq. (8.7) can be used to expand $c^{(b)}(\vec{r}_1, \vec{r}_2)$ in powers of $\rho^{(b)}(\vec{r}) = \rho(\vec{r}) - \rho$ to give,

\begin{center}
\begin{align}
c^{(b)}(\vec r_1,\vec r_2;[\rho]) 
&= \int d\vec r_3 \, c^{(0)}_3(\vec r_1,\vec r_2,\vec r_3;\rho_0)
\rho^{(b)}(\vec r_3) \nonumber \\
&\quad + \frac{1}{2}\int d\vec r_3 \int d\vec r_4 \,
c^{(0)}_4(\vec r_1,\vec r_2,\vec r_3,\vec r_4;\rho_0)
\rho^{(b)}(\vec r_3)\rho^{(b)}(\vec r_4) \nonumber \\
&\quad + \cdots
\tag{8.22a}
\end{align}
\begin{align}
\equiv
\raisebox{-0.6cm}{\includegraphics[width=1.5cm]{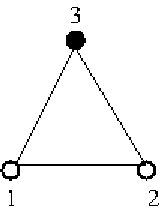}}+\frac{1}{2}\raisebox{-0.6cm}{\includegraphics[width=1.5cm]{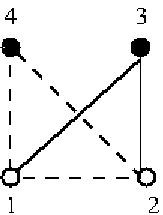}}+\frac{1}{2}\raisebox{-0.6cm}{\includegraphics[width=1.5cm]{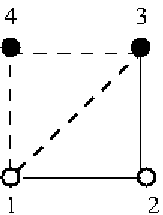}}+\frac{1}{2}\raisebox{-0.6cm}{\includegraphics[width=1.5cm]{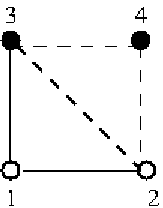}}+\dots,
\tag{8.22b}
\end{align}
\end{center}
where black circles represent integration over all configurations of these particles and each carries weight $\rho^{(b)}(\mathbf{r}_i) =
\sum_{G} \rho_G e^{i \mathbf{G}\cdot \mathbf{r}_i}$, whereas each white circle carries weight unity.

Note that Eq (8.22) satisfies the condition that $c^{(b)}$ vanishes in the isotropic phase and depends on the magnitude (order parameter). and phase factor of the density waves. These density waves affect the nature and magnitude of the positional ordering. While each wave contributes independently in the first term of Eq. (8.22), interference between the waves contributes in the second term and so on. The usefulness of the series of Eq. (8.22), however, depends on its convergence.

\subsubsection{Evaluation of first term of Eq.~(8.22)}

Substituting the value of $[\rho^{(b)}(\vec{r})$ and using the notations $r = \vec{r}_2 - \vec{r}_1$, $r' = \vec{r}_3 - \vec{r}_1$, $r_c = \frac{1}{2}(\vec{r}_1 + \vec{r}_2)$, we find
\begin{align}
\raisebox{-0.6cm}{\includegraphics[width=1.5cm]{3body11.eps}}
\equiv
c^{(b,1)}(\vec r_1,\vec r_2)
&= \sum_G \rho_G e^{i\vec G\cdot \vec r_c} 
t(r)e^{-\frac{1}{2}i\vec G\cdot \vec r}  \notag \\
&\quad \times \int d\vec r^{\,\prime}\,
t(r^{\prime})t(|\vec r^{\,\prime}-\vec r|)
e^{i\vec G\cdot \vec r^{\,\prime}} .
\tag{8.23}
\end{align}
This is solved to give~\cite{Bharadwaj2013,SinghSingh2009} 
\begin{align}
c^{(b,1)}(\vec r_1,\vec r_2)
= \sum_G e^{i\vec G\cdot \vec r_c}
\sum_{lm} c^{(G,1)}_{l}(r)
Y_{lm}(\hat r)Y^*_{lm}(\hat G),
\tag{8.24}
\end{align}
where
\begin{align}
c^{(G,1)}_{l}(r)
= \rho_G \sum_{l_1}\sum_{l_2}
\Lambda_1(l_1,l_2,l)
\, j_{l_2}\!\left(\tfrac{1}{2}Gr\right)
B_{l_1}(r,G),
\tag{8.25}
\end{align}
\begin{align}
\Lambda_1(l_1,l_2,l)
&= (i)^{l_1+l_2}(-1)^{l_2}
\left[
\frac{(2l_1+1)(2l_2+1)}{(2l+1)}
\right]^{1/2}  \notag \\
&\quad \times
\left[C_g(l_1,l_2,l;0,0,0)\right]^2 .
\tag{8.26}
\end{align}
and
\begin{equation}
B_{l_1}(r,G) = 8\,t(r)
\int dk\,k^{2} t(k) j_{l_1}(kr) \times
\int dr' \, r'^{2} t(r') j_{l_1}(kr') j_{l_1}(Gr')
\tag{8.27}
\end{equation}
In Eq.~(8.26) $C_g$ is the Clebsch--Gordan coefficient. The crystal symmetry dictates that $l$ and $l_1 + l_2$ are even and for a cubic crystal, $m = 0, \pm 4$.
\begin{figure}
\centering
\includegraphics[width=12cm]{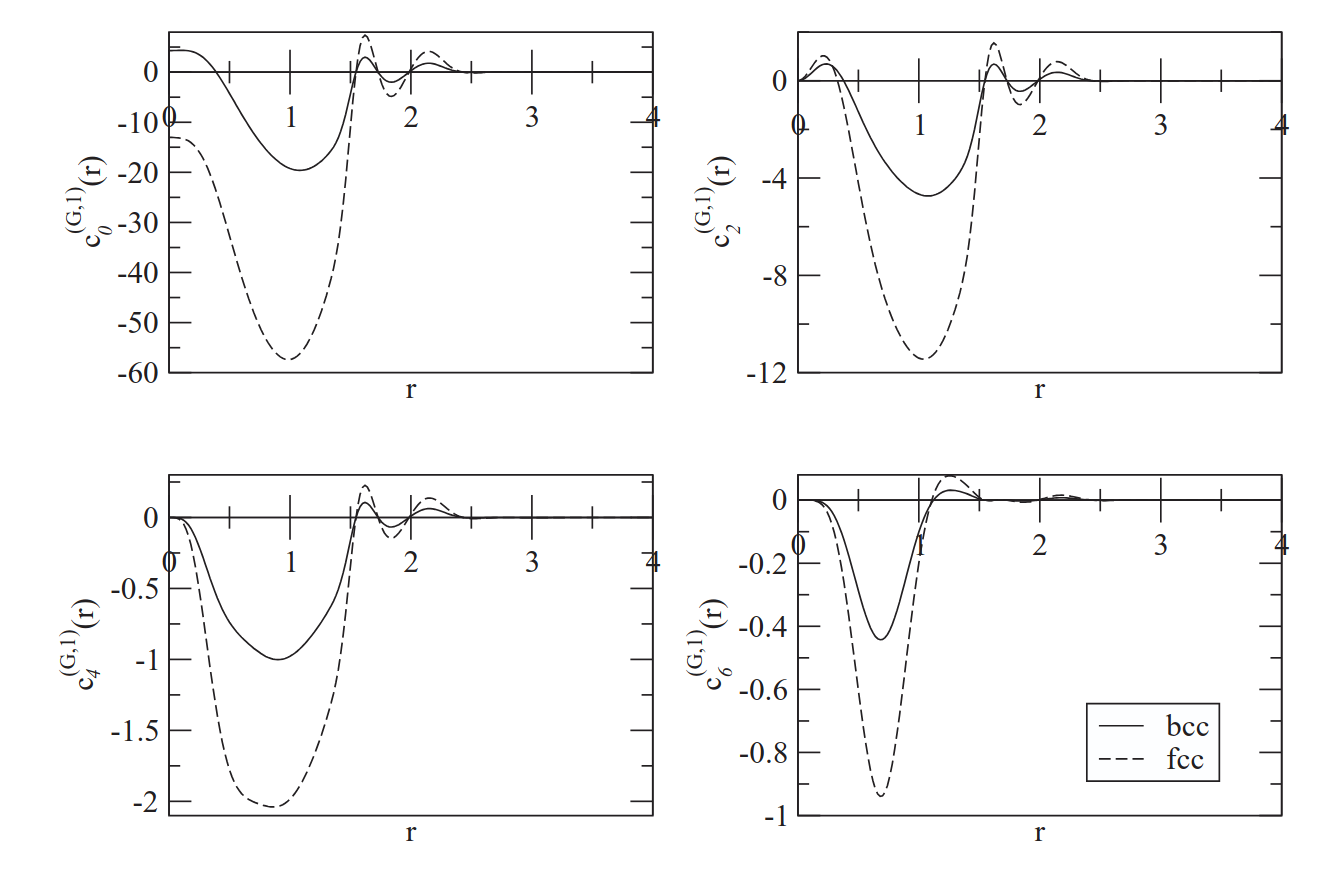}
\caption*{\textbf{Fig. 8.4} Comparison of values of $c^{(1)}_{l}(G,r)$ as a function of $r$ for
a $\mathbf{G}$ vector of the first set of fcc and bcc lattices for
$n = 6$, $\gamma_s = 2.32$, $\alpha_{\mathrm{fcc}} = 32$, and
$\alpha_{\mathrm{bcc}} = 18$. The distance $r$ is in units of
$a_0 = \left(\frac{3}{4\pi\rho}\right)^{1/3}$ and
$\mu_G = e^{-G^2/4\alpha}$. The dashed curve represents values of
the fcc structure while the full curve represents the bcc structure. Reproduced with permission from Ref.~\cite{Bharadwaj2013}}
\end{figure}
Values of $c_l^{G,1}(r)$ depends on parameter $\rho_G = \rho \mu_G$ where $\mu_G = e^{-G^2/4\alpha}$ and on magnitude of $\mathbf{G}$. In Fig.(8.4) and Fig.(8.5) values of  $c_l^{(G,1)}(r)$ are plotted and compared for bcc and fcc crystals of soft spheres for $n = 6$, $\gamma_s = 2.32$, $\alpha_{\mathrm{bcc}} = 18$, and $\alpha_{\mathrm{fcc}} = 32$
. The values given in these figures are for the first and second sets of RLVs. As expected, the values are far from negligible and differ considerably for the two structures.

\begin{figure}
\centering
\includegraphics[width=12cm]{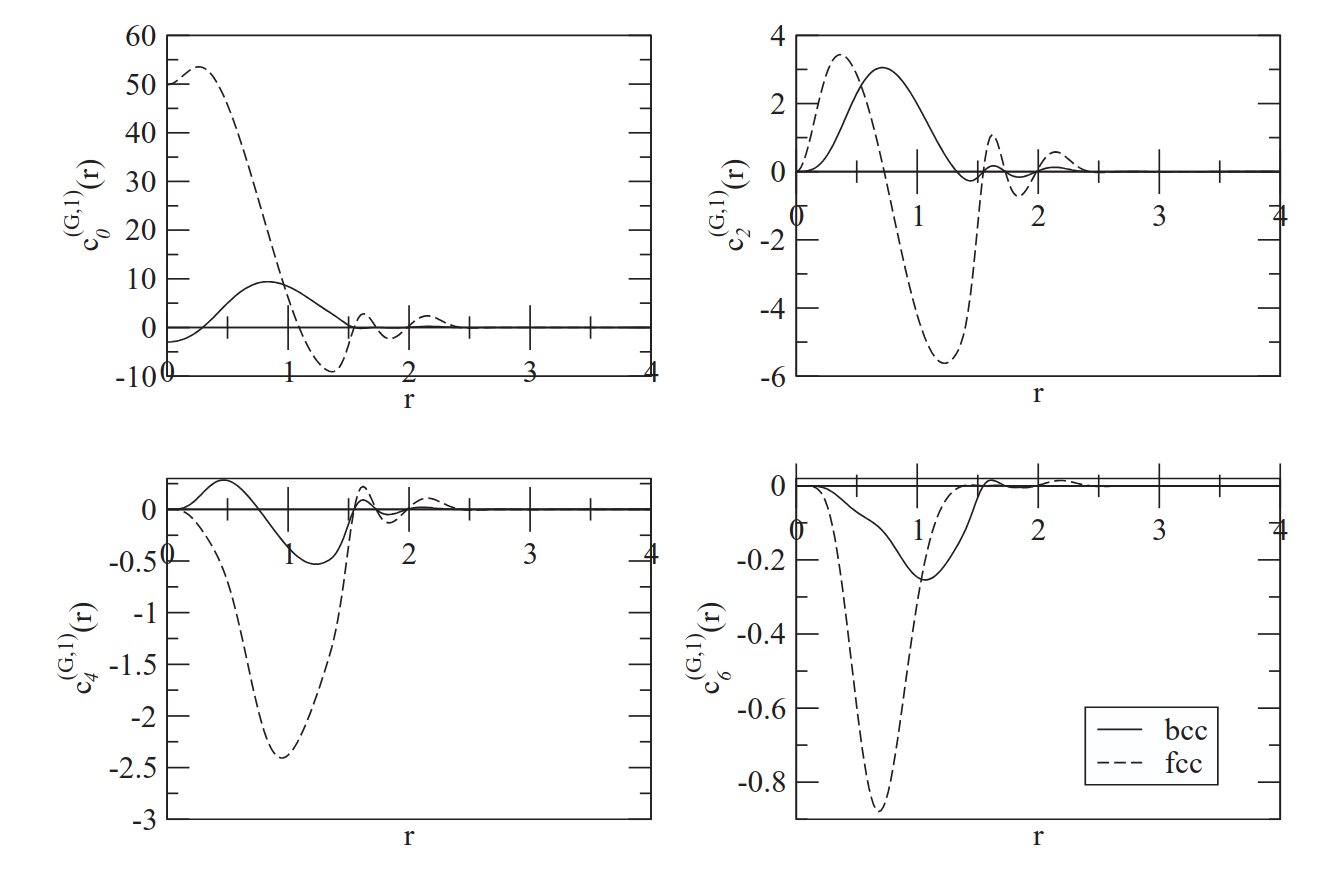}
\caption*{\textbf{Fig. 8.5} Comparison of values of $c^{(1)}_{l}(G,r)$ as a function of $r$ for a
$\mathbf{G}$ vector of the second set of fcc (dashed curve) and bcc
(full curve) lattices for $n = 6$, $\gamma_s = 2.32$,
$\alpha_{\mathrm{fcc}} = 32$, and $\alpha_{\mathrm{bcc}} = 18$.
Other notations are the same as in Fig.~8.3. Reproduced with permission from Ref.~\cite{Bharadwaj2013}}
\end{figure}

\begin{figure}
\centering
\includegraphics[width=12cm]{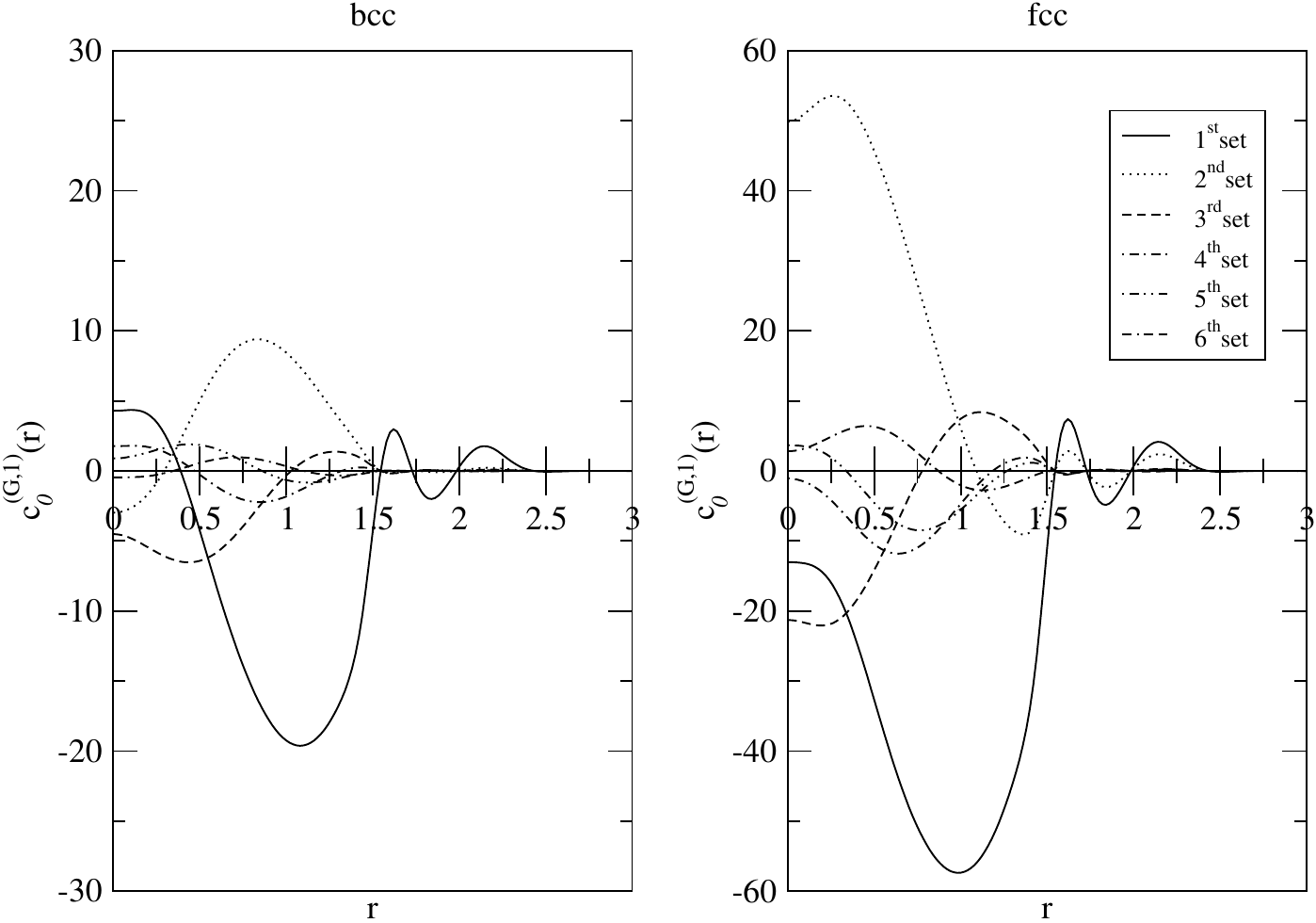}
\caption*{\textbf{Fig. 8.6} Comparison of values of $c^{(1)}_{0}(G,r)$ as a function of $r$ for a
$\mathbf{G}$ vector of the first six sets of fcc and bcc lattices.
The distance $r$ is in units of
$a_0 = \left(\frac{3}{4\pi\rho}\right)^{1/3}$. Reproduced with permission from Ref.~\cite{Bharadwaj2013}}
\end{figure}

Value  of $c_l^{(G,1)}$ is found to decrease rapidly as the value of $l$ is increased; the maximum contribution comes from $l = 0$. We also find, as shown in Fig.~8.6, that the value decreases rapidly as the magnitude of the $\mathbf{G}$ vector increases; the maximum contribution comes from the first two sets of RLVs. The other point to be noted is that, at a given point $r$, values of $c^{(1)}_{l}(G,r)$ are positive for some $\mathbf{G}$ vectors while for others the values are negative, leading to mutual cancellation in a quantity where summation over $\mathbf{G}$ is involved.

\subsubsection{Evaluation of the second term of Eq. (8.22)}

The contribution arising from the second term of Eq.~(8.22)
is the sum of three diagrams in which the last two contributions
are equal. Thus,

\begin{equation}
c^{(b,2)}(\mathbf{r}_1,\mathbf{r}_2) 
=\frac{1}{2}\raisebox{-0.6cm}{\includegraphics[width=1.5cm]{4body11.eps}}+\raisebox{-0.6cm}{\includegraphics[width=1.5cm]{4body21.eps}}
\tag{8.28}
\end{equation}
If we write $\mathbf{r}'' = \mathbf{r}_4 - \mathbf{r}_1$ and
$\mathbf{r}_4 = \mathbf{r}'' + \mathbf{r}_c - \frac{1}{2}\mathbf{r}$
and use other notations defined above, the first diagram can be written as

\begin{equation}
\begin{aligned}
\frac{1}{2}\raisebox{-0.6cm}{\includegraphics[width=1.5cm]{4body11.eps}} \equiv c^{(b,2,1)}(\mathbf{r}_1,\mathbf{r}_2)
&= \frac{1}{2} s(r)
\sum_{\mathbf{G}_1} \sum_{\mathbf{G}_2}
\rho_{G_1}\rho_{G_2}
e^{i(\mathbf{G}_1+\mathbf{G}_2)\cdot(\mathbf{r}_c-\frac{1}{2}\mathbf{r})} \\
&\times
\int d\mathbf{r}' \,
t(r')t(|\mathbf{r}'-\mathbf{r}|)
e^{i\mathbf{G}_1\cdot\mathbf{r}'} \\
&\times
\int d\mathbf{r}'' \,
s(r'')s(|\mathbf{r}''-\mathbf{r}|)
e^{i\mathbf{G}_2\cdot\mathbf{r}''}.
\end{aligned}
\tag{8.29}
\end{equation}
This is solved to give ~\cite{Bharadwaj2013}

\begin{equation}
c^{(b,2,1)}(\mathbf{r}_1,\mathbf{r}_2)
=
\sum_{\mathbf{G}} e^{i\mathbf{G}\cdot\mathbf{r}_c}
\sum_{lm}\sum_{l'm'}
c^{(G,2,1)}_{lm,l'm'}(r)
Y^{*}_{l'm'}(\hat{G})\,Y_{lm}(\hat{r}),
\tag{8.30}
\end{equation}
where
\begin{equation}
c^{(G,2,1)}_{lml'm'}(r)
=
\sum_{G_1}\rho_{G_1}\rho_{K}
\sum_{l_1m_1}\sum_{l_2m_2}
\Lambda^{\,l\,l'\,l_1\,l_2}_{m\,m'\,m_1\,m_2}
M_{l_1}(r,G_1)M_{l_2}(r,K)
j_{l'}\!\left(\frac{1}{2}Gr\right)
Y^{*}_{l_1m_1}(\hat{\mathbf G}_1)
Y^{*}_{l_2m_2}(\hat{\mathbf K}),
\tag{8.31}
\end{equation}
Here $\mathbf K = \mathbf G - \mathbf G_1$,

\begin{align}
\Lambda^{\,l\,l'\,l_1\,l_2}_{m\,m'\,m_1\,m_2}
&=
16 \sum_{l_3m_3}
(i)^{\,l_1+l_2+l'}
(-1)^{l'}
\left[
\frac{(2l_1+1)(2l_2+1)(2l'+1)}
{(2l+1)}
\right]^{1/2}
\nonumber\\
&\quad\times
C_g(l_1,l_2,l_3;0,0,0)
C_g(l',l_3,l;0,0,0)
\nonumber\\
&\quad\times
C_g(l_1,l_2,l_3;m_1,m_2,m_3)
C_g(l',l_3,l;m',m_3,m),
\tag{8.32}
\end{align}

\begin{equation}
M_{l_1}(r,G_1)
=
\int dr' r'^2 j_{l_1}(G r')t(r')
\int dk\,k^2 t(k)
j_{l_1}(kr) j_{l_1}(kr'),
\tag{8.33}
\end{equation}
and
\begin{equation}
M_{l_2}(r,K)
=
\int dr'' r''^{2} j_{l_2}(K r'') s(r'')
\int dk\,k^2 s(k)
j_{l_2}(kr) j_{l_2}(kr'').
\tag{8.34}
\end{equation}
The crystal symmetry dictates that all $l_i$ are even and for a
cubic crystal all $m_i$ are $0$ and $\pm 4$. From the second diagram of Eq.~(8.28) we get

\begin{align}
\raisebox{-0.6cm}{\includegraphics[width=1.5cm]{4body21.eps}} \equiv
c^{(b,2,2)}(\mathbf r_1,\mathbf r_2)
&=
t(r)
\sum_{G_1}\sum_{G_2}
\rho_{G_1}\rho_{G_2}
\,e^{i(\mathbf G_1+\mathbf G_2)\cdot
(\mathbf r_c-\frac{1}{2}\mathbf r)}
\nonumber\\
&\quad\times
\int d\mathbf r' \,
s(r')\,t(|\mathbf r'-\mathbf r|)
\,e^{i\mathbf G_1\cdot \mathbf r'}
\nonumber\\
&\quad\times
\int d\mathbf r''\,
s(r'')\,s(|\mathbf r''-\mathbf r|)
\,e^{i\mathbf G_2\cdot \mathbf r''}.
\tag{8.35}
\end{align}
This is solved to give ~\cite{Bharadwaj2013}

\begin{equation}
c^{(b,2,2)}(\vec r_1,\vec r_2)
=
\sum_{G} e^{i\vec G \cdot \vec r_c}
\sum_{lm}\sum_{l'm'}
c^{(G,2,2)}_{lm,l'm'}(r)
Y^{*}_{l'm'}(\hat G) Y_{lm}(\hat r),
\tag{8.36}
\end{equation}
where
\begin{align}
c^{(G,2,2)}_{lm,l'm'}(r)
&=
\sum_{G_1} \rho_{G_1}\rho_{K}
\sum_{l_1 m_1}\sum_{l_2 m_2}\sum_{l_3 m_3}
\Lambda^{\,ll'l_1l_2l_3}_{mm'm_1m_2m_3}
N_{l_1,l_2,l_3}(r,G,G_1)
\nonumber \\
&\qquad \times
j_{l'}\!\left(\frac{1}{2}Gr\right)
Y^{*}_{l_1m_1}(\hat G_1)
Y^{*}_{l_2m_2}(\hat K),
\tag{8.37}
\end{align}

\begin{align}
\Lambda^{\,ll'l_1l_2l_3}_{mm'm_1m_2m_3}
&=
32 (i)^{l_1+l_2+l'}(-1)^{l'}
\left[
\frac{(2l_1+1)(2l_2+1)(2l'+1)}
{(2l+1)}
\right]^{1/2}
\nonumber \\
&\quad \times
C_g(l_1,l_2,l_3;0,0,0)
C_g(l',l_3,l;0,0,0)
\nonumber \\
&\quad \times
C_g(l_1,l_2,l_3;m_1,m_2,m_3)
C_g(l',l_3,l;m',m_3,m),
\tag{8.38}
\end{align}

\begin{equation}
N_{l_1,l_2,l_3}(r,G,G_1)
=
t(r)
\int dr'\, r'^2 s(r') j_{l_1}(G_1 r')
B_{l_2}(r',K) A_{l_3}(r,r'),
\tag{8.39}
\end{equation}

\begin{equation}
A_{l_3}(r,r')
=
\int dk\,k^2 t(k) j_{l_3}(kr) j_{l_3}(kr'),
\tag{8.40}
\end{equation}
and
\begin{align}
B_{l_2}(r',K)
&=
\int dr''\, r''^{2} j_{l_2}(K r'') s(r'')
\nonumber \\
&\qquad \times
\int dk\,k^2 s(k) j_{l_2}(kr') j_{l_2}(kr'').
\tag{8.41}
\end{align}
The total contribution arising from the second term of
Eq.~(8.22) is
\begin{align}
c^{(b,2)}(\vec r_1,\vec r_2)
&=
\sum_{G} e^{i\vec G \cdot \vec r_c}
\sum_{lm}\sum_{l'm'}
\Big[
c^{(G,2,1)}_{lm,l'm'}(r)
+
c^{(G,2,2)}_{lm,l'm'}(r)
\Big]
\nonumber \\
&\qquad \times
Y_{lm}(\hat r) Y^{*}_{l'm'}(\hat G).
\tag{8.42}
\end{align}
\begin{figure}
\centering
\includegraphics[width=0.7\columnwidth]{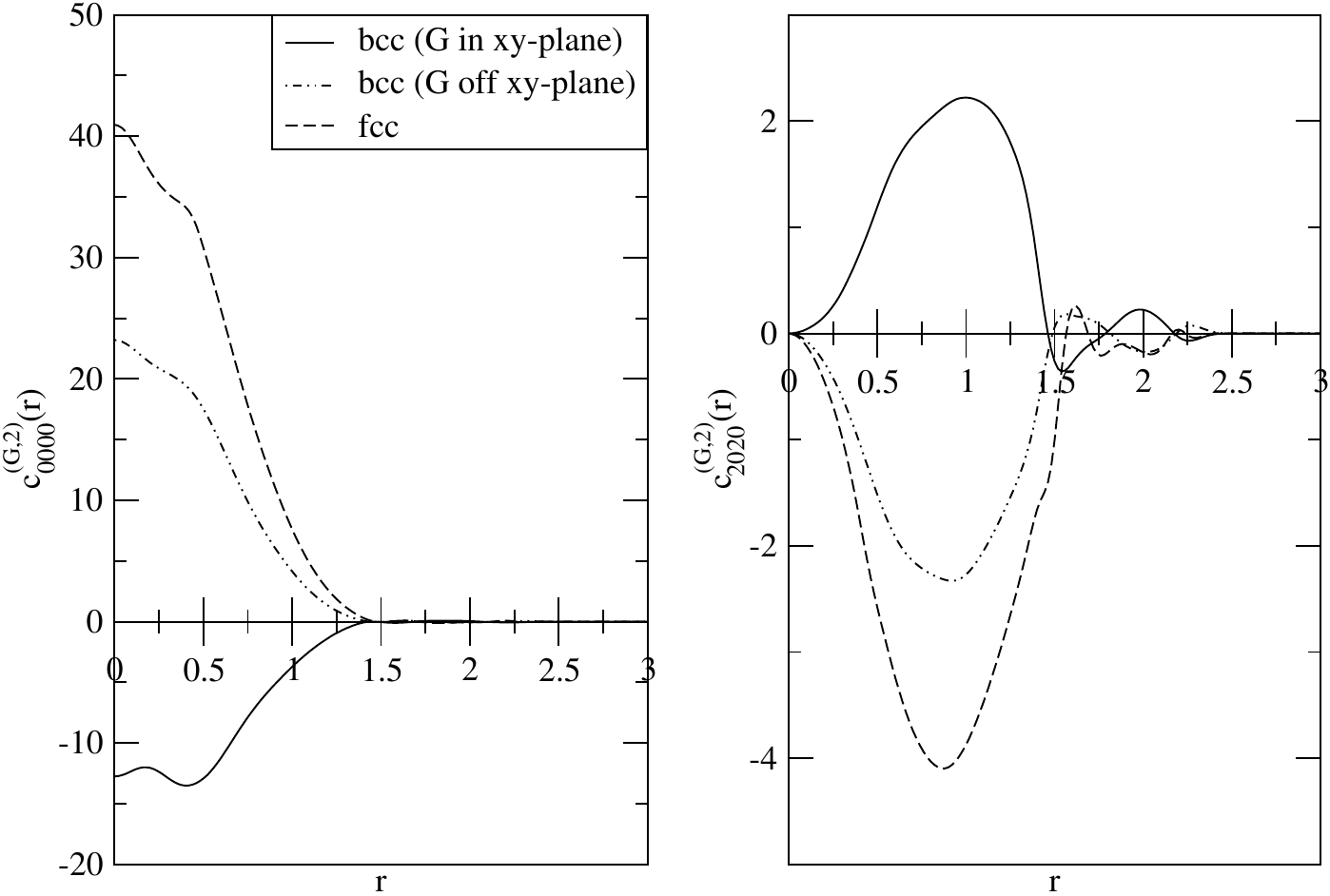}
\caption*{\textbf{Fig. 8.7} Comparison of values of $c^{(G,2)}_{lml'm'}(r)$ as a function of $r$ for a $\mathbf{G}$ vector of the first set of fcc and bcc lattices for $n=6$ at $\gamma_s = 2.32$, $\alpha_{fcc}=32$, and $\alpha_{bcc}=18$. The distance $r$ is in unit of $a_0=\left(\frac{3}  {4\pi\rho}\right)^{1/3}$. There are two sets of values for the bcc lattice, one for $\mathbf{G}$ vectors lying in $x$-$y$ plane and the other for the rest of the $\mathbf{G}$ vectors of the first set. There is only one set of values for the fcc lattice. Reproduced with permission from Ref.~\cite{Bharadwaj2013}}
\end{figure}
where $l,l'$ are even and $m = 0, \pm 4$ for cubic lattices. In Figs.~8.7 and 8.8 we plot values of
\begin{equation}
c^{(G,2)}_{lml'm'}(r)
=
c^{(G,2,1)}_{lml'm'}(r)
+
c^{(G,2,2)}_{lml'm'}(r),
\tag{8.43}
\end{equation}
as a function of $r$ for bcc and fcc structures for $n=6$, $\gamma_s = 2.32$, $\alpha_{bcc}=18$, and $\alpha_{fcc}=32$. The values given in these figures are for the first two sets of RLVs for $l=l' = 0$ and $2$ and $m=m' = 0$. These are the terms which mostly contribute to $c^{(G,2)}_{lml'm'}(r)$; the contributions from terms $l \ne l'$ and $m \ne m'$ are approximately an order of magnitude smaller. For a bcc lattice we find two sets of values, one for $\mathbf{G}$ vectors lying in the $x$-$y$ plane
and the other for the rest of the vectors. Since all vectors of the first set of RLVs of a fcc lattice are from the $x$-$y$ plane we get only one set of values.
For the second set of RLVs of a fcc lattice, two sets of values are found but they are close, unlike in the case of the bcc lattice where the two sets of values differ not only in magnitude but also in sign. The values differ considerably for the two cubic structures. The value of $c^{(G,2)}_{lml'm'}(r)$ decreases rapidly for both bcc and fcc structures as the magnitude of $\mathbf{G}$ vectors increases, as was found in the case of $c^{(G,1)}_{l}(r)$. Furthermore, the values of $c^{(G,2)}_{lml'm'}(r)$ at a given value of $r$ is positive for
some $\mathbf{G}$ vector and negative for others.

\begin{figure}
\centering
\includegraphics[width=0.7\columnwidth]{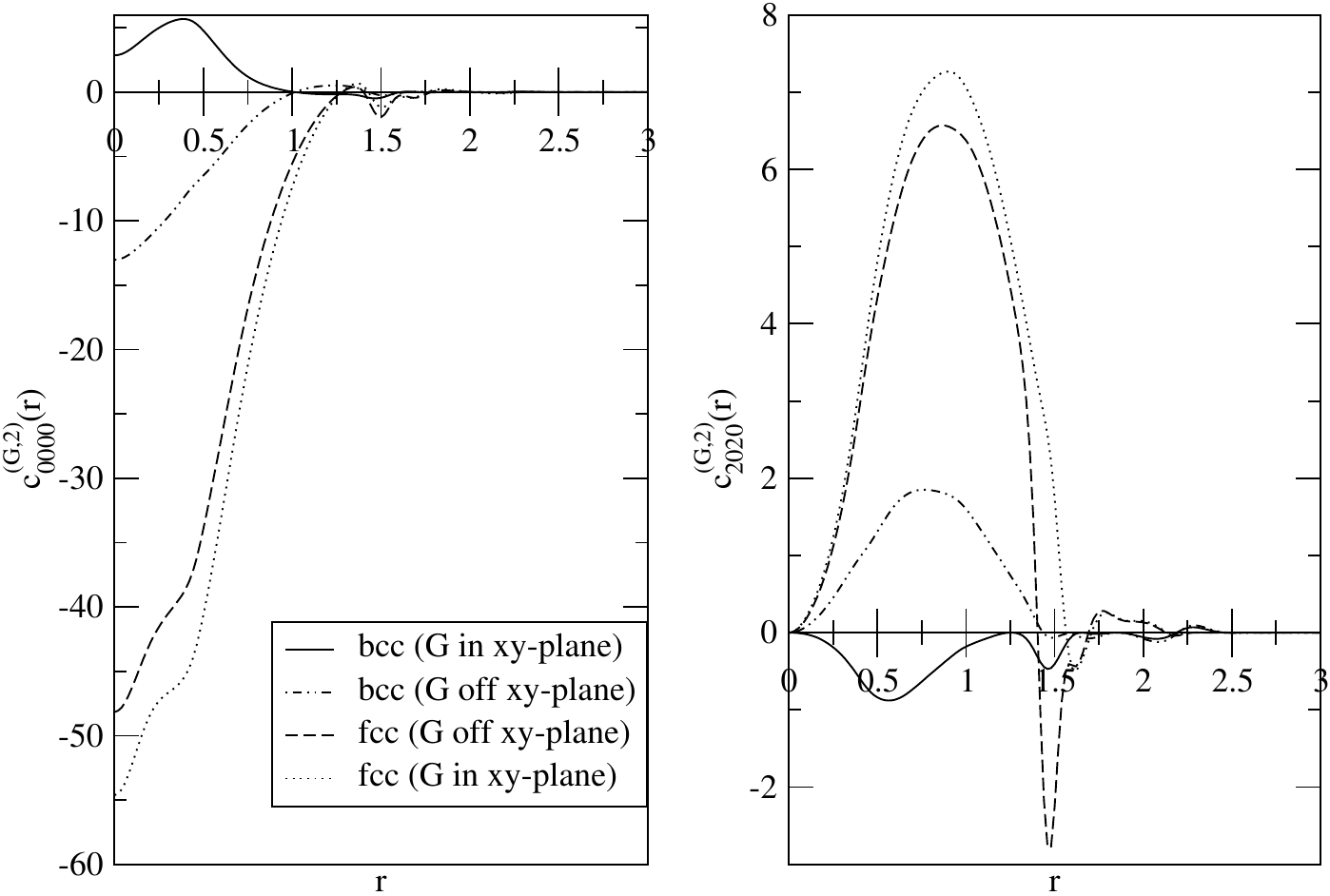}
\caption*{\textbf{Fig. 8.8} Comparison of the values of $c^{(G,2)}_{lml'm'}(r)$ as a function of $r$ for a $\mathbf{G}$ vector of the second set of fcc and bcc lattices. Other notations are same as in Fig.~8.6, except that there is now two sets of
values shown by dashed and dotted curves for fcc lattice (see text). Reproduced with permission from Ref.~\cite{Bharadwaj2013}}
\end{figure}

\subsection{Convergence of the series of Eq.~(8.22)}

In order to compare magnitude of contributions made by the first and second terms of Eq. (8.22) Bharadwaj and Singh ~\cite{Bharadwaj2013} calculated $\hat{c}^{(G,1)}(k, \theta_k, \phi_k)$ and $\hat{c}^{(G,2)}(k, \theta_k, \phi_k)$ defined as,
\begin{align}
\hat{c}^{(G,1)}(k,\theta_k,\phi_k)
&= \rho \sum_{l m} \int d\vec{r}\, c^{(G,1)}_l(r)\,
e^{i\vec{k}\cdot\vec{r}}\, Y_{lm}(\hat{r})\, Y^*_{lm}(\hat{G}) \nonumber \\
&= 4\pi \rho \sum_{l m} i^l\, Y_{lm}(\hat{k})\, Y^*_{lm}(\hat{G})
\int_0^\infty dr\, r^2\, c^{(G,1)}_l(r)\, j_l(kr)
\tag{8.44a}
\end{align}
and
\begin{align}
\hat{c}^{(G,2)}(k,\theta_k,\phi_k)
&= \rho \sum_{l m} \sum_{l' m'} \int d\vec{r}\,
c^{(G,2)}_{l m\, l' m'}(r)\,
e^{i\vec{k}\cdot\vec{r}}\, Y_{lm}(\hat{r})\, Y^*_{l' m'}(\hat{G}) \nonumber \\
&= 4\pi \rho \sum_{l m} \sum_{l' m'} i^l\, Y_{lm}(\hat{k})\, Y^*_{l' m'}(\hat{G})
\int_0^\infty dr\, r^2\, c^{(G,2)}_{l m\, l' m'}(r)\, j_l(kr);
\tag{8.44b}
\end{align}
as a function of $cos\theta_k$ and $cos\phi_k$ for some values of $ka_0$ (see fig.10 and 11 of Ref.~\cite{Bharadwaj2013}). Comparison of values of $\hat c^{(G,1)}$ and $\hat c^{(G,2)}$ show that the contribution made by the second term to $c^{(b)}(\vec{r}_1, \vec{r}_2)$ is nearly negligible compared to the first term, indicating fast convergence of the series. Moreover, since in the expression of free energy functional $c^{(b)}(\vec{r}_1, \vec{r}_2)$ is averaged over density and order parameters and also there is summation over $\vec{G}$ vectors. As a consequence the contribution made by the second term of Eq. (8.22) to free-energy reduces to an order of magnitude smaller than the first term even in cases where it is expected to be important.

\section{Two-dimensional crystals: Pair correlation functions}

There are only five distinct types of Bravais lattices in two-dimensions, the oblique lattice and the four special lattices. The symmetry operations used in the two-dimensional ``crystallographic'' point groups are 1-, 2-, 3-, 4- and 6-fold rotations about a point and mirror reflection along a line. The hexagonal lattice is of a special interest as it is a lattice with the highest rotational symmetry and is the only closed packed lattice in two-dimensions.

Jaiswal et al.~\cite{Jaiswal2013} studied freezing of two-dimensional fluids interacting via the inverse power potentials which in case of 2-dimensions is expressed as $\beta u(r) = \frac{\Gamma}{r^n}$ where $\Gamma = \gamma^{n/2}$ and $\gamma=(\rho\sigma^2)(\beta\epsilon)^{2/n}=\rho^*T^{*-2/n}$. The distance $r$ is measured in units of $a_0=(1/\rho)^{1/2}$. Such repulsive potentials can be realized in colloidal suspensions. One such a two-dimensional system has been provided by paramagnetic colloidal particles in a pendent water droplet,  which are confined to the air--water interface ~\cite{Zahn1999,vonGrunberg2004}. By applying an external magnetic field perpendicular to the interface, a magnetic moment is induced in the particles resulting in a tunable mutual dipolar repulsion between them. The pair interaction thus created is repulsive and proportional to $r^{-3}$. The crystallization of this system has been investigated by von Teeffelen and coworkers ~\cite{vonTeeffelen2006,vonTeeffelen2008} using different versions of approximate DFT described in Section 5. Another example where short range repulsion between molecules is found is microgel spheres whose diameter could be temperature tuned ~\cite{Han2008}. Most computer simulation studies on these systems suffer from the finite-size effect. However, in the case of hard and soft disks a large scale Monte Carlo simulation has been performed ~\cite{Bernard2011,Antlanger2014}. The result confirms two-step transitions from liquid to solid with the intermediate \textit{hexatic phase}. However, the fluid--hexatic transition, in contrast to the prediction of KTHNY theory ~\cite{Kosterlitz1973,Halperin1978,Young1979}, is found to be first order while the exatic--solid transition is second order. The DFT predicts the fluid--solid transition to be first order. In this section, we describe the calculation of pair correlation function of a hexagonal crystal and the fluid--solid transition in Section 10.

\subsection{Calculation of broken symmetry contribution to DPCF}

The method for determining $c_m^{(0)}(\vec{r})(\vec{r_1},\vec{r_2},...,\vec{r_m})$ and $c^{(b)}(\vec{r}_1,\vec{r}_2)$ detailed in the preceding section has been used in Ref.~\cite{Jaiswal2013} to find values of direct pair correlation functions of a two-dimensional crystal. Taking only the first term of the series given by Eq. (8.22), the expression for $c^{(b)}(\vec{r}_1, \vec{r}_2)$ in terms of the function $t(r)$ (see Eq. 8.10) can be written as,

\begin{equation}
    c^{(b)}(\vec{r}_1, \vec{r}_2) = \sum_G \rho_G \int d\vec{r}_3 t(|\vec{r}_3 - \vec{r}_1|) t(|\vec{r}_3 - \vec{r}_2|) e^{i\vec{G} \cdot \vec{r}_3} \tag{9.1}
\end{equation}

Substituting the following relations in Eq. (9.1)

\begin{equation}
    \vec{r}_3 = \frac{1}{2}(\vec{r}_1 + \vec{r}_2) + \vec{r}' - \frac{1}{2}\vec{r} = \vec{r}_c + \vec{r}' - \frac{1}{2}\vec{r} \tag{9.2}
\end{equation}

one gets

\begin{equation}
    c^{(b)}(\vec{r}_1, \vec{r}_2) = \sum_G e^{i\vec{G} \cdot \vec{r}_c} c^{(G)}(\vec{r}) \tag{9.3}
\end{equation}

where

\begin{equation}
    c^{(G)}(\vec{r}) = \rho_G t(r) e^{-\frac{1}{2} i \vec{G} \cdot \vec{r}} \int d\vec{r}' t(r') e^{i\vec{G} \cdot \vec{r}'} t(|\vec{r}' - \vec{r}|) \tag{9.4}
\end{equation}

Using the relation $e^{i\vec{G} \cdot \vec{r}} = \sum_m (i)^m J_m(Gr) e^{im(\phi_G - \phi_r)}$ where $J_m(Gr)$ is the Bessel function of the first kind of integral order $m$, $\phi_G$ and $\phi_r$ are angles of vectors $\vec{G}$ and $\vec{r}$ with respect to a space fixed coordinate frame, one gets

\begin{equation}
c^{(G)}(\vec{r}) = \sum_{M} (i)^M c_M^{(G)}(r)\, e^{i m \phi_G}\, e^{-i m \phi_r}
\tag{9.5}
\end{equation}
where
\begin{equation}
c_M^{(G)}(r) = \rho_G \ t(r) \sum_{m} B_m(r,G)\, J_{m+M}(\frac{1}{2}G r)
\tag{9.6}
\end{equation}
and
\begin{equation}
B_m(r,G) = \int dk \, k\, t(k)\, J_m(k r) \int dr'\, r'\, J_m(k r')\,J_m(G r')\,  t(r')
\tag{9.7}
\end{equation}

For hexagonal lattice $M= 0, \pm 6$. Jaiswal et al.~\cite{Jaiswal2013} calculated values for harmonic coefficients $c_0^{(G)}(r)/\mu_G$ and $c_6^{(G)}(r)/\mu_G$ for a system interacting via inverse power potential  $\beta u(r) = \frac{\Gamma}{r^n}$. Results are reported for $n=3, 6$ and $12$. In Fig.~9.1--9.3 we reproduce results for RLV's of the first four sets, respectively. For a different set of RLV's, $c_M^{(G)}(r)$ varies with $r$ in a different way, the value in all cases become negligible for  (measured in units of $a_0 = (\frac{1}{\rho})^\frac{1}{2}$) $r>1.5$. As the magnitude of $G$ increases, value of $c_M^{(G)}(r)$ decreases and after  sixth set the value becomes negligible.

\begin{figure}
\centering
\includegraphics[width=11.5cm]{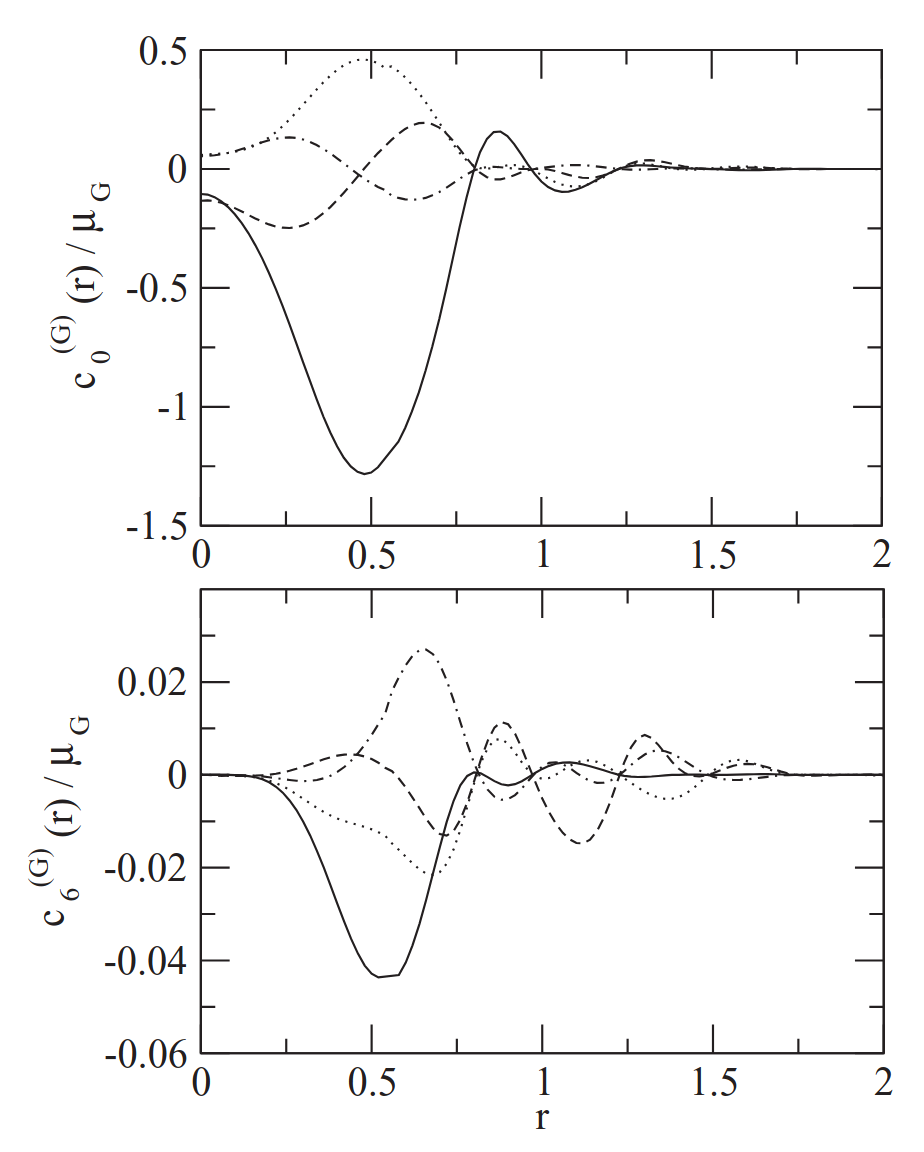}
\caption*{\textbf{Fig. 9.1} Harmonic coefficients $c_M^{(G)}(r)/\mu_G$ for RLVs of the first four sets for $n=3$, $\gamma=4.42$ ($\Gamma=9.32$). Notations are as follows: Full line represents values of the first set, the dotted line of the second set, the dashed line of the third set, and the dash--dotted line of the fourth set. The distance $r$ is expressed in units of $a_0$, where $a_0=(1/\rho)^{1/2}$. Reproduced with permission from Ref.~\cite{Jaiswal2013}}
\end{figure}
\begin{figure}
\centering
\includegraphics[width=11.5cm]{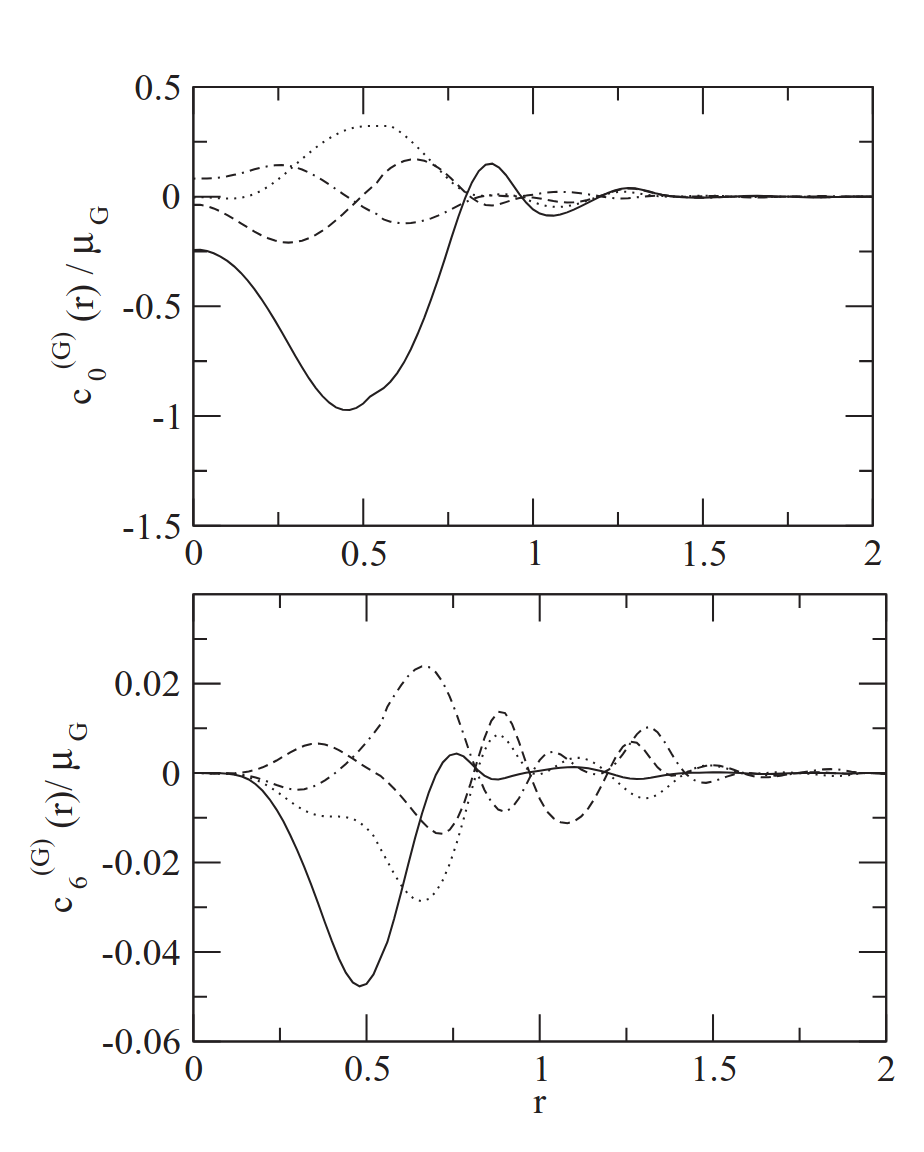}
\caption*{\textbf{Fig. 9.2}
 Harmonic coefficients $c_M^{(G)}(r)/\mu_G$ for RLVs of the first four sets for  $n=6$, $\gamma=1.52$ ($\Gamma=3.52$). Notations are same as in Fig.9.1. Reproduced with permission from Ref.~\cite{Jaiswal2013}}
\end{figure}

\begin{figure}
\centering
\includegraphics[width=11.5cm]{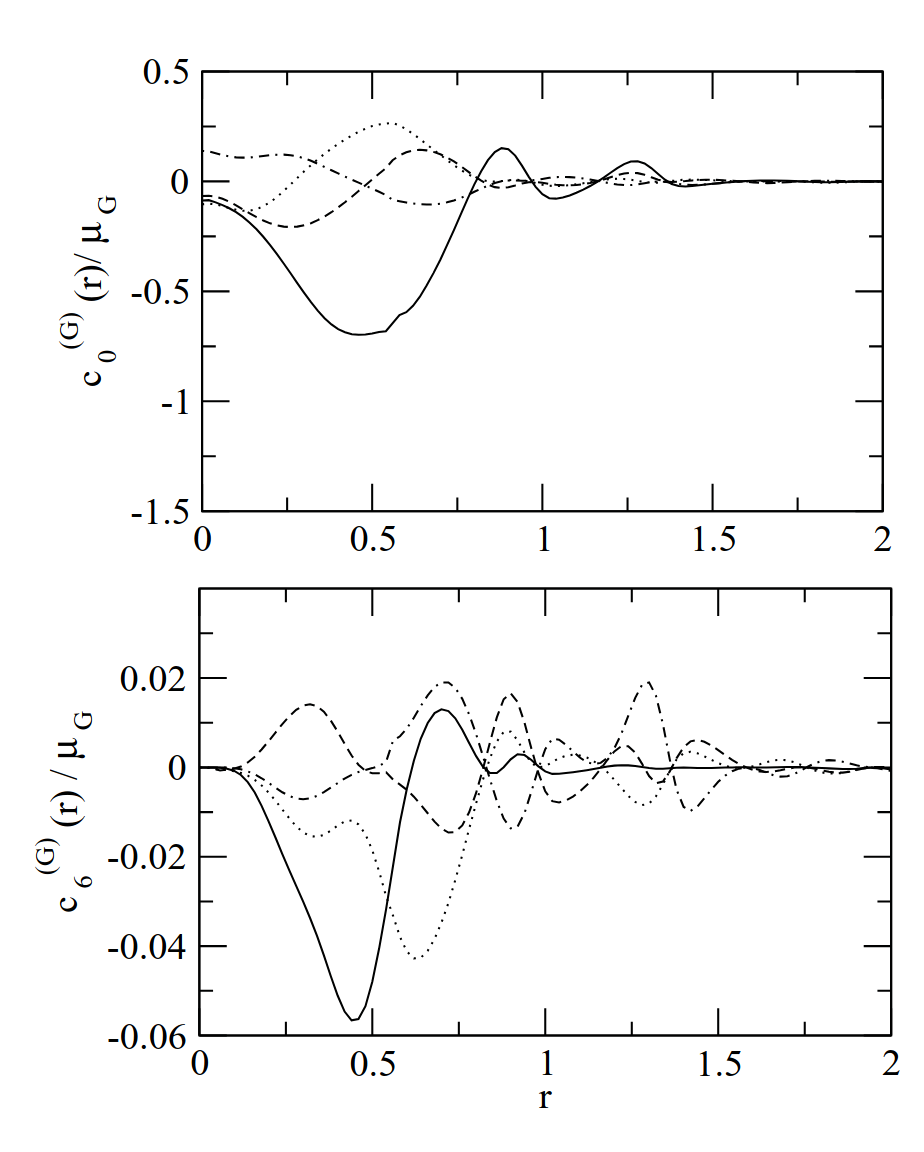}
\caption*{\textbf{Fig. 9.3}
 Harmonic coefficients $c_M^{(G)}(r)/\mu_G$ for RLVs of the first four sets for  $n=12$, $\gamma=1.05$ ($\Gamma=1.34$). Notations are same as in Fig. 9.1. Reproduced with permission from Ref.~\cite{Jaiswal2013}}
\end{figure}

\subsection{Broken symmetry contribution to total pair correlation function}

The OZ relation (6.5) can be rewritten as\cite{Jaiswal2014},

\begin{equation}
h^{(b)}(\vec{r}_1, \vec{r}_2) = H^{(b)}(\vec{r}_1, \vec{r}_2)
+ \int d\vec{r}_3 \, h^{(b)}(\vec{r}_1, \vec{r}_3)\,\rho(\vec{r}_3)
\left[ c^{(0)}(|\vec{r}_3 - \vec{r}_2|) + c^{(b)}(\vec{r}_3, \vec{r}_2) \right]
\label{eq:9.4}
\tag{9.8}
\end{equation}
where
\begin{equation}
H^{(b)}(\vec{r}_1, \vec{r}_2)
= c^{(b)}(\vec{r}_1, \vec{r}_2)
+\int d\vec{r}_3 \, h^{(0)}(|\vec{r}_3- \vec{r}_1|)\,\left[\rho^{(b)}(\vec{r}_3)
 c^{(0)}(|\vec{r}_3 - \vec{r}_2|) + \rho(\vec{r}_3)c^{(b)}(\vec{r}_3, \vec{r}_2) \right]
\label{eq:9.4}
\tag{9.9}
\end{equation}
Using Eq.~(8.21) for $h^{(b)}$ and $c^{(b)}$ and the relation

\begin{equation}
\rho(\vec{r}) = \rho + \rho^{(b)}(\vec{r})
\label{eq:9.6}
\tag{9.10}
\end{equation}
the following expression is found ~\cite{Jaiswal2014} for $h^{(G)}(\vec{r})$,

\begin{figure}
\centering
\includegraphics[width=\columnwidth]{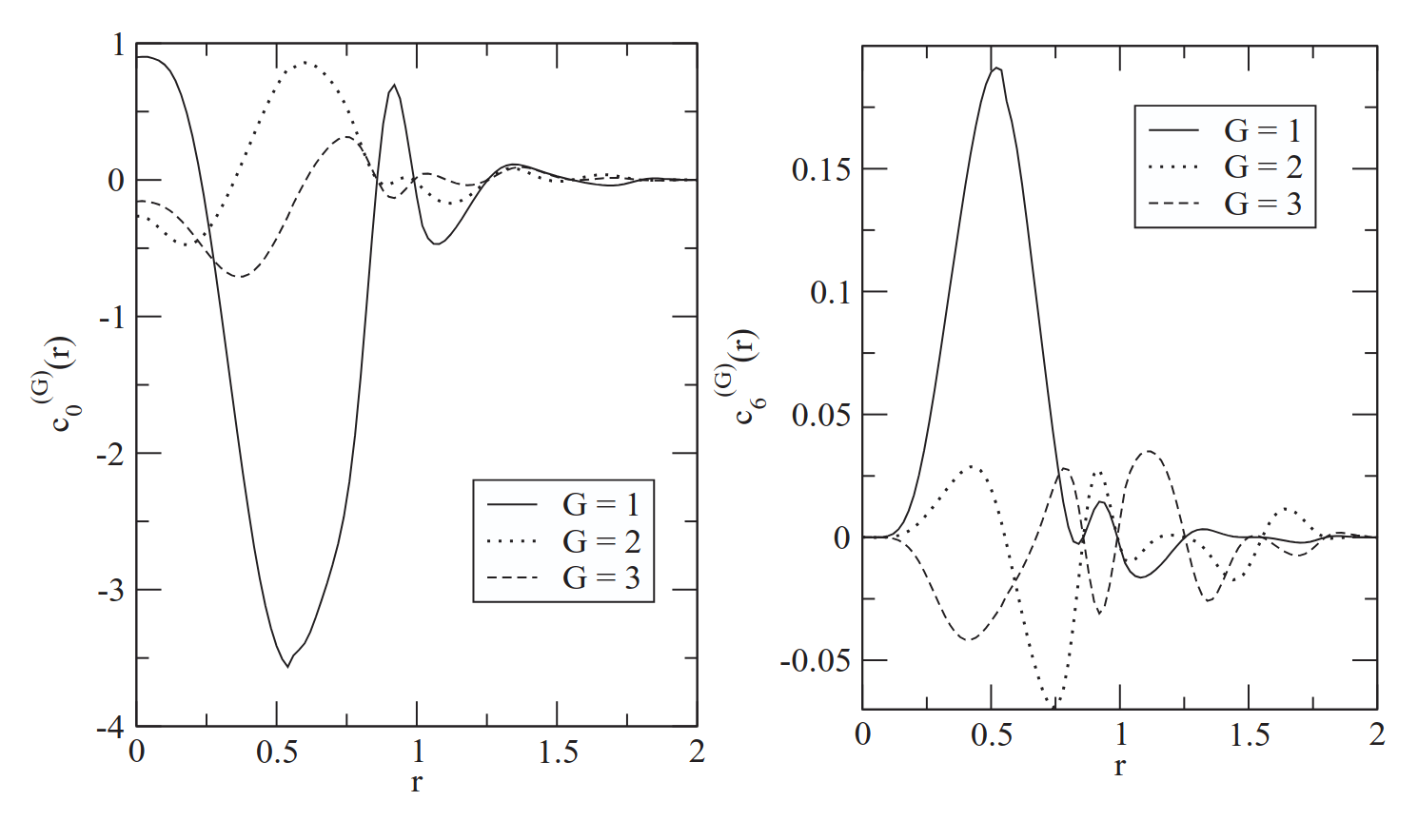}
\caption*{\textbf{Fig. 9.4}
 Harmonic coefficients $c_M^{(G)}(r)$ for RLVs of the first three sets. Full line represents values of the first set, the dotted line of the second set, the dashed line of the third set. The distance $r$ is expressed in units $a_0$, where $a_0=(1/\rho)^{1/2}$, for a system of soft disks interacting via the potential $\beta u(r) = \frac{\Gamma}{r^3},(\Gamma = 12)$. Reproduced with permission from Ref.~\cite{Jaiswal2014}.
}
\end{figure}

\begin{align}
h^{(G)}(\vec{r}) ={}&
\frac{1}{V} \sum_{\vec{k}} H^{(G)}(\vec{k}) 
e^{i \vec{k}\cdot \vec{r}}
+\frac{\rho}{V} 
\sum_{\vec{G}_1} \sum_{\vec{k}}
h^{(G_1)}\!\left(
\vec{k}
- \frac{1}{2}\vec{G}
- \frac{1}{2}\vec{G}_1
\right)
\nonumber\\[6pt]
&\times
\Bigg[
\mu_{G-G_1}\,
c^{(0)}\!\left(
-\left| \vec{k}
+ \frac{1}{2}\vec{G} \right|
\right)
+ \sum_{\vec{G}_2}
\mu_{G-G_1-G_2}\,
c^{(G_2)}\!\left(
\frac{1}{2}\vec{G}_2
- \frac{1}{2}\vec{G}
- \vec{k}
\right)
\Bigg]
e^{i \vec{k}\cdot \vec{r}} .
\tag{9.11}
\end{align}

where

\begin{align*}
H^{(G)}(\vec{k}) = c^{(G)}(\vec{k}) 
+ \rho \, \mu_G \, c^{(0)}\!\left( \left| \vec{k} - \frac{1}{2}\vec{G} \right| \right)
h^{(0)}\!\left( -\left| \vec{k} + \frac{1}{2}\vec{G} \right| \right)\\
+ \rho \sum_{\vec{G}_1} \mu_{G_1-G} \, c^{(G_1)}\!\left( \vec{k} - \frac{1}{2}\vec{G} + \frac{1}{2}\vec{G}_1 \right)
h^{(0)}\!\left( -\left| \vec{k} + \frac{1}{2}\vec{G}_1 \right| \right) .
\tag{9.12}
\end{align*}

\begin{figure}
\centering
\includegraphics[width=\columnwidth]{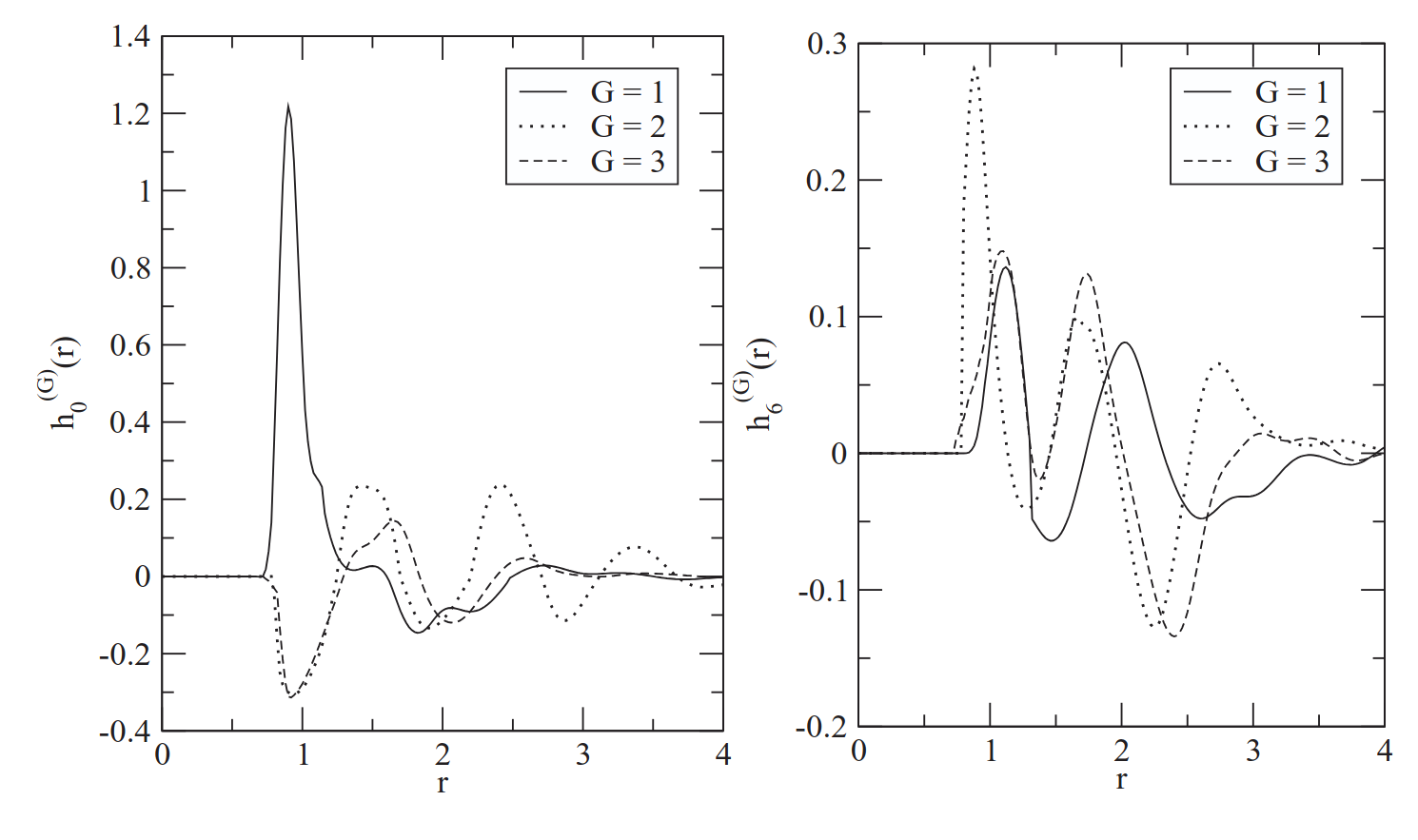}
\caption*{\textbf{Fig. 9.5}
Harmonic coefficients $h_M^{(G)}(r)$ for RLVs of the first three sets. Notations are same as in Fig. 9.4. Reproduced with permission from Ref. ~\cite{Onsager1949}
}
\end{figure}

Values of $h^{(G)}(\vec{r})$ were calculated for a system of soft disks interacting via the potential $\beta u(r)=(\Gamma/r^3)$ for the softness parameter $n=3$ and $\Gamma=12$. In the calculation, values of $c^{(G)}(\vec{r})$ plotted in Fig.~9.4 were used. Values of $h_M(r)$ calculated from the relation  
\[ h^{(G)}(\vec{r}) = \sum_M (i)^M h_M^{(G)}(r) e^{iM\phi_G} e^{-iM\phi_r}, \tag{9.13} \]
is plotted in Fig. 9.5 for the first three sets of \( \vec G \) for \( \Gamma = 12 \), $\mu_G=exp(-G^2/4\alpha)$ and \( \alpha = 100 \). The values of \( h_M^{(G)}(r) \) depend on values of order parameters \( \mu_G \) and on values of \(\vec G \) vectors. For \( G \) belonging to different sets, values of \( h_M^{(G)}(r) \) as a function of \( r \) oscillate about zero in different ways; the maxima and minima are located at different values of \( r \). Values are found to decay rapidly and become almost zero for \( r > 5 \) in all cases. Unlike \( c_M^{(G)}(r) \), the values of maxima and minima of \( h_0^{(G)}(r) \) and \( h_6^{(G)}(r) \) are comparable. As the magnitude of \( G \) increases, the value of $h_M^{(G)}(r)$ decreases, and as in the case of $c_M^{(G)}(r)$ after the sixth set, the values become negligible.

The accuracy of $h^{(G)}(r)$ can be tested using a Ward identity, which in this case is represented by the Born-Green-Yvon (YGB) equation ~(2.42),

\begin{equation}
\nabla_1 \ln \rho(\vec{r}_1) = -\beta \int d\vec{r}_2 \, \rho(\vec{r}_2)
\left[ g^{(0)}(|\vec{r}_2 - \vec{r}_1|) + h^{(b)}(\vec{r}_1,\vec{r}_2) \right]
\nabla_1 u(|\vec{r}_2 - \vec{r}_1|),
\tag{9.14}
\end{equation}
where $u(|\vec{r}_2 - \vec{r}_1|)$ is the pair potential. It is solved to give ~\cite{Antlanger2014}

\begin{align*}
1 = - \frac{\rho}{2 \alpha r_1} \sum_G i J_1(G r_1)
\int d\vec{r} \left( \frac{\vec{G}}{|\vec{G}|} \cdot \frac{\vec{r}}{|\vec{r}|} \right)
\frac{\delta(\beta u(r))}{\delta r}
\Bigg[
\mu_{G}g^{(0)}(r) e^{i \vec{G} \cdot \vec{r}}
+ h^{(G)}(r) e^{\frac{i}{2} \vec{G} \cdot \vec{r}}\\
+ \sum_{G_1} \mu_{G-G_1} h^{(G_1)}(r) e^{i(\vec{G} - \frac{1}{2}\vec{G}_1)\cdot \vec{r}}
\Bigg]
\tag{9.15}
\end{align*}

The values calculated for several values of $r_1$ are found to lie between $0.92-1.10$, indicating that the value of $h_M^{(G)}(r)$ given in Fig.~9.5 is reasonably accurate. In computer simulations ~\cite{Antlanger2014} and in experiments ~\cite{Han2008} $g(\vec{r}_1,\vec{r}_2)$ is reduced to a function of one variable,

\begin{equation}
G(r) = \frac{1}{\rho^2 V} \int_0^{2\pi} \frac{d\phi_r}{2\pi}
\int d\vec{r}_1 \rho(\vec{r}_1)\rho(\vec{r}_2)
\left[ g^{(0)}(|\vec{r}_2 - \vec{r}_1|) + h^{(b)}(\vec{r}_1,\vec{r}_2) \right],
\tag{9.16}
\end{equation}
where $\phi_r$ is the angle of vector $\vec{r} = \vec{r}_2 - \vec{r}_1$. In the fluid $G(r)$ reduces to the radial distribution function $g^{(0)}(r)$. The values of $G(r)$ plotted in Fig.~9.6 for $\Gamma = 12$ as a function of $r$ are in good qualitative agreement with the values found from simulations and experiments for soft colloidal particles  ~\cite{Han2008,Antlanger2014}, as the pair potential in these work are different we cannot expect quantitative agreement. In the figure value of $g^{(0)}(r)$ and the value of $G^{(0)}(r)$ found without including the contribution of $h^{(b)}(\vec{r_1},\vec{r_2})$ are also shown. In the figure, it is also seen that the effect of $h^{(b)}(\vec{r}_1,\vec{r}_2)$ is limited to the first two peaks of $G(r)$ only.

\begin{figure}
\centering
\includegraphics[width=\columnwidth]{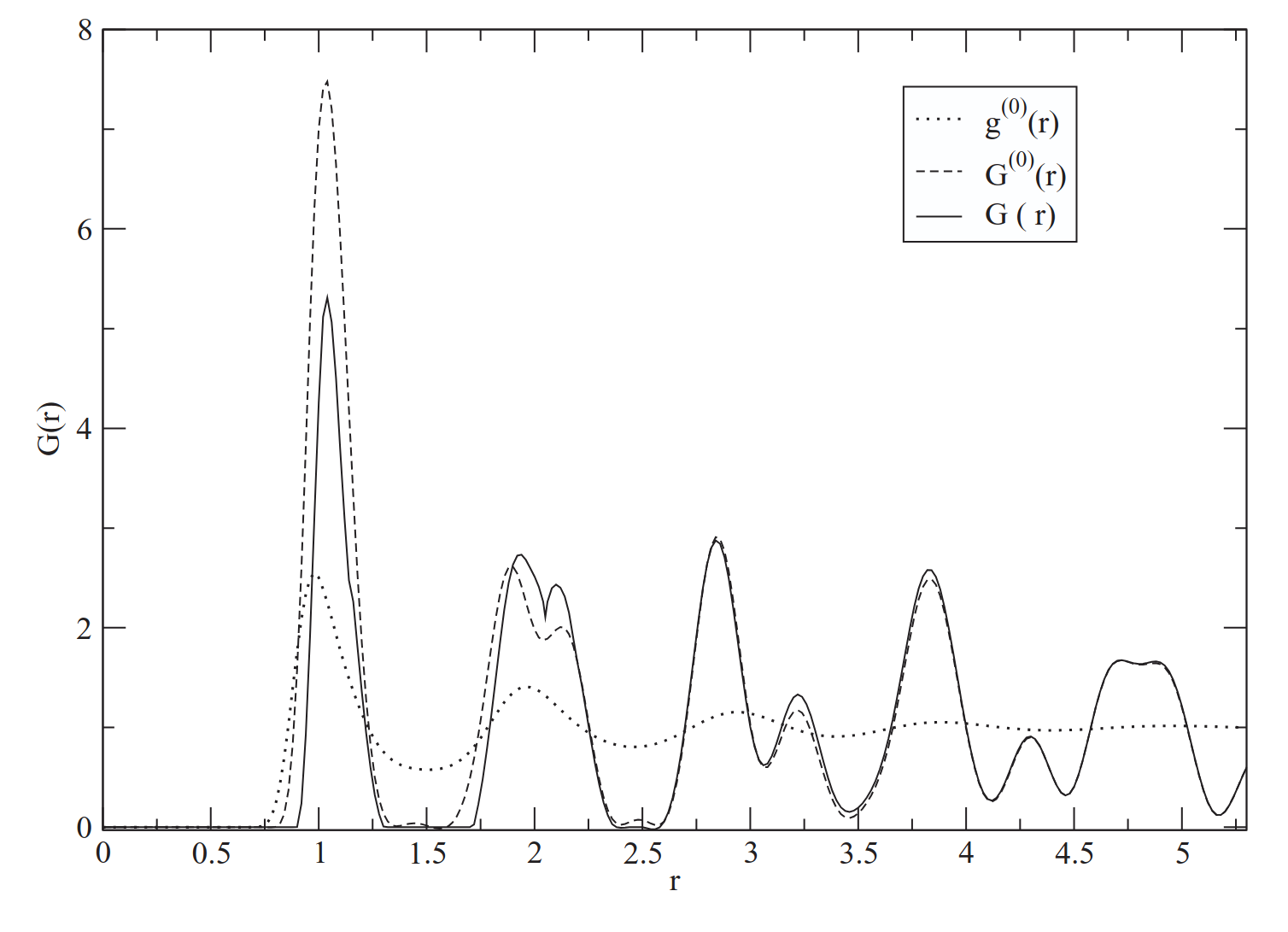}
\caption*{\textbf{Fig. 9.6}
Correlation function $G(r)$ (see Eq. (9.12)) as a function of $r$. $G^{(0)}$ refers to values found without symmetry breaking part of $h(\vec r_1, \vec r_2 )$. $g^{(0)}(r)$ is the radial distribution function in a fluid of the average density of the crystal. Reproduced with permission from Ref. ~\cite{Jaiswal2014}
}
\end{figure}

\section{Fluid-solid phase transition}

The freezing of fluids into crystalline solids is a basic phenomenon, the most inevitable of all phase changes. At the freezing point, the continuous symmetry of translations and rotations of fluid is broken into one of the Bravais lattices. Efforts have been made for the last several decades to find a first principle theory which can answer questions regarding at what density, pressure, and temperature does a particular fluid freeze? What is the change in thermodynamic variables like entropy and density upon freezing? Which of the Bravais lattices emerges at the freezing point for a given system, and what are the values of order parameters?

The computational study of a variety of model systems  ~\cite{Prestipino2005,Agrawal1995,Davidchack2005,Tan2011,likos2007,Zhu2011,Laird1991} has led researchers to derive a fundamental link between macroscopic phase behavior and interparticle interactions. In the case of particles interacting via spherically symmetric potentials, it is found that the crystalline structure that emerges at the freezing point is predominantly controlled by the nature of the repulsive component of the interparticle potential. Harsh repulsion favors a compact structure such as fcc or the hexagonal-closed-packed (hcp) structures, whereas soft repulsion favors a relatively open structure such as bcc. Many real as well as model systems are known to have more than one crystalline solid phase  ~\cite{Young1991}. The phase diagrams of such systems have one or more solid-solid phase boundaries in addition to the fluid-solid boundary. The well known example is water, whose pressure-temperature (P-T) phase diagram has several solid-solid phase boundaries and triple points  ~\cite{Soper2002}. Most metallic elements on the left-hand side of the periodic table together with all lanthanides and actinides are found to freeze at low pressures in a bcc structure and transform to other structures at higher pressures  ~\cite{Donohue1974}. Model systems with sufficiently soft repulsive interactions show similar behavior  ~\cite{Monson2007}. In this section, EDFT is used to describe the fluid-solid phase transitions and in the next section, the solid-solid transition. The results of other theories are compared.

The fluid-solid transition point is determined by the condition $\Delta W = W_f - W_c = 0$, where the subscripts $f$ and $c$ refer, respectively, to the fluid and the crystal. From Eq.\ (6.14) we have

\begin{equation}
\Delta W = \Delta W_{\mathrm{id}} + \Delta W_{0} + \Delta W_{b}, \tag{10.1}
\end{equation}
where

\begin{equation}
\Delta W_{\mathrm{id}} = \int d\vec{r} \left[\rho_c(\vec r) \ln\!\left( \frac{\rho_c(\vec{r})}{\rho_f} \right) - \left( \rho_c(\vec{r}) - \rho_f \right) \right], \tag{10.2}
\end{equation}

\begin{equation}
\Delta W_{0} = -\frac{1}{2} \int d\vec{r}_1 d\vec{r}_2 \left( \rho_c(\vec{r}_1) - \rho_f \right) \left( \rho_c(\vec{r}_2) - \rho_f \right) \, \bar{c}^{(0)}(|\vec{r}_2 - \vec{r}_1|), \tag{10.3}
\end{equation}
and

\begin{equation}
\Delta W_{b} = -\frac{1}{2} \int d\vec{r}_1 d\vec{r}_2 \left( \rho_c(\vec{r}_1) - \rho_c \right) \left( \rho_c(\vec{r}_2) - \rho_c \right) \, \bar{c}^{(b)}(\vec{r}_1,\vec{r}_2), \tag{10.4}
\end{equation}
Quantities $\bar{c}^{(0)}$ and $\bar{c}^{(b)}$ appearing, respectively, in (10.3) and (10.4) are defined (see Sec.\ 6) as

\begin{equation}
\bar{c}^{(0)}(|\vec{r}_2 - \vec{r}_1|) = 2 \int_0^1 d\lambda \, (1-\lambda) \, c^{(2)}\!\left( |\vec{r}_2 - \vec{r}_1| ; \rho_f + \lambda(\rho_c - \rho_f) \right), \tag{10.5}
\end{equation}

\begin{equation}
\bar{c}^{(b)}(\vec{r}_1,\vec{r}_2) = 4 \int_0^1 d\lambda \, (1-\lambda) \int_0^1 d\xi \, (1-\xi) \,
c^{(b)}\!\left( \vec{r}_1, \vec{r}_2 ; \lambda \rho_c, \xi \rho_G \right). \tag{10.6}
\end{equation}
In above equations, $\rho_f$ is density of the coexisting fluid, $\rho_c(\vec{r})$ is the one-particle density of the crystal, $\rho^{(b)}(r)=\rho_c(\vec r)-\rho_c$ and $\rho_c$ is the averaged crystal density at the freezing point, where $\rho_c = \rho_f (1+\Delta\rho^*)$, $\Delta\rho^*$ being the change in density on crystallization.

Since in most cases $(\rho_c - \rho_f)\rho_f \ll \rho_f$, it is plausible to replace
$\bar{c}^{(0)}(r)$ in Eq.\ (10.5) by $c^{(2)}(r,\rho_f)$, which is DPCF of the
coexisting fluid. As the order parameters that appear are linear in $c^{(b,1)}(\vec{r}_1,\vec{r}_2)$ (see Eqs.\ (8.24) and (8.25)) and quadratic in
$c^{(b,2)}(\vec{r}_1,\vec{r}_2)$ (see Eqs.\ (8.28)--(8.41)), the integration over
$\xi$ variables in Eq.\ (10.6) is performed analytically, leading to Eq. ~\cite{Bharadwaj2013}.

\begin{align}
\bar{c}^{(b)}(\vec{r}_1,\vec{r}_2)
&= \sum_{\vec{G}} e^{i \vec{G}\cdot \vec{r}_c}
\Bigg[
\sum_{l m} \bar{c}^{(G,1)}_{l}(r)\, Y^*_{l m}(\hat{G})\, Y_{l m}(\hat{r})
\nonumber\\
&\qquad\qquad
+ \sum_{l m} \sum_{l' m'}
\left( \bar{c}^{(G,2,1)}_{l m, l' m'}(r) + \bar{c}^{(G,2,2)}_{l m, l' m'}(r) \right)
Y^*_{l' m'}(\hat{G})\, Y_{l m}(\hat{r})
\Bigg] .
\tag{10.7}
\end{align}

\noindent where
\begin{equation}
\bar{c}^{(G,1)}_{l}(r)
= \frac{1}{3}\,\rho_G \sum_{l_1} \sum_{l_2}
\Lambda_{1}(l_1,l_2,l)\,
j_{l_2}\!\left( \tfrac{1}{2} G r \right)\,
\bar{B}_{l_1}(r,G) ,
\tag{10.8}
\end{equation}

\begin{align}
\bar{c}^{(G,2,1)}_{l m, l' m'}(r)
&= \frac{1}{6} \sum_{G_1} \rho_{G_1}\rho_{K}
\sum_{l_1 m_1} \sum_{l_2 m_2}
\Lambda^{\,l\,l'\,l_1\,l_2}_{m\,m'\,m_1\,m_2}\,
j_{l}\!\left( \tfrac{1}{2} G r \right)
\nonumber\\
&\qquad \times
\bar{Q}_{l_1 l_2}(r, G, G_1)\,
Y^*_{l_1 m_1}(\hat{G}_1)\,
Y^*_{l_2 m_2}(\hat{K}) ,
\tag{10.9}
\end{align}

\begin{align}
\bar{c}^{(G,2,2)}_{l m, l' m'}(r)
&= \frac{1}{6} \sum_{G_1} \rho_{G_1}\rho_{K}
\sum_{l_1 m_1} \sum_{l_2 m_2} \sum_{l_3 m_3}
\Lambda^{\,l\,l'\,l_1\,l_2\,l_3}_{m\,m'\,m_1\,m_2\,m_3}\,
j_{l'}\!\left( \tfrac{1}{2} G r \right)
\nonumber\\
&\qquad \times
\bar{N}_{l_1 l_2 l_3}(r, G, G_1)\,
Y^*_{l_2 m_2}(\hat{G}_1)\,
Y^*_{l_3 m_3}(\hat{K}) ,
\tag{10.10}
\end{align}

with

\begin{align*}
\bar{B}_{l_1}(r,G)
&= 2 \int_{0}^{1} d\lambda\,(1- \lambda) \,
B_{l_1}(r,G;\lambda\rho),
\tag{10.11}
\\[6pt]
\bar{Q}_{l_1 l_2}(r,G,G_1)
&= 2 \int_{0}^{1} d\lambda\, (1-\lambda) \,
Q_{l_1 l_2}(r,G,G_1;\lambda\rho),
\\[6pt]
Q_{l_1 l_2}(r,G,G_1;\rho)
&= M_{l_1}(r,G_1;\rho)\, M_{l_2}(r,K;\rho),
\tag{10.12}
\\[6pt]
\end{align*}
and
\begin{align*}
\bar{N}_{l_1 l_2 l_3}(r,G,G_1)
&= 2 \int_{0}^{1} d\lambda\, (1-\lambda) \,
N_{l_1 l_2 l_3}(r,G,G_1;\lambda\rho).
\tag{10.13}
\end{align*}

The quantities $B_{l_1}(r,G)$, $M_{l_1}(r,G_1)$, $M_{l_2}(r,K)$, and
$N_{l_1 l_2 l_3}(r,G,G_1)$ are defined by Eqs.~(8.27), (8.33), (8.34), and
(8.39), respectively. The integrations over $\lambda$ and $\lambda'$ have
been performed numerically by varying them from 0 to 1 on a fine grid and
evaluating the functions $B_{l_1}$, $Q_{l_1 l_2}$, and $N_{l_1 l_2 l_3}$ on
these densities. Since these functions vary smoothly with density and their
values have been evaluated at closely spaced values of density, the result
found for $\bar{c}^{(b)}(\vec{r}_1,\vec{r}_2)$ is expected to be accurate.
 
The ideal part $\Delta W_{id}$  is calculated using expression for $\rho(\vec r)$  given by Eqs.~(4.30) which yields, 

\begin{equation}
\frac{\Delta W_{\mathrm{id}}}{N_c}
= \frac{1}{1+\Delta \rho^*}
+ \left[ \frac{d}{2}\ln\!\left(\frac{\alpha_c}{\pi}\right)
- \frac{d+2}{2} - \ln \rho_f \right],
\tag{10.14}
\end{equation}

where $d$ is dimension of space. For calculating $\frac{\Delta W_0}{N_c}$ one uses for $\rho(r)$ expression given by Eq.~(4.32). This leads to

\begin{equation}
\frac{\Delta W_0}{N_c}
= -\frac{1}{2}\rho_c
\left( \frac{\Delta \rho^*}{1+\Delta \rho^*} \right)^2
\hat{c}^{(0)}(0)
- \frac{1}{2\rho_c} \sum_{\vec{G}} |\rho_{\vec{G}}|^2
\hat{c}^{(0)}(G),
\tag{10.15}
\end{equation}

where $N_c = \rho_c V$ is the number of particles in the crystal and $\rho_G=\rho_c \mu_G$.

For $\Delta W_b$ we write

\begin{equation}
\Delta W_b = \Delta W_b^{(1)} + \Delta W_b^{(2)},
\tag{10.16}
\end{equation}

where ~\cite{Bharadwaj2013}

\begin{equation}
\frac{\Delta W_b^{(1)}}{N_c}
= -\frac{1}{2}\rho_f (1+\Delta \rho^*)^2
\sum_{\vec{G}}' \sum_{\vec{G}_2}
\mu_{\vec{G}_2}\, \mu_{\vec{-G}-\vec{G}_2}\,
\hat{\bar c}^{(\vec G,1)}\!\left(\vec{G}_2 + \frac{1}{2}\vec{G}\right),
\tag{10.17}
\end{equation}

\begin{equation}
\frac{\Delta W_b^{(2)}}{N_c}
= -\frac{1}{2}\rho_f (1+\Delta \rho^*)^2
\sum_{\vec{G}}' \sum_{\vec{G}_2}'
\mu_{\vec{G}_2}\, \mu_{\vec{-G}-\vec{G}_2}\,
\hat{\bar c}^{(\vec G,2)}\!\left(\vec{G}_2 + \frac{1}{2}\vec{G}\right).
\tag{10.18}
\end{equation}

The prime on summations in (10.17) and (10.18) indicates the condition
$\vec{G} \neq 0$, $\vec{G}_2 \neq 0$ and $\vec{G}+\vec{G}_2 \neq 0$ and 

\begin{equation}
\hat{\bar c}^{(0)}(G) = \int d\vec{r}\, \bar c^{(0)}(r,\rho_f)\, e^{i\vec{G}\cdot\vec{r}},
\tag{10.19}
\end{equation}

\begin{equation}
\hat{\bar c}^{(G,1)}\!\left(\vec{G}_2 + \frac{1}{2}\vec{G}\right)
= \frac{1}{3}\mu_G \sum_{l_1}\sum_{l_2} \Lambda_1(l_1,l_2,l)\, Y_{lm}^*(\hat{G})
\int d\vec{r}\, j_{l_2}\!\left(\frac{1}{2}Gr\right)\, \bar{B}_{l_1}(r,G)\,
e^{i(\vec{G}_2+\frac{1}{2}\vec{G})\cdot\vec{r}}\, Y_{lm}(\hat{r}),
\tag{10.20}
\end{equation}

\begin{equation*}
\hat{c}^{(G,2)}\!\left(\vec{G}_2 + \frac{1}{2}\vec{G}\right)
= \hat{c}^{(G,2,1)}\!\left(\vec{G}_2 + \frac{1}{2}\vec{G}\right)
+ 2\,\hat{c}^{(G,2,2)}\!\left(\vec{G}_2 + \frac{1}{2}\vec{G}\right),
\end{equation*}

\begin{equation*}
\hat{\bar c}^{(G,2,1)}\!\left(\vec{G}_2 + \frac{1}{2}\vec{G}\right)
= \frac{1}{6}\sum_{G_1}\mu_{G_1}\mu_K
\sum_{lm}\sum_{l'm'}\sum_{l_1 m_1}\sum_{l_2 m_2}
\Lambda^{ll'l_1l_2}_{mm'm_1m_2}\,
Y_{l'm'}^*(\hat{G})\, Y_{l_1 m_1}^*(\hat{G}_1)\, Y_{l_2 m_2}^*(\hat{K})
\end{equation*}

\begin{equation}
\qquad \times \int d\vec{r}\, j_{l_2}\!\left(\frac{1}{2}Gr\right)\,
\bar{Q}_{l_1 l_2}(r,G,G_1)\, e^{i(\vec{G}_2+\frac{1}{2}\vec{G})\cdot\vec{r}}\, Y_{lm}(\hat{r}),
\tag{10.21}
\end{equation}

\begin{equation*}
\hat{\bar c}^{(G,2,2)}\!\left(\vec{G}_2 + \frac{1}{2}\vec{G}\right)
= \frac{1}{6}\sum_{G_1}\mu_{G_1}\mu_K
\sum_{lm}\sum_{l'm'}\sum_{l_1 m_1}\sum_{l_2 m_2}\sum_{l_3 m_3}
\Lambda^{ll'l_1l_2 l_3}_{mm'm_1m_2m_3}\,
Y_{l'm'}^*(\hat{G})\, Y_{l_2 m_2}^*(\hat{G}_1)\, Y_{l_3 m_3}^*(\hat{K})
\end{equation*}

\begin{equation}
\qquad \times \int d\vec{r}\, j_{l_2}\!\left(\frac{1}{2}Gr\right)\,
\bar{N}_{l_1 l_2 l_3}(r,G,G_1)\, e^{i(\vec{G}_2+\frac{1}{2}\vec{G})\cdot\vec{r}}\, Y_{lm}(\hat{r}),
\tag{10.22}
\end{equation}

The terms $\Delta W_0/N$, $\Delta W_b^{(1)}/N$, and $\Delta W_b^{(2)}/N$ are,
respectively, second, third, and fourth orders in order parameters.

\subsection{ Results for systems interacting via inverse power potentials in 3-dimensions}

The fluid-solid transition for a family of inverse power potentials defined by Eq.~(8.2) have been investigated using EDFT in 3-dimensions in Refs.~\cite{Bharadwaj2013,Jaiswal2014,Bharadwaj2017} and for 2-dimensions in Ref. ~\cite{Jaiswal2013} for several values of softness parameter $n$.

The function $\Delta W_{fc}/N_c$  (where $N_c = \rho_c V$) was minimized with respect to $\gamma_f$, $\Delta \rho^{*}(\Delta \gamma)$ and $\alpha_c$. It was found that when $\gamma_f$ is close to the transition, $\Delta W_{fc}/\rho_c V$ develops a minimum at some value of $\alpha_c$. The value of minimum of $\Delta W_{fc}/\rho_c V$ and of $\alpha_c$ at which this minimum occurs depend on value of $\Delta \rho^{*}$. The variation of $\Delta W_{fc}/N_c$ (where $N_c = \rho_c V$) with $\gamma_f$, $\alpha_c$ and $\Delta \rho^{*}$ is shown in Fig. 10.1. The lowest value of $\gamma_f$ for which the condition $\Delta W_{fc}/N_c = 0$ is satisfied and the corresponding values of $\Delta \rho^{*}$ and $\alpha_c$ are taken as the fluid-solid transition parameters (see Fig. 10.1c). The result satisfies transition conditions, $
\frac{\partial}{\partial \alpha} \left( \frac{\Delta W_{fc}}{N_c} \right) = 0$,
 $\frac{\partial}{\partial \Delta \rho^{*}} \left( \frac{\Delta W_{fc}}{N_c} \right) = 0$ along with $\Delta W_{fc}/N_c = 0$.

In Table 10.1 values of different terms of $\Delta W/N_c$ at freezing point are listed for potentials with $n = 4, 6, 6.5, 7, 12$ and $\infty$. The contribution made by the broken symmetry part to the grand potential is seen to be substantial and its importance increases with the softness of the potential. For example, while for $n = \infty$, the contribution of broken symmetry is about $8\%$ of the contribution made by symmetry-conserved part, it increases to $45\%$ for $n = 4$. Without this contribution the theory strongly overestimates the stability of the fluid phase. This explains why the SODFT (RY theory) gives reasonable result for hard sphere potentials but fails for potentials that have a soft-core and attractive tail.

\begin{figure}
\centering
\includegraphics[width=10.0cm]{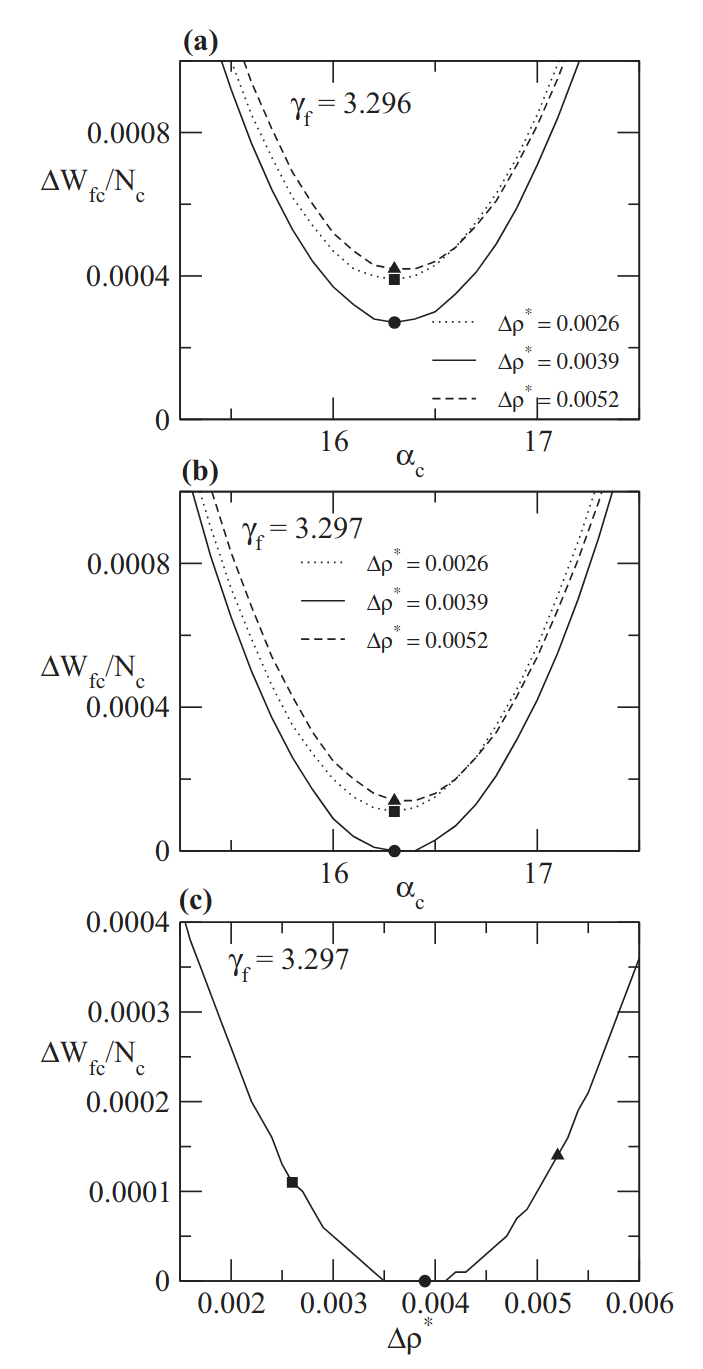}
\caption*{\textbf{Fig. 10.1} Variation of $\frac{\Delta W_{fc}}{N_c}$ as a function of $\alpha_c$ and $\Delta \rho^{*}$ for $1/n = 0.20$ is shown. The minimum of $\frac{\Delta W_{fc}}{N_c}$ and the value of $\alpha_c$ at which this minimum occurs depend on $\Delta \rho^{*}$. In (a) the variation is shown for $\gamma_f = 3.296$ for three values of $\Delta \rho^{*}$; in all cases the minimum is greater than zero; the lowest minimum is for $\Delta \rho^{*} = 0.0039$. In (b) we show the plot for $\gamma_f = 3.297$ where the condition $\frac{\Delta W_{fc}}{N_c} = 0$ is satisfied for $\Delta \rho^{*} = 0.0039$ and $\alpha_c = 16.3$. In (c) the variation of minimum of $\frac{\Delta W_{fc}}{N_c}$ for $\gamma_f = 3.297$ as a function $\Delta \rho^{*}$ is shown. Reproduced with permission from Ref.\cite{Bharadwaj2017}
}
\end{figure}

In Table 10.2 we reproduce results reported in Ref. ~\cite{Bharadwaj2013} for freezing parameters $\gamma_f$, $\gamma_c$, $\Delta \gamma = (\gamma_s - \gamma_f)/\gamma_f$ and the Lindemann $L_n$, and $P \sigma^3 / \epsilon$, where $P$ is pressure at the transition point and compared with those found from computer simulations and with results found by others using approximate free energy functionals. The Lindemann parameter is defined as the ratio of the average displacement of particle from its lattice point to the nearest-neighbor distance in the crystal. For the fcc crystal with the Gaussian density profile of Eq. (4.30), it is expressed ~\cite{Bharadwaj2013},
\begin{equation}
L_n = \left( \frac{3}{a_{\mathrm{fcc}}^{2} \alpha} \right)^{1/2}
\tag{10.23}
\end{equation}
where $a_{\mathrm{fcc}} = \left( \frac{4}{\rho_c} \right)^{1/3}$ is the fcc lattice constant.\\
For the bcc crystal [20],
\begin{equation}
L_n = \left( \frac{2}{a_{\mathrm{bcc}}^{2} \alpha} \right)^{1/2}
\tag{10.24}
\end{equation}
where $a_{\mathrm{bcc}} = \left( \frac{2}{\rho_c} \right)^{1/3}$ is the bcc lattice constant.

The comparison shows that EDFT gives very good account of freezing transition for all values of softness parameter $n$. The fluid for $n > 6.5$ freezes into fcc structure while for $n \le 6$ it freezes into a bcc structure. The fluid--bcc--fcc triple point is found at $\frac{1}{n} = 0.158$. The energy-difference between the two cubic structures at the transition is found to be small in agreement with simulation results ~\cite{Prestipino2005,Agrawal1995,Davidchack2005,Tan2011}

\begin{table}[htbp]
\centering
\caption*{{Table 10.1:} Freezing parameters $\gamma_l$, $\Delta \gamma$ and the contributions of ideal, symmetry-conserving and symmetry-broken parts arising from first and second terms of Eq.~(10.16) to $\Delta W/N$ at the transition point. Reproduced with permission from Ref.~\cite{Bharadwaj2013}.}
\label{tab:freezing}
\begin{tabular}{c c c c c c c c}
\hline
\hline
$n$ &  Lattice &  $\gamma_l$ &  $\Delta \gamma$ &  $\Delta W_{\mathrm{id}}/N$ &  $\Delta W_{0}/N$ &  $\Delta W_{b}^{(1)}/N$ &  $\Delta W_{b}^{(2)}/N$ \\
\hline
4.0  & bcc & 5.57 & 0.007 & 2.86 & $-2.09$ & $-0.94$ & 0.17 \\
     & fcc & 5.60 & 0.008 & 3.52 & $-2.64$ & $-1.03$ & 0.15 \\
6.0  & bcc & 2.30 & 0.011 & 2.56 & $-1.99$ & $-0.58$ & 0.01 \\
     & fcc & 2.32 & 0.012 & 3.48 & $-2.75$ & $-0.72$ & 0.002 \\
6.5  & bcc & 2.04 & 0.011 & 2.38 & $-1.89$ & $-0.50$ & 0.001 \\
     & fcc & 2.03 & 0.013 & 3.34 & $-2.69$ & $-0.66$ & 0.001 \\
7.0  & bcc & 1.86 & 0.015 & 2.29 & $-1.85$ & $-0.44$ & 0.000\\
     & fcc & 1.84 & 0.014 & 3.39 & $-2.76$ & $-0.63$ & 0.000 \\
12.0 & fcc & 1.17 & 0.034 & 3.71 & $-3.14$ & $-0.57$ & 0.000 \\
$\infty$ & fcc & 0.937 & 0.106 & 4.44 & $-4.10$ & $-0.34$ & 0.000 \\
\hline
\hline
\end{tabular}
\end{table}

\begin{table*}[htbp]
\centering
\caption*{{Table 10.2:}Comparison of the parameters $\gamma_l$, $\gamma_s$, and $\Delta\gamma$, the Lindemann parameter $L$, and the pressure $P$ at the coexistence found from different free-energy functional and computer simulations. MWDA denotes the modified weighted density approximation, SODFT denotes the Ramakrishnan-Yussouff density-functional theory, MHNC denotes the modified hypernetted-chain closure relation, and MSMC denotes the Mayer sampling Monte Carlo.Reproduced with permission from Ref. \cite{Bharadwaj2013}}
\label{tab:comparison_fcc}
\begin{tabular}{c c l c c c c c}
\toprule
$n$ & Lattice & Theory/Simulation & $\gamma_l$ & $\gamma_s$ & $\Delta\gamma$ & $L$ & $P\sigma^3/\epsilon$ \\
\midrule

\multirow{8}{*}{$\infty$}
& \multirow{8}{*}{fcc}
& EDFT \cite{Bharadwaj2013,SinghSingh2009} & 0.937 & 1.036 & 0.106 & 0.09 & 11.46 \\
& & MWDA-static \cite{Wong1999} & 0.863 & 0.964 & 0.115 & 0.13 & --- \\
& & MWDA \cite{Wong1999} & 0.906 & 1.044 & 0.116 & 0.10 & --- \\
& & SODFT \cite{Barrat1987} & 0.980 & 1.146 & 0.174 & 0.06 & --- \\
& & Simulation \cite{Alder1968} & 0.942 & 1.041 & 0.105 & --- & --- \\
& & MC simulation \cite{Agrawal1995} & 0.940 & 1.041 & 0.107 & 0.12 & 11.70 \\
& & MC simulation \cite{Davidchack2005} & 0.939 & 1.037 & 0.104 & --- & 11.57 \\

\midrule

\multirow{7}{*}{12}
& \multirow{7}{*}{fcc}
& EDFT \cite{Bharadwaj2013} & 1.17 & 1.21 & 0.034 & 0.11 & 23.67 \\
& & MWDA-static \cite{Wong1999} & 1.12 & 1.16 & 0.037 & 0.14 & --- \\
& & MWDA/MHNC \cite{Laird1987,Laird1990} & 1.19 & 1.25 & 0.046 & 0.10 & --- \\
& & SODFT \cite{Barrat1987} & 1.28 & 1.37 & 0.07 & 0.07 & --- \\
& & MC simulation$^{a}$ \cite{Agrawal1995} & 1.17 & 1.22 & 0.042 & 0.14 & 23.64 \\
& & MSMC \cite{Tan2011} & 1.16 & 1.20 & 0.037 & --- & 23.24 \\
& & MC simulation \cite{Davidchack2005} & 1.16 & 1.21 & 0.037 & --- & 23.41 \\

\midrule

\multirow{5}{*}{7}
& fcc & EDFT \cite{Bharadwaj2013,Bharadwaj2017} & 1.84 & 1.87 & 0.014 & 0.12 & 64.97 \\
& & MC simulation$^{a}$ \cite{Agrawal1995} & 1.85 & 1.88 & 0.017 & 0.15 & 64.98 \\
& & MC simulation \cite{Davidchack2005} & 1.84 & 1.87 & 0.016 & --- & 64.22 \\
& bcc & EDFT \cite{Bharadwaj2013,Bharadwaj2017} & 1.86 & 1.89 & 0.015 & 0.18 & 67.12 \\
& & MC simulation \cite{Davidchack2005} & 1.83 & 1.86 & 0.015 & --- & 63.88 \\

\midrule

\multirow{4}{*}{6.5}
& fcc & EDFT \cite{Bharadwaj2013,Bharadwaj2017} & 2.03 & 2.06 & 0.013 & 0.12 & 80.11 \\
& & MC simulation$^{a}$ \cite{Agrawal1995} & 2.04 & 2.07 & 0.014 & 0.15 & 80.40 \\
& bcc & EDFT \cite{Bharadwaj2013,Bharadwaj2017} & 2.04 & 2.07 & 0.014 & 0.17 & 78.98 \\
& & MC simulation$^{a}$ \cite{Prestipino2005,Saija2006} & 2.03 & 2.05 & 0.010 & 0.18 & 78.40 \\

\midrule

\multirow{6}{*}{6}
& fcc & EDFT \cite{Bharadwaj2013,Bharadwaj2017} & 2.32 & 2.35 & 0.012 & 0.12 & 103.7 \\
& & MWDA-static \cite{Wong1999} & 2.33 & 2.35 & 0.007 & 0.17 & --- \\
& & MWDA/MHNC \cite{Laird1990} & 2.67 & 2.72 & 0.02 & 0.07 & --- \\
& & SODFT \cite{Barrat1987} & 3.43 & 3.52 & 0.026 & 0.07 & --- \\
& & MC simulation$^{a}$ \cite{Agrawal1995} & 2.34 & 2.37 & 0.012 & 0.15 & 104.5 \\
& & MC simulation \cite{Davidchack2005} & 2.32 & 2.35 & 0.012 & --- & 103.0 \\

\multirow{5}{*}{}
& bcc & EDFT \cite{Bharadwaj2013,Bharadwaj2017}& 2.30 & 2.33 & 0.011 & 0.16 & 101.22 \\
& & MC simulation$^{a}$ \cite{Agrawal1995} & 2.32 & 2.35 & 0.011 & 0.17 & 103.6 \\
& & MSMC \cite{Tan2011} & 2.30 & 2.32 & 0.011 & --- & 100.1 \\
& & MC simulation \cite{Davidchack2005} & 2.30 & 2.33 & 0.012 & --- & 100.0 \\
& & MC simulation$^{a}$ \cite{Prestipino2005,Saija2006} & 2.29 & 2.31 & 0.009 & 0.18 & 99.34 \\

\midrule

\end{tabular}
\end{table*}

\begin{table*}[htbp]
\begin{tabular}{c c c c c c c c}
\multirow{5}{*}{4}
& fcc & EDFT \cite{Bharadwaj2013,Bharadwaj2017}& 5.60 & 5.63 & 0.008 & 0.12 & 565.6 \\
& & MWDA-static \cite{Wong1999} & 5.22 & 5.26 & 0.008 & 0.13 & --- \\
& & MWDA/MHNC \cite{Laird1990} & 8.18 & 8.24 & 0.07 & 0.07 & --- \\
& & SODFT \cite{Barrat1987} & 12.3 & 12.47 & 0.014 & 0.07 & --- \\
& & MC simulation \cite{Agrawal1995} & 5.68 & 5.71 & 0.005 & 0.17 & 637.0 \\

\multirow{3}{*}{}
& bcc & EDFT \cite{Bharadwaj2013,Bharadwaj2017} & 5.57 & 5.61 & 0.007 & 0.16 & 561.2 \\
& & MWDA-static reference \cite{Barrat1987} & 5.05 & 5.09 & 0.008 & 0.18 & --- \\
& & MC simulation \cite{Davidchack2005} & 5.73 & 5.75 & 0.004 & 0.18 & 648.0 \\

\bottomrule
\end{tabular}
\end{table*}

\subsection{Results for systems interacting via inverse power potential in 2-dimensions}

The fluid-solid transition in a two-dimensional system interacting via IPP [Eq.(8.2)] was studied by Jaiswal et al. ~\cite{Jaiswal2013}. In their calculation values of broken symmetry contribution to DPCF, $c^{(b)}(\vec{r}_1,\vec{r}_2)$, described in Section 9 was used. Substituting value of  $c^{(b)}(\vec{r}_1, \vec{r}_2)$ given by Eqs.(9.3) - (9.7) into Eq. (10.6) one gets

\begin{align}
\bar{c}^{(b)}(\vec{r}_1, \vec{r}_2)
&= 4 \sum_{\vec{G}} e^{i\vec{G} \cdot \vec{r}_c}
   \sum_M \sum_m (i)^M e^{iM\phi_G} e^{-iM\phi_r}
   J_{M+m}\!\left(\frac{1}{2} G r\right) \nonumber \\
&\quad \times \int_0^1 d\xi \,(1-\xi)\, \rho_G
   \int_0^1 d\lambda \,(1-\lambda)\, L_m(r, G, \lambda \rho)
\end{align}

where $L_m(r, G, \rho) = t(r, \rho) B_m(r, G, \rho)$ depends on density $\rho$ which is scaled from zero to $\rho_c$ by varying $\lambda$ between 0 and 1. The integration over $\xi$ is done analytically leading to  ~\cite{Jaiswal2013},

\begin{equation}
\bar c^{(b)}(\vec{r}_1,\vec{r}_2)
= \frac{1}{3} \sum_{\vec{G}} \rho_G e^{i \vec{G}\cdot \vec{r}_c}
\sum_{m} \sum_{n} i^{m} e^{i m \phi_G} e^{-i n \phi_r}
J_{m+n}\!\left(\frac{1}{2} G r \right) I_m(r,G,\rho_c),
\tag{10.25}
\end{equation}

where $\rho_G = \rho_c \mu_G$ and

\begin{equation}
I_m(r,G,\rho_c) = 2 \int_0^1 d \lambda \, (1-\lambda) \, L_m(r,G,\lambda \rho_c),
\tag{10.26}
\end{equation}

These integrals were evaluated numerically for a given density $\rho_c$ from known values of $L_m(r,G,\rho)$ for $\lambda$ varying from zero to one.

When the value of $c^{(b)}(\vec r_1,\vec r_2)$ is substituted in the expression of $\Delta W_b$  given by Eq.~(10.4) one gets,

\begin{equation}
\frac{\Delta W_b}{N_c}
= -\frac{1}{2} \rho_f (1+\Delta \gamma)^2
\sum ^{'}_{\vec{G}} \sum ^{'}_{\vec{G}_1}
\mu_{G_1} \mu_{-G-G_1} \, \hat{\bar c}\!\left(\vec{G}_1 + \frac{1}{2}\vec{G}\right).
\tag{10.27}
\end{equation}
where 

\begin{equation*}
\bar c^{(b)}(\vec{G_1}+\frac{1}{2}\vec{G})
= \int {d\vec{r}} \, \bar c^{(a)}(\vec{r},\vec{r}_c) \, exp[{-i(\vec{G_1}+\frac{1}{2}\vec{G}) . \vec r }]
 \end{equation*}
and
\begin{equation*}
    \Delta \gamma = \frac{\gamma_c-\gamma_f}{\gamma_f}
\end{equation*}
The prime on a summation in Eq(10.27) indicates the condition $\vec G\neq 0$, $\vec G_1 \neq 0$ and $\vec G+ \vec G_1 \neq 0$.

Values of freezing parameters found in Ref.\cite{Jaiswal2013} are reproduced in Table 10.3. The experimental results given in the table is from real-space microscopy measurement of magnetic colloids confined to an air-water interface \cite{Zahn1999,vonGrunberg2004}. The numerical simulation values are from \cite{Lowen1996,Haghgooie2005}.

\begin{table}[htbp]
\centering
\caption*{{Table 10.3:}Freezing parameters $\Gamma_f(= \gamma_f^{n/2})$, $\Gamma_c(= \gamma_s^{n/2})$, the width of coexistence region $\Delta \Gamma = \Gamma_c - \Gamma_f$, and the relative displacement parameter $\xi (\simeq 2/\alpha)$ at the coexistence obtained from various density functional schemes. The MWDA stands for modified weighted density approximation, EMA stands for extended modified weighted density approximation.}

\label{tab:freezing_params}
\begin{tabular}{lcccc}
\hline\hline
 & $\Gamma_f$ & $\Gamma_s$ & $\Delta \Gamma$ & $\xi$ \\
\hline
EDFT \cite{Jaiswal2013}      & 11.04 & 11.46 & 0.42 & 0.020 \\
MWDA  \cite{vonTeeffelen2008}           & 41.07 & 41.13 & 0.06 & 0.017 \\
EMA \cite{vonTeeffelen2008}            & 23.00 & 23.08 & 0.09 & 0.020 \\
Simulation \cite{Haghgooie2005}             & 12.0  & 12.25 & 0.25 & ---   \\
Experiment \cite{vonGrunberg2004}             & 10.0  & 10.75 & 0.75 & 0.038 \\
\hline\hline
\end{tabular}
\end{table}
\newpage
\subsection{Fluid--solid transition in systems interacting via purely repulsive Weeks--Chandler--Anderson and the full Lennard--Jones potentials}
Bharadwaj and Singh \cite{Bharadwaj2015} used the theory to calculate freezing parameters of systems interacting via the 12--6 Lennard--Jones potential,
\begin{equation*}
U(r) = 4\varepsilon \left[ \left( \frac{\sigma}{r} \right)^{12} - \left( \frac{\sigma}{r} \right)^{6} \right],
\label{eq:10.28}
\tag{10.28}
\end{equation*}
where $\varepsilon$ and $\sigma$ are potential parameters. Also, in order to estimate the role played by the attractive and repulsive parts of the LJ potential in formation of crystalline structure at the freezing point, they considered the purely repulsive Weeks--Chandler--Anderson (WCA) reference potential defined as
\begin{equation}
u_R(r) =
\begin{cases}
u(r) + \varepsilon, & r \le r_m, \\
0, & r > r_m,
\end{cases}
\label{eq:10.29}
\tag{10.29}
\end{equation}
where $r_m = 2^{1/6}\sigma$ is the value of $r$ at which the LJ potential has its minimum value. In Ref.\cite{Bharadwaj2015} the WCA potential is referred to as a reference Lennard--Jones (RLJ) potential. While the LJ potential mimics characteristics of interaction potential of spherical molecules, the RLJ potential is used to model interactions in polymers \cite{Kroger2004} and dendrimers \cite{Basko2006}. Using Eqs.~(10.14)--(10.16) freezing parameters were calculated.
\begin{figure}
\centering
\includegraphics[width=12.0cm]{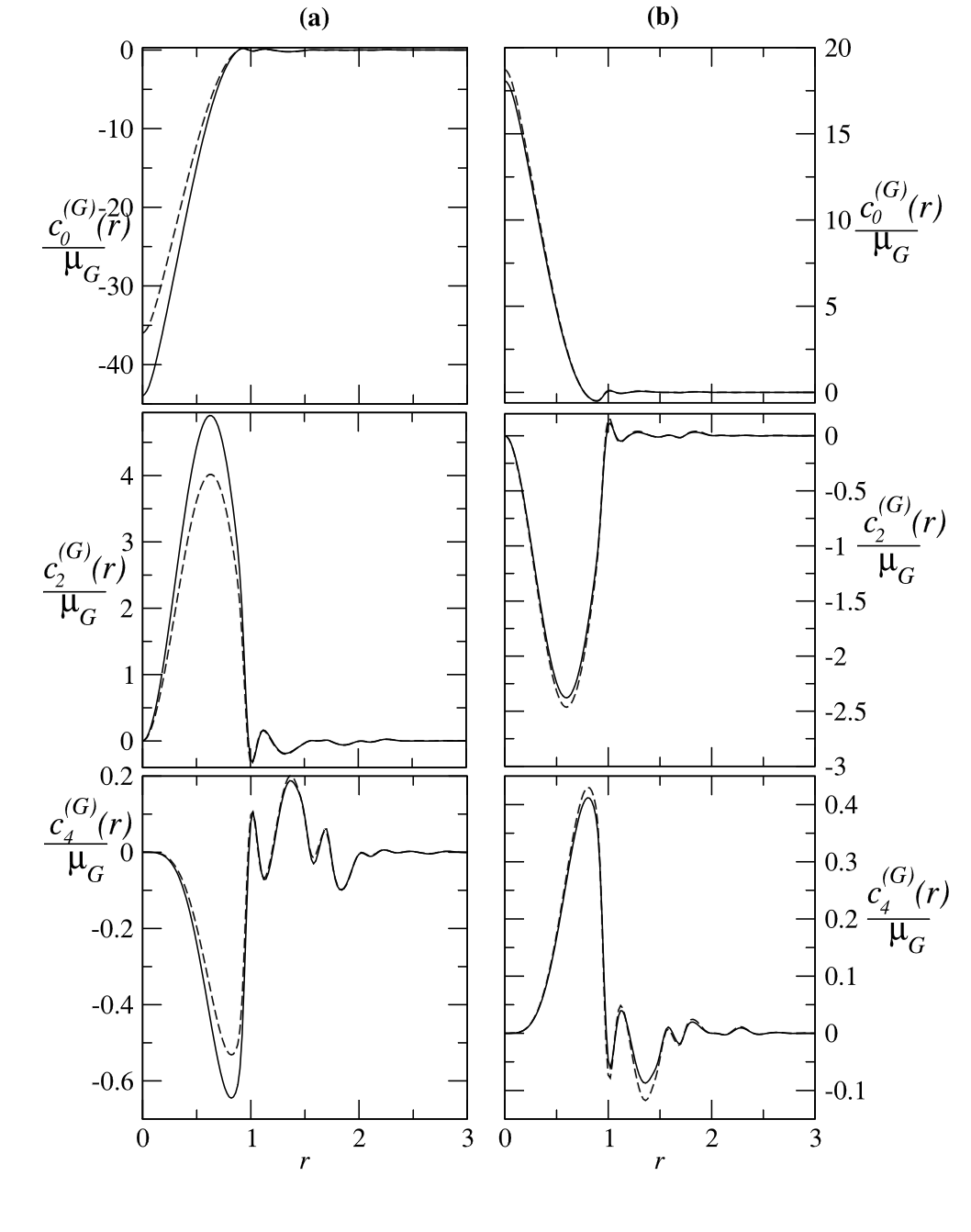}
\caption*{\textbf{Fig. 10.2} Comparison of values of $c_l^{(G)}(r)/\mu_G$ as a function of $r$ (measured in units of $\sigma$) for $l = 0, 2,$ and $4$, for reciprocal lattice vectors (RLVs) of the first (a) and second (b) sets of an fcc lattice at $\rho_s^* = 1.10$ and $T^* = 1.50$; the full and dashed lines correspond to LJ and RLJ potentials, respectively. Reproduced with permission from Ref.\cite{Bharadwaj2015}
}
\end{figure}

The symmetry conserving components $h^{(0)}$ and $c^{(0)}$ of PCFs and their derivatives with respect to $\rho$ were calculated using IET. The closure relation of Zerah and Hansen (ZH) \cite{Zerah1986} (see Eq. (A.14) in Appendix A) was used. In the ZH relation, pair potential $u(r)$ is divided into reference part $u_0(r)$ and perturbation part $u_p(r)$. This decision is done using two schemes; one is called WCAS and the other, ODS\cite{bomont2001}. The details of these schemes are given in the Appendix A (see Eqs. (A.15) and (A.16)).

In Ref \cite{Bharadwaj2015} $c^{(b)}(\vec{r}1, \vec{r}_2)$ was approximated by the first term of the series (8.22). This is justified on the ground that, as already observed in the case of IPPs, the contribution of the second term of the series is negligible unless the potential repulsion varies as $r^{-n}$ with $n \le 6$, and that the contribution made by the attractive part of LJ potential to PCFs at crystal density is small. In Fig 10.2 a comparison of values of $c^{(G)}\ell(r)$ as a function of $r$ (measured in units of $\sigma$) for $\ell = 0, 2$ and 4 for RLV's of first and second sets of a fcc lattice is made for LJ and RLJ potentials. To calculate the quantity $c^{(G)}_\ell(r)$, Eq. (8.25) has been used. The difference between the values corresponding to LJ and RLJ potentials is found to be maximum for the first set of $\vec G$ vectors and  negligible for other sets. The comparison made in Fig 10.2 shows that the attractive interaction contribution to broken symmetry component of correlation function is small compared to that of the repulsive interaction.

In Tables 10.4 and 10.5 we reproduce results reported in Ref.\cite{Bharadwaj2015}. In Table~10.4 values of freezing parameters $\rho_f^*$, $\rho_c^*$, $\Delta \rho^*$, the Lindemann parameter $L$ and $P^* = P\sigma^3/\varepsilon$ found from the EDFT are compared with those found from SODFT and modified weighted density approximation (MWDA)\cite{deKuijper1990} and simulations \cite{Baus1980,deKuijperr1990} for the RLJ potential for few selected temperatures. In Fig.~10.3 fluid--solid phase boundary  $p^*$--$T^*$ is drawn: The dashed line represents values formed from the EDFT \cite{Bharadwaj2015}, full line from SODFT and open circle and triangles from simulations. The spread in simulation values is due to different approaches used in locating the transition and also due to different sizes used in the calculation. It is, however, clear that while  EDFT gives phase boundary and  the  freezing parameters which are in very good agreement with simulation values. The SODFT values are far away from the ``exact'' (simulation) values.
\begin{table}[h]
\centering
\caption*{{Table 10.4:}Comparison of freezing parameters $\rho_l^*, \rho_s^*, \Delta \rho^*$, Lindemann parameter $L$, and pressure $P^* = P\sigma^3/\epsilon$ found from the simulations and with the SODFT and the $MWDA$ for the RLJ potential at several values of $T^*$. Reproduced with the permission from Ref.\cite{Bharadwaj2015}}
\begin{tabular}{c l c c c c c}
\hline\hline
$T^*$ & Simulation/theory group & $\rho_l^*$ & $\rho_s^*$ & $\Delta \rho^*$ & $L$ & $P\sigma^3/\epsilon$ \\
\hline
0.80 &  EDFT \cite{Bharadwaj2015} & 0.930 & 0.988 & 0.062 & 0.092 & 9.68 \\
     &  SODFT \cite{Bharadwaj2015}        & 0.988 & 1.058 & 0.070 & 0.077 & 12.48 \\
     &  Simulation \cite{Baus1980}    & 0.920 & 0.990 & 0.076 &       & 9.60 \\
\hline
1.00 & EDFT \cite{Bharadwaj2015}& 0.957 & 1.016 & 0.061 & 0.093 & 12.52 \\
     & SODFT   \cite{Bharadwaj2015}      & 1.022 & 1.093 & 0.069 & 0.076 & 16.43 \\
     & MWDA theory \cite{deKuijper1990}   & 0.905 & 1.015 & 0.120 & 0.103 & 10.40 \\
     & Simulation \cite{Baus1980}    & 0.950 & 1.016 & 0.069 &       & 12.57 \\
     & Simulation  \cite{deKuijperr1990}   & 0.952 & 1.023 & 0.075 &       & 12.60 \\
\hline
1.50 & EDFT\cite{Bharadwaj2015} & 1.019 & 1.075 & 0.055 & 0.095 & 20.52 \\
     & SODFT  \cite{Bharadwaj2015}       & 1.095 & 1.167 & 0.066 & 0.076 & 27.57 \\
     & Simulation  \cite{deKuijperr1990}    & 1.010 & 1.080 & 0.069 &       & 20.60 \\
\hline
2.74 & EDFT\cite{Bharadwaj2015} & 1.128 & 1.187 & 0.052 & 0.098 & 42.91 \\
     & SODFT \cite{Bharadwaj2015}        & 1.236 & 1.311 & 0.060 & 0.077 & 62.23 \\
     \hline
5.00 & EDFT \cite{Bharadwaj2015} & 1.271 & 1.331 & 0.047 & 0.101 & 92.49 \\
     & SODFT  \cite{Bharadwaj2015}       & 1.416 & 1.494 & 0.055 & 0.077 & 142.8 \\
     & MWDA theory  \cite{deKuijper1990}  & 1.275 & 1.350 & 0.060 & 0.110 & 93.90 \\
     & Simulation  \cite{deKuijperr1990}    & 1.304 & 1.370 & 0.051 &       & 104.5 \\
\hline
10.00& EDFT \cite{Bharadwaj2015} & 1.478 & 1.539 & 0.041 & 0.107 & 227.8 \\
     & SODFT \cite{Bharadwaj2015}        & 1.671 & 1.758 & 0.052 & 0.077 & 369.7 \\
     \hline\hline
\end{tabular}
\end{table}
In Table~10.5 values of freezing parameters for the LJ potential at few temperatures ranging from $0.8$ to $10.0$  found from the EDFT \cite{Bharadwaj2015} are given and compared with values found from other theories. In Fig.~10.4 fluid--solid phase boundaries in the $p^*$--$T^*$ plane found from EDFT and SODFT are compared. In the figure, symbols represent simulation data found from Refs.\cite{Agrawal1995,Ahmed2009,Sousa2012}. Results show that the SODFT is unable to predict correctly the phase boundary of LJ fluid and crystal, while EDFT results are in very good agreement with simulation results.
\begin{figure}
\centering
\includegraphics[width=12.0cm]{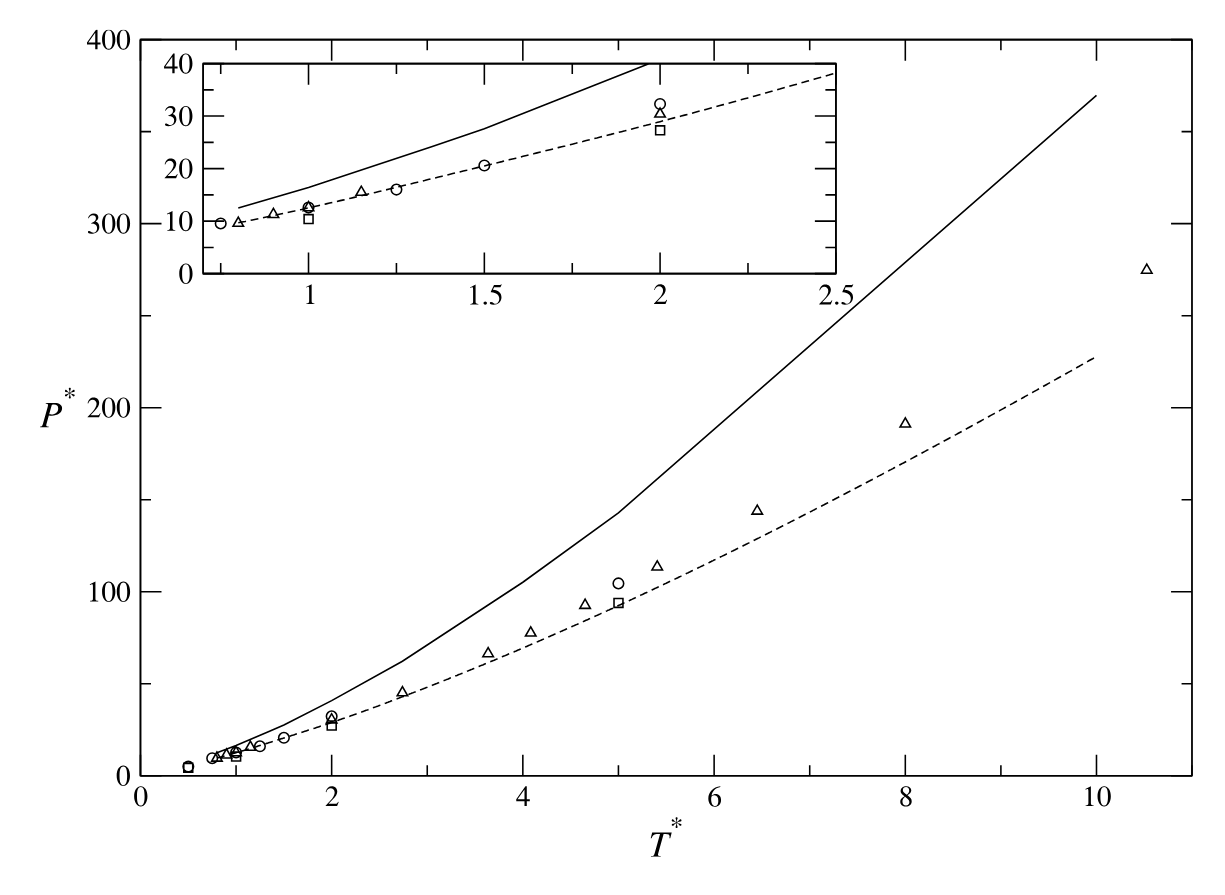}
\caption*{\textbf{Fig. 10.3} Pressure $P^*$ vs temperature $T^*$ for RLJ potential. Dashed line represents data from EDFT, full line represents data found from SODFT. Squares represent values from MWDA, triangles and circles represent simulation data found from Refs. \cite{Agrawal1995,Baus1980,deKuijperr1990} respectively. Reproduced with permission from Ref.\cite{Bharadwaj2015}
}
\end{figure}

\begin{table}
\centering
\caption*{{Table 10.5:}Comparison of freezing parameters $\rho_l^*$, $\rho_s^*$, $\Delta \rho^*$, Lindemann parameter $L$, and pressure $P^* = P\sigma^3/\epsilon$ found from the  simulations$^{26,29}$ and with the SODFT and the MWDA $^6$ for the LJ potential at different values of $T^*$. Reproduced with permission from Ref.\cite{Bharadwaj2015}}
\begin{tabular}{c l l c c c c c}
\hline\hline
$T^*$ & Simulation/theory group &  & $\rho_l^*$ & $\rho_s^*$ & $\Delta \rho^*$ & $L$ & $P\sigma^3/\epsilon$ \\
\hline
0.80 & EDFT \cite{Bharadwaj2015} & ODS  & 0.918 & 0.976 & 0.063 & 0.091 & 2.57 \\
     &                & WCAS & 0.892 & 0.960 & 0.076 & 0.091 & 1.65 \\
     & SODFT \cite{Bharadwaj2015}       & ODS  & 1.009 & 1.073 & 0.064 & 0.074 & 6.39 \\
     &                & WCAS & 0.963 & 1.033 & 0.073 & 0.078 & 3.85 \\
     & Simulation  \cite{Ahmed2009}   &      & 0.891 & 0.983 & 0.103 &       & 1.65 \\
\hline
1.00 & EDFT \cite{Bharadwaj2015}& ODS  & 0.957 & 1.011 & 0.056 & 0.093 & 5.56 \\
     &                & WCAS & 0.929 & 0.994 & 0.069 & 0.092 & 4.18 \\
     & SODFT   \cite{Bharadwaj2015}      & ODS  & 1.053 & 1.115 & 0.059 & 0.075 & 11.14 \\
     &                & WCAS & 1.001 & 1.073 & 0.072 & 0.077 & 7.33 \\
     & MWDA+MF theory  \cite{deKuijper1990} &      & 0.880 & 1.025 & 0.160 & 0.100 & 3.20 \\
     & Simulation \cite{Ahmed2009}    &      & 0.923 & 1.008 & 0.092 &       & 4.05 \\
     & Simulation \cite{Sousa2012}    &      & 0.920 & 1.007 & 0.095 &       & 3.94 \\
\hline
1.50 & EDFT\cite{Bharadwaj2015} & ODS  & 1.028 & 1.082 & 0.052 & 0.095 & 13.70 \\
     &                & WCAS & 1.001 & 1.064 & 0.063 & 0.094 & 11.35 \\
     & SODFT  \cite{Bharadwaj2015}       & ODS  & 1.137 & 1.202 & 0.057 & 0.075 & 24.25 \\
     &                & WCAS & 1.085 & 1.157 & 0.066 & 0.077 & 17.92 \\
     & Simulation \cite{Ahmed2009}    &      & 0.993 & 1.069 & 0.077 &       & 11.20 \\
     & Simulation  \cite{Sousa2012}  &      & 1.002 & 1.076 & 0.074 &       & 11.75 \\
\hline
2.74 & EDFT \cite{Bharadwaj2015}& ODS  & 1.152 & 1.207 & 0.047 & 0.098 & 37.86 \\
     &                & WCAS & 1.126 & 1.188 & 0.055 & 0.098 & 33.25 \\
     & SODFT   \cite{Bharadwaj2015}      & ODS  & 1.288 & 1.356 & 0.053 & 0.075 & 64.31 \\
     &                & WCAS & 1.235 & 1.307 & 0.058 & 0.077 & 51.18 \\
     & Simulation  \cite{Ahmed2009}   &      & 1.144 & 1.211 & 0.059 &       & 36.91 \\
     & Simulation  \cite{Sousa2012}  &      & 1.116 & 1.181 & 0.058 &       & 33.20 \\
     & Simulation  \cite{Hansen1969}   &      & 1.113 & 1.179 & 0.059 &       & 32.20 \\
\hline
5.00 & EDFT\cite{Bharadwaj2015} & ODS  & 1.297 & 1.354 & 0.044 & 0.102 & 89.53 \\
     &                & WCAS & 1.275 & 1.336 & 0.048 & 0.102 & 82.04 \\
     & SODFT   \cite{Bharadwaj2015}      & ODS  & 1.471 & 1.546 & 0.051 & 0.074 & 155.5 \\
     &                & WCAS & 1.420 & 1.495 & 0.053 & 0.077 & 130.6 \\
     & MWDA+MF theory\cite{deKuijper1990} &      & 1.270 & 1.350 & 0.060 & 0.110 & 79.80 \\
     & Simulation \cite{Ahmed2009}    &      & 1.300 & 1.366 & 0.051 &       & 91.20 \\
     & Simulation \cite{Sousa2012}   &      & 1.279 & 1.349 & 0.055 &       & 86.00 \\
\hline
10.0 & EDFT\cite{Bharadwaj2015} & ODS  & 1.506 & 1.562 & 0.037 & 0.107 & 231.3 \\
     &                & WCAS & 1.485 & 1.547 & 0.042 & 0.107 & 215.9 \\
     & SODFT \cite{Bharadwaj2015}        & ODS  & 1.729 & 1.812 & 0.048 & 0.074 & 409.6 \\
     &                & WCAS & 1.668 & 1.756 & 0.053 & 0.077 & 347.3 \\
     & MWDA+MF theory\cite{deKuijper1990} &      & 1.530 & 1.580 & 0.040 & 0.104 & 242.0 \\
     & MWDA theory \cite{deKuijper1990}   &      & 1.520 & 1.570 & 0.030 & 0.114 & 237.0 \\
     & Simulation   \cite{Hansen1969}    &      & 1.500 & 1.572 & 0.048 &       & 231.0 \\
\hline\hline
\end{tabular}
\end{table}

\begin{figure}
\centering
\includegraphics[width=12.0cm]{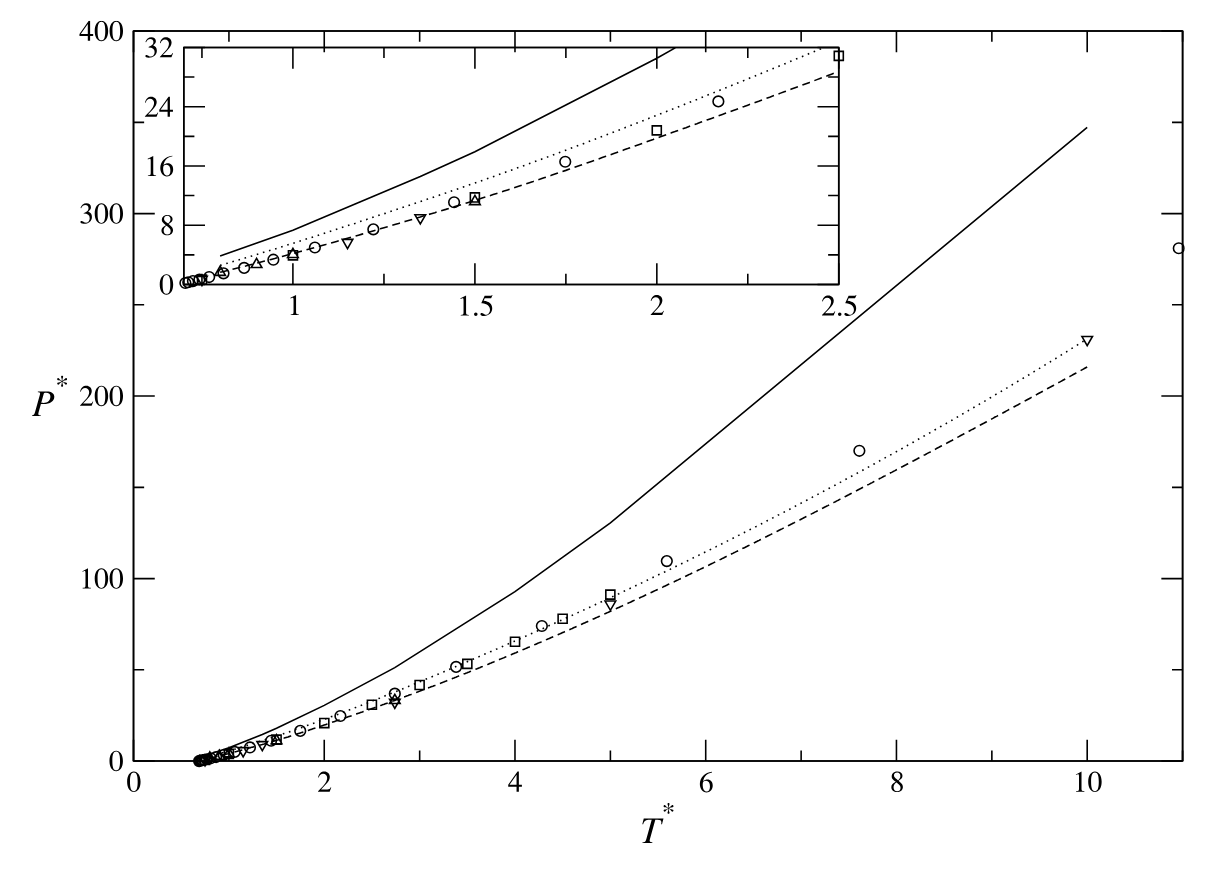}
\caption*{\textbf{Fig. 10.4} Pressure  $P^*$ vs temperature $T^*$ for LJ potential. Dashed and dotted lines represent EDFT data found from WCAS and ODS division of LJ
potential, respectively, and full line represents data found from SODFT.
Symbols represent simulation data found from Refs.\cite{Agrawal1995,Ahmed2009,Sousa2012}; notations are
same as in Fig. 10.3. Reproduced with permission from Ref. \cite{Bharadwaj2015}}
\end{figure}

\newpage
\section{Solid-solid phase transition}
The EDFT was used by Bharadwaj and Singh \cite{Bharadwaj2017} to investigate the solid--solid phase transition. The description given below is based on Ref. \cite{Bharadwaj2017}. At the solid-solid phase boundary, the phase that has lower density will be referred to as low density crystal (LDC) and the one that has higher density as high density crystal (HDC); letters \textit{l} (for LDC) and \textit{h} (for HDC) will be used to indicate quantities of these phases.

Let the single particle density distribution of the LDC be expressed in the Fourier series and in the superposition of normalized Gaussian as
\begin{align}
\rho_{l}(\vec{r}) = \rho_{l} + \sum_{K\neq0} \rho_{K} e^{i\vec{K}\cdot\vec{r}} \quad \text{and} \quad \rho_{l}(\vec{r}) = \left(\frac{\alpha_{l}}{\pi}\right)^{3/2}  \sum_{i} e^{-\alpha_{l}\left(\vec{r}-\vec{R}_{i}^{(l)}\right)^{2}} \quad \text{with} \quad \rho_{K} = \rho_{l} e^{-\frac{K^{2}}{4\alpha_{l}}} ,
\tag{11.1}
\end{align}
and for the HDC as
\begin{align}
\rho_{h}(\vec{r}) = \rho_{h} + \sum_{G\neq0} \rho_{G} e^{i\vec{G}\cdot\vec{r}} \quad \text{and} \quad \rho_{h}(\vec{r}) = \left(\frac{\alpha_{h}}{\pi}\right)^{3/2}  \sum_{i} e^{-\alpha_{h}\left(\vec{r}-\vec{R}_{i}^{(h)}\right)^{2}} \quad \text{with} \quad \rho_{G} = \rho_{h} e^{-\frac{G^{2}}{4\alpha_{h}}} ,
\tag{11.2}
\end{align}
where $ \rho_{h} = \rho_{l}(1+\Delta\rho^{*}) $, $ \rho_{l} $ and $ \rho_{h} $ are average densities of the LDC and the HDC, respectively and $ \Delta \rho^{*} $ is relative change in density at the transition. $ \vec{K} $ and $ \vec{G} $ are RLVs and $ \rho_{K} $ and $ \rho_{G} $ are order parameters. $ \alpha_{l} $ and $ \alpha_{h} $ are localization parameters. The difference in the grand thermodynamic potential of these phases is expressed as
\begin{align}
\Delta W_{lh} = W_{l}-W_{h} = A[\rho_{h}]-A[\rho_{l}] -\beta\mu_{h}\int d\vec{r} \rho_{h}(\vec{r})+\beta\mu_{l}\int d\vec{r} \rho_{l}(\vec{r})
\tag{11.3}
\end{align} 
where $ A[\rho_{h}] $ and $ \mu_{h} $ are reduced free energy and chemical potential of the HDC and $ A[\rho_{l}] $ and $ \mu_{l} $ are those of the LDC.

Since in the present case there is no coexisting fluid to act as a reference system, a homogeneous system of density $\rho_l$ (the same as average density of the LDC), chemical potential $\mu_0$ and its free energy $A(\rho_l)$ is taken as reference system to calculate the symmetry-conserving contribution to the intrinsic free energy of the crystal. The chemical potential $\mu_0$ of this reference system may, in general, be different from $\mu_l$ and $\mu_h$. The free energy of the LDC  $A(\rho_l)$ is expressed as,

\begin{align}
A[\rho_{l}] - A(\rho_{l}) =& \int d\vec{r} \left[\rho_{l}(\vec{r}) \left.ln\frac{\rho_{l}(\vec{r}) }{\rho_{l}}\right. - \left(\rho_{l}(\vec{r})-\rho_{l}\right)\right] + \beta \mu_{0} \int d\vec{r} \left(\rho_{l}(\vec{r})-\rho_{l}\right) \nonumber \\
&-\frac{1}{2}  \int d\vec{r_{1}} \int d\vec{r_{2}} \left(\rho_{l}(\vec{r_{1}})-\rho_{l}\right) \left(\rho_{l}(\vec{r_{2}})-\rho_{l}\right) c^{(0)}\left(r,\rho_{l}\right) \nonumber \\
&-\frac{1}{2}\int d\vec{r_{1}} \int d\vec{r_{2}} \left(\rho_{l}(\vec{r_{1}})-\rho_{l}\right) \left(\rho_{l}(\vec{r_{2}})-\rho_{l}\right) \bar{c}_{l}^{(b)}\left(\vec{r_{1}},\vec{r_{2}}\right),
\tag{11.4}
\end{align}  
where
\begin{align}
\bar{c}_{l}^{(b)}\left(\vec{r_{1}},\vec{r_{2}}\right) = 4 \int_{0}^{1}d\lambda \left(1-\lambda\right) \int_{0}^{1}d\xi \left(1-\xi\right) c_{l}^{(b)}\left(\vec{r_{1}},\vec{r_{2}}; \lambda \rho_{l}, \xi \rho_{K}\right). 
\tag{11.5}
\end{align}
In writing Eq.(11.4) use has been made of the relation 
\begin{align}
ln(\rho_{l}\Lambda^3) - c^{(1)}(\rho_{l}) = \beta \mu_{0},
\tag{11.6}
\end{align}
where $ c^{(1)}(\rho_{l}) $ is one particle direct correlation function of the reference system and $ \Lambda $ is the thermal wavelength associated with a particle. A similar expression can also be written for $ A[\rho_{h}] $.

Substituting expressions of $ A[\rho_{l}] $  and $ A[\rho_{h}] $ in Eq.~(11.3) we get 
\begin{align}
\Delta W_{lh} =& \int d\vec{r} \left[\rho_{h}(\vec{r}) \left.ln\frac{\rho_{h}(\vec{r}) }{\rho_{l}}\right. - \rho_{l}(\vec{r}) \left.ln\frac{\rho_{l}(\vec{r}) }{\rho_{l}}\right. - \left(\rho_{h}(\vec{r})-\rho_{l}(\vec{r})\right)\right] - \beta\left(\mu - \mu_{0}\right)\int d\vec{r} \left(\rho_{h}(\vec{r})-\rho_{l}(\vec{r})\right) \nonumber \\
&-\frac{1}{2}  \int d\vec{r_{1}} \int d\vec{r_{2}} \left[ \left(\rho_{h}(\vec{r_{1}})-\rho_{l}\right) \left(\rho_{h}(\vec{r_{2}})-\rho_{l}\right)-\left(\rho_{l}(\vec{r_{1}})-\rho_{l}\right) \left(\rho_{l}(\vec{r_{2}})-\rho_{l}\right)\right] c^{(0)}\left(\mid\vec{r}_{2}-\vec{r}_{1}\mid,\rho_{l}\right) \nonumber \\
&- \frac{1}{2}  \int d\vec{r_{1}} \int d\vec{r_{2}} \left(\rho_{h}(\vec{r_{1}})-\rho_{h}\right) \left(\rho_{h}(\vec{r_{2}})-\rho_{h}\right) \bar{c}_{h}^{(b)}\left(\vec{r_{1}},\vec{r_{2}}\right) \nonumber \\
&+\frac{1}{2}  \int d\vec{r_{1}} \int d\vec{r_{2}} \left(\rho_{l}(\vec{r_{1}})-\rho_{l}\right) \left(\rho_{l}(\vec{r_{2}})-\rho_{l}\right) \bar{c}_{l}^{(b)}\left(\vec{r_{1}},\vec{r_{2}}\right),
\tag{11.7}
\end{align}  
where $ \bar{c}_{l}^{(b)}\left(\vec{r_{1}},\vec{r_{2}}\right) $ is defined by Eq. (11.5) and
\begin{align}
\bar{c}_{h}^{(b)}\left(\vec{r_{1}},\vec{r_{2}}\right) = 4 \int_{0}^{1}d\lambda \left(1-\lambda\right) \int_{0}^{1}d\xi \left(1-\xi\right) c_{h}^{(b)}\left(\vec{r_{1}},\vec{r_{2}}; \lambda \rho_{h}, \xi \rho_{G}\right) .
\tag{11.8}
\end{align}

In writing Eq.~(11.7) we assumed $ \mu_{h} = \mu_{l} = \mu$ which happens only at the phase boundary. The quantity $ \beta(\mu-\mu_{0}) $ that appears in Eq.~(11.7) is found from following relation ~\cite{Hansen2006},
\begin{align}
\beta\left(\mu-\mu_{0}\right) = \Delta f(\rho_{l},\alpha_{l}) + \rho_{l}\left.\dfrac{\partial \Delta f(\rho_{l},\alpha_{l})}{\partial \rho_{l}}\right\vert_{T,V} ,
\tag{11.9}
 \end{align}
where
\begin{align}
\Delta f(\rho_{l},\alpha_{l}) = Minimum \left[\frac{1}{N_{l}}\left(A[\rho_{l}]-A(\rho_{l})\right)\right]
\tag{11.10}
\end{align}

\begin{figure}
\centering
\includegraphics[width=10.0cm]{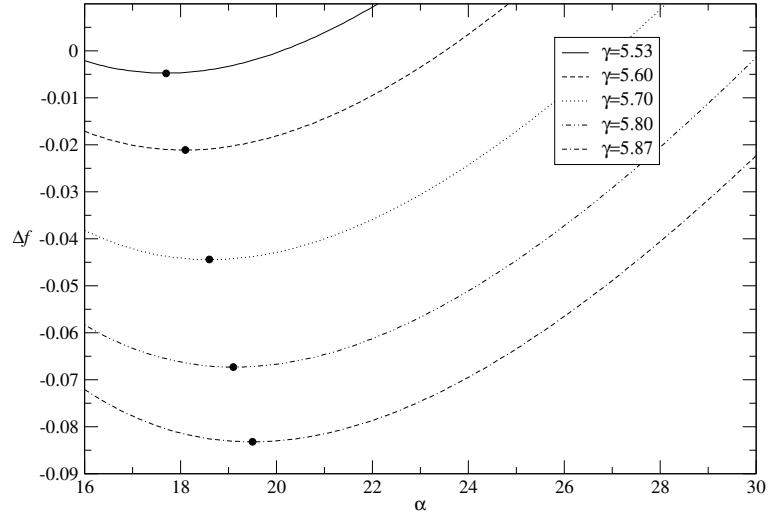}
\caption*{\textbf{Fig. 11.1} Variation of $\Delta f$ as a function of $\alpha_l$ for different $\gamma_l$. The minimum found at a value of $\alpha_l$ for each $\gamma_l$ (shown in figure by full circle) is $\Delta f(\gamma_l,\alpha_l)$ [see Eq. (11.10)].  Reproduced with permission from Ref.\cite{Bharadwaj2017}
}
\end{figure}

and $ N_{l} = V\rho_{l} $, $ V $ being the volume of the system. The minimum in Eq.~(11.10) is found for a given $ \rho_{l} $ by minimizing $ \frac{1}{N_{l}}(A[\rho_{l}]-A(\rho_{l})) $ by varying $ \alpha_{l} $. The minimum found at a particular value of $ \alpha_{l} $ for given $ \rho_{l} $ is $ \Delta f(\rho_{l},\alpha_{l}) $ (see Fig.11.1).

In the preceding section fluid--solid phase transition is described and the results for transition parameters for systems interacting via IPPs are given. It was shown that for the softness parameter $n<7.0 $ the systems freeze into bcc and on increasing the density, transition from bcc to fcc takes place. In this section we describe calculation and results found for solid (bcc) to solid (fcc) phase transition.

One first uses Eq.(11.4) to calculate $ \beta(\mu - \mu_{0}) $ and $ \alpha_{l} $. When the Gaussian and the Fourier forms of $\rho(\vec r)$ are substituted Eq.~(11.4) reduces to,
\begin{align}
\Delta f = \frac{A[\rho_{l}]-A(\rho_{l})}{N_{l}} =& \frac{3}{2} \left(ln\left(\frac{\alpha_{l}}{\pi}\right)-1\right)-ln\rho_{l} - \frac{1}{2}\rho_{l}\sum_{K} \mid\mu_{K}\mid^{2}\hat{c}^{(0)}\left(\mid\vec{K}\mid\right) \nonumber \\
 &-\rho_{l}^{2}\sum_{K}\sum_{K_{1}} \mu_{K_{1}} \mu_{\mid-\vec{K}-\vec{K_{1}}\mid} \hat{\bar{c}}_{l}^{(K)}\left(\vec{K_{1}} + \frac{1}{2}\vec{K}\right).
\tag{11.11}
\end{align}
Note that in this case $ \Delta \rho^{*} = 0 $. In Fig.11.1 we show for a few values of $ \gamma_{l} $ the variation of $ \Delta f $ as a function of $ \alpha_{l} $. The value of $ \Delta f(\gamma_{l}, \alpha_{l}) $ defined by Eq(11.10) corresponds to minimum value of $ \Delta f $ (shown by full circle in Fig.11.1) for each $ \gamma_{l} $. The value of $ \beta(\mu - \mu_{0}) $ is determined from Eq.(11.9). This provides us values of $ \beta(\mu - \mu_{0}) $ and $ \alpha_{l} $ for each $ \gamma_{l} $ which are used in Eq.(11.7). 

The bcc-fcc phase boundary is found from Eq.(11.7) which is rewritten as 
\begin{align*}
\frac{\Delta W_{l-h}}{N_{h}} =& \frac{3}{2}\left[(ln\left(\frac{\alpha_{h}}{\pi}\right) - \frac{1}{(1+\Delta \rho^{*})} ln\left(\frac{\alpha_{l}}{\pi}\right)\right] - \frac{\Delta \rho^{*}}{(1+\Delta \rho^{*})}\left[ln\frac{\rho_{h}}{(1+\Delta \rho^{*})} + \Delta \mu + \frac{5}{2}\right] \nonumber \\
&-\frac{1}{2}\left[\frac{{\Delta \rho^{*}}^{2} \rho_{h}}{(1+\Delta \rho^{*})^{2}}\hat{c}^{(0)}(0) + \rho_{h}\sum_{G} \mid\mu_{G}\mid^{2}\hat{c}^{(0)}\left(\mid\vec{G}\mid\right) - \frac{\rho_{l}}{(1+\Delta \rho^{*})}\sum_{K} \mid\mu_{K}\mid^{2}\hat{c}^{(0)}\left(\mid\vec{K}\mid\right)\right] \nonumber \\
&-\rho_{h}^{2}\sum_{G}\sum_{G_{1}}  \mu_{G_{1}} \mu_{\mid-\vec{G}-\vec{G_{1}}\mid} \hat{\bar{c}}_{h}^{(G)}\left(\vec{G_{1}} + \frac{1}{2}\vec{G}\right) \\
&+\frac{\rho_{l}^{2}}{(1+\Delta \rho^{*})}\sum_{K}\sum_{K_{1}} \mu_{K_{1}} \mu_{\mid-\vec{K}-\vec{K_{1}}\mid} \hat{\bar{c}}_{l}^{(K)}\left(\vec{K_{1}} + \frac{1}{2}\vec{K}\right)
\tag{11.12}
\end{align*}
where $ \Delta\mu = \beta\left(\mu-\mu_{0}\right) $ and $ N_{l} = N_{h}/(1+\Delta \rho^{*}) $. $ N_{l} $ and $ N_{h} $ are number of particles in volume $ V $, respectively, in the LDC and the HDC.

\begin{figure}
\centering
\includegraphics[width=10.0cm]{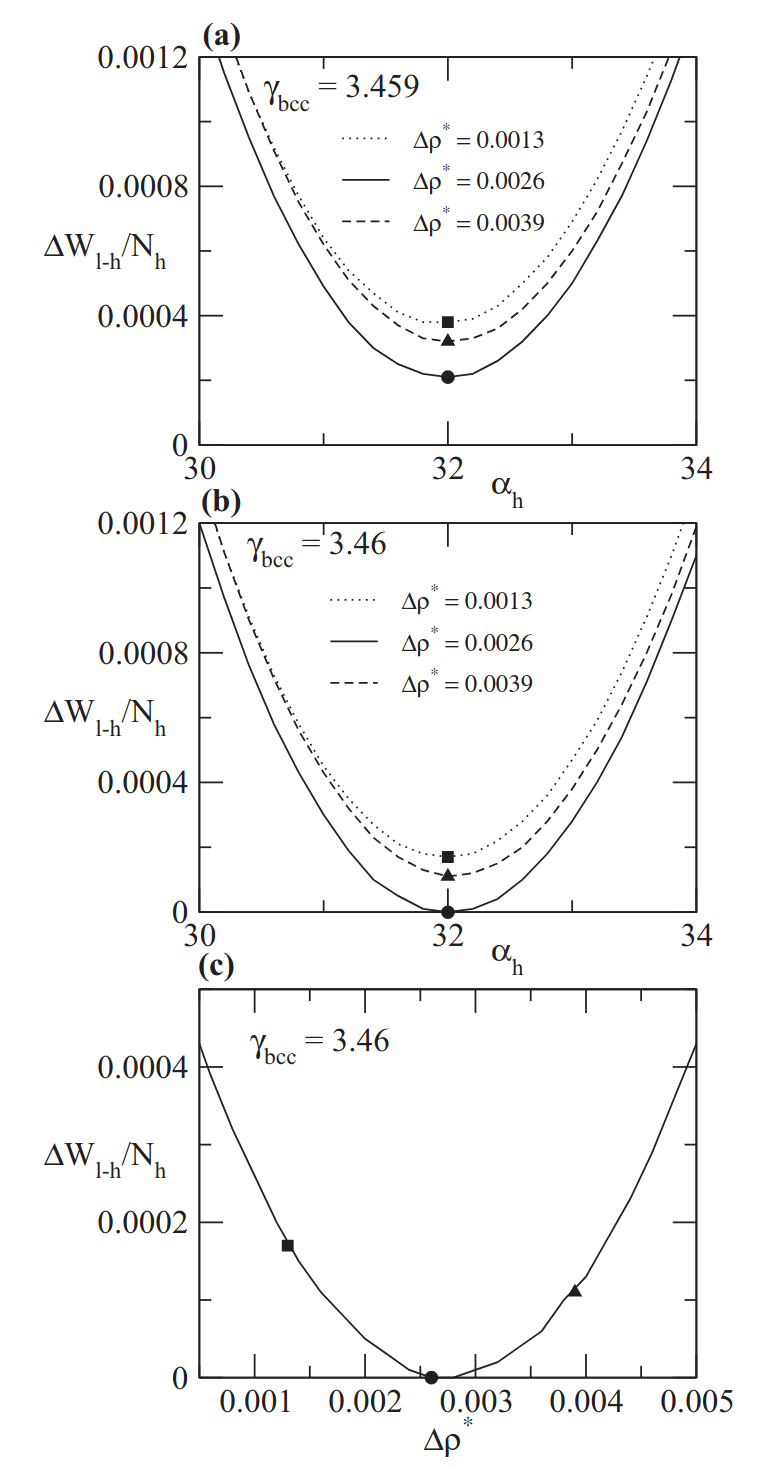}
\caption*{\textbf{Fig. 11.2} Variation of $\dfrac{\Delta W_{l-h}}{N_h}$ as a function of
$\alpha_h$ and $\Delta \rho^\ast$ for $1/n = 0.20$ is shown.
In (a) the variation is shown for
$\gamma_{bcc}=3.459$
for three values of $\Delta \rho^\ast$.
The lowest minimum occurs at
$\Delta \rho^\ast = 0.0026$.
In (b) the variation is for
$\gamma_{bcc}=3.46$,
where the condition
$\dfrac{\Delta W_{l-h}}{N_h}=0$
is satisfied for
$\Delta \rho^\ast = 0.0026$
and $\alpha_h = 32.0$.
In (c) the variation of
$\dfrac{\Delta W_{l-h}}{N_h}$
for $\gamma_{bcc}=3.46$
as a function of $\Delta \rho^\ast$ is shown. Reproduced with permission from Ref.\cite{Bharadwaj2017}
}
\end{figure}

The function  $ \frac{\Delta W_{l-h}}{N_{h}} $ is minimized with respect to $ \gamma_{l} $, $ \Delta \rho^{*} $ and $ \alpha_{h} $ as value of $ \alpha_{l} $ is already known. The minimization procedure is same as described above. In Fig.11.2 we show the variation of $ \frac{\Delta W_{l-h}}{N_{h}} $ as a function of $ \alpha_{h} $ for different $ \Delta \rho^{*} $ for $ 1/n = 0.2 $. At the transition point $ \frac{\Delta W_{l-h}}{N_{h}} = 0 $(see Fig.11.2(c)).

\begin{figure}
\centering
\includegraphics[width=06.0cm,angle=270]{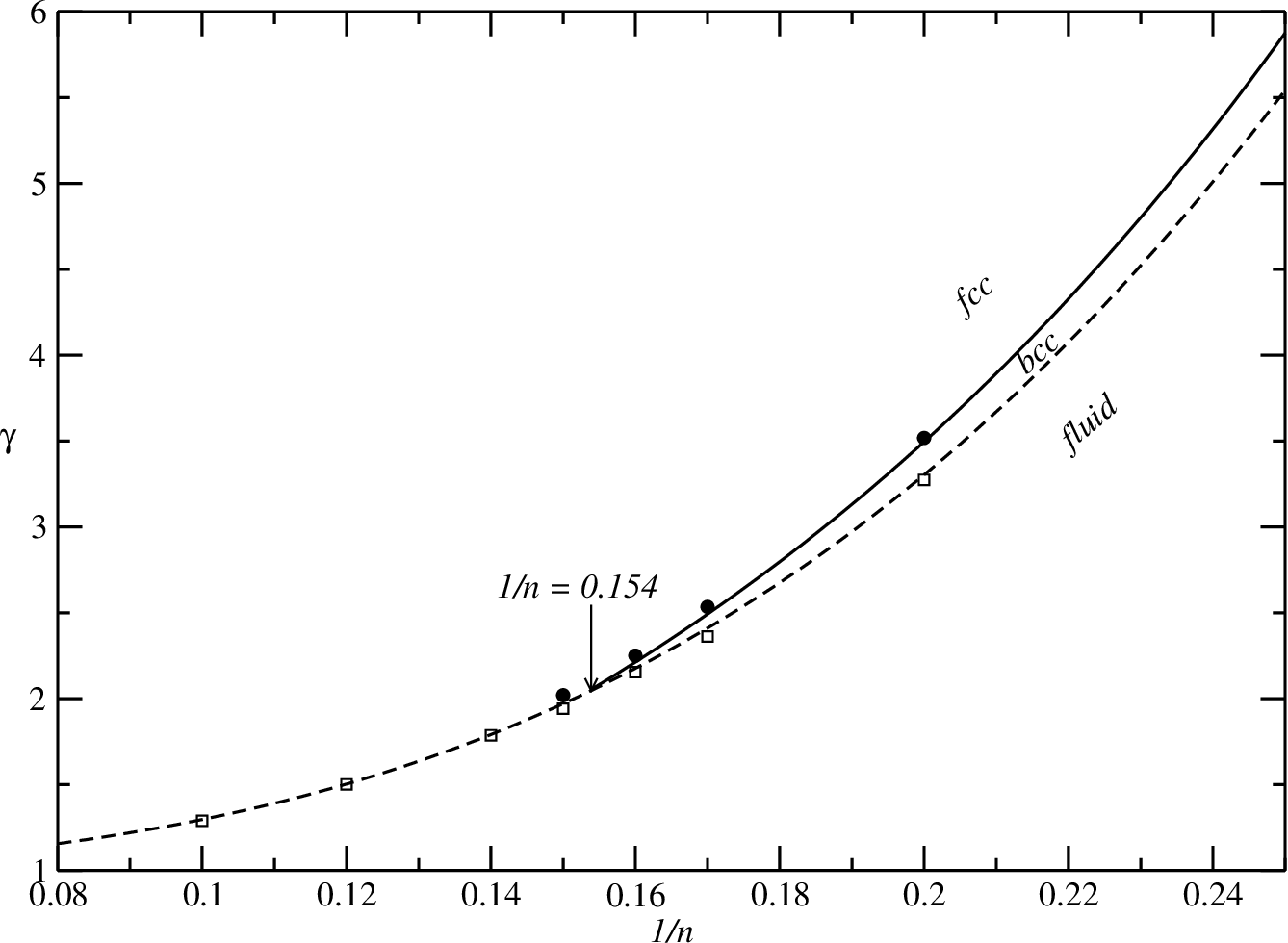}
\caption*{\textbf{Fig. 11.3} Phase diagram of systems interacting via IPPs.
The dashed line represents the theoretical values for the fluid--solid
phase boundary and the full line represents those for the bcc--fcc
phase boundary.
Open squares and full circles represent, respectively,
simulation values \cite{Bharadwaj2015} for the fluid--solid (fcc or bcc)
and the bcc--fcc transition points. Reproduced with permission from Ref.\cite{Bharadwaj2017}
}
\end{figure}

The phase diagram in a plane of softness parameter $ 1/n $ and the scaled quantity $ \gamma $ is plotted in Fig.11.3 and compared with simulation results \cite{Prestipino2005}. From the figure we see the agreement between the two values is good for both the fluid-solid and the solid-solid transition densities. In Fig.11.4 we plot for $ 1/n = 0.2 $ the fluid-bcc and bcc-fcc phase boundaries in $ \rho^{*} - T^{*}$ plane. For this we used the relation $ \gamma = \rho^{*}{T^{*} }^{-3/5} $. It is seen from the figure that the stability region of the bcc phase is somewhat underestimated by the theory. The density of fluid at the fluid-solid transition is marginally higher and the density of the bcc phase at the bcc-fcc transition is lower than simulation values. However, as the estimated error in the phase transition densities due to limited statistics in simulation is quoted as $ 3\times10^{-2} $\cite{Prestipino2005}, the difference in the theoretical and simulation values of densities falls within the error bounds. It may also be noted that the change in density at the bcc-fcc transition is considerably smaller than at the fluid-solid transition. This suggests that the bcc-fcc transition is weak first-order transition compared to the fluid-solid transition. In case of weak first-order transition the finite size effects need careful examination for the sample size as large as considered in Ref.\cite{Prestipino2005}. 

\begin{figure}
\centering
\includegraphics[width=06.0cm,angle=270]{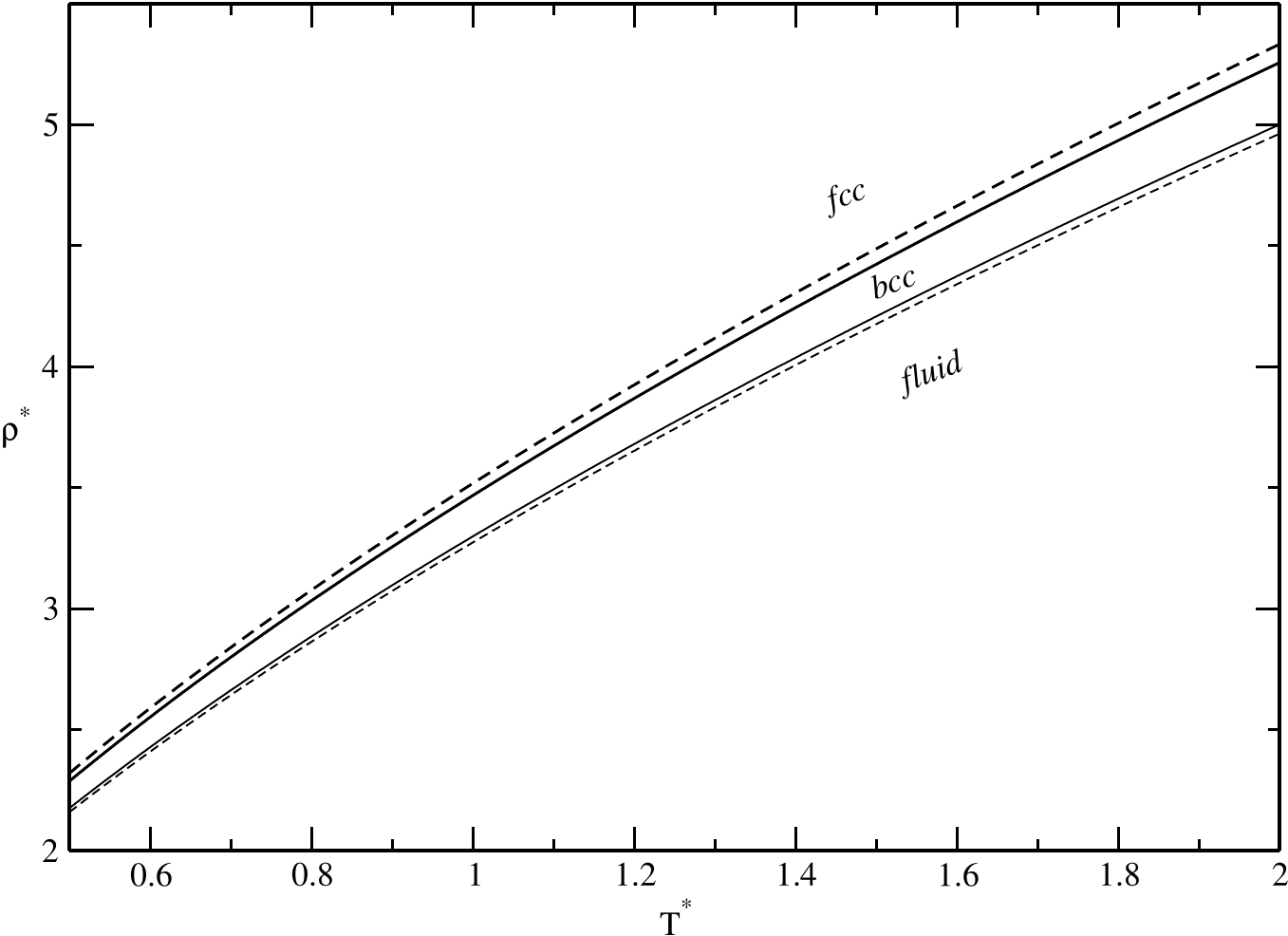}
\caption*{\textbf{Fig. 11.4} The density--temperature phase diagram of the IPP for
$n=5$ (or $1/n=0.20$).
The dashed lines represent simulation values \cite{Laird1990}
and the full lines theoretical values.  Reproduced with permission from Ref.\cite{Bharadwaj2017}
}
\end{figure}

\newpage
\section{Summary and Outlook}
Results reported in Sections 7, 10 and 11 show that the EDFT formulated in Section 6 predicts accurately boundaries separating phases of different symmetries and values of parameters associated with phase transitions. The expression for free energy functional is found directly by double functional integrations of a relation that connects the second functional derivative of $A[\rho]$ with respect to $\rho(\mathbf{x})$. The path of integration in the density space covers regions of all phases that appear between isotropic and the phase of interest and is characterized by parameters $\lambda$ and $\xi$ which, when varied from $0$ to $1$, raise respectively, density and order parameters from zero to their final values. This approach differs in a fundamental way from those which involve functional Taylor expansion or use weighted density approximations.

When a phase of higher symmetry freezes into a phase of lower symmetry, a qualitatively new contribution to correlation functions arises at the freezing point. As a consequence, correlation functions of the emergent phase have two distinct components: one that preserves the symmetry of the higher symmetry phase and the one that breaks it. The broken symmetry components contain the structural specificity of the emergent phase and play a fundamental role in its stability. Methods for calculating PCFs of ordered phases with broken rotational and translational symmetries are described in Sections 7--9.

The density functional approach provides an efficient framework for the theoretical study of a large variety of problems of soft and hard condensed matter ~\cite{Singh1991,Hansen2006,DeLasHeras2011,Lutsko2018,TeVrugt2020,Schmidt2022,Yadav2022,Yadav2024}. However, in most of the cases, particularly in those where breaking of symmetry and phase transitions are involved, the approximate versions of DFT have largely failed. Accurate description of all these problems can now be found using the EDFT.

Molecular systems, depending upon the shape of constituent particles and anisotropy in intermolecular interactions, exhibit a number of broken symmetry phases. Associated with each broken symmetry are distortions (elastic moduli), topological defects, and low-frequency dynamical modes that provide a path to restore the symmetry of the original phase of maximum symmetry. The properties of each broken symmetry phase are controlled by these distortions, defects, and modes. The EDFT provides a theoretical tool at the first-principle level to investigate these properties. The applicability of the theory can further be extended by making the order parameters in the expression of free-energy density position dependent so that  fluctuations can appropriately be addressed.

\addcontentsline{toc}{section}{Appendices}
\section*{\underline{Appendix A}}

\section*{Integral equation theory for pair correlation functions of a homogeneous system}

One of the most commonly used approaches to find the structure and thermodynamics of the fluid phase is the integral equation theory (IET). The theory is about solving the Ornstein--Zernike (OZ) equation (Eq.~2.40) of the homogeneous isotropic system with the help of a relation known as the closure relation. Such a relation can be found using the formalism described in Section~3 and an idea suggested by Percus \cite{percus1962} . 

The Percus idea is that the quantity $\rho\, g({r}, \vec\Omega, \vec\Omega_0)$ can be treated as the one-particle density $\rho({r}, \vec\Omega)$ at position ${r}$ and orientation $\vec\Omega$, when a molecule of the system is frozen at the origin with orientation $\vec\Omega_0$. The frozen molecule creates a potential field equal to the pair potential 
$u({r}, \vec\Omega, \vec\Omega_0)$, that acts on other molecules of the system as an "external" potential $u^{(e)}({r}, \vec\Omega)$ . Thus,

\begin{equation}
u^{(e)}(\vec{x}) = u(\vec{x_0},\vec{x}),
\tag{A.1}
\end{equation}

where $\vec{x}_0$ denotes the coordinates of the molecule fixed at the origin and $\vec{x}$ is the coordinates of any other molecule of the system.

Substituting the relation (A.1) into Eq.~(2.30b), we get

\begin{equation}
\rho(\vec{x}) = \frac{1}{\Lambda^3}
\exp\left[-\beta \mu - \beta u(\vec{x_0}, \vec{x}) + c^{(1)}(\vec{x})\right],
\tag{A.2}
\end{equation}

For the quantity $c^{(1)}(\vec{x})$, which is a functional of $\rho(\vec{x})$, we use Eq.~(3.2) to write

\begin{equation}
\begin{aligned}
c^{(1)}(\vec{x}) =\;& c^{(1)}(\rho_0) 
+ \int d\vec{x}_2\, \, \Delta \rho(\vec{x}_2)\,
c^{(0)}(\vec{x}_1, \vec{x}_2;\rho_0) \\
&+ \int d\vec{x}_2\,\Delta \rho(\vec{x}_2,)
\left[
\int_0^1 d\lambda \,
c^{(0)}(\vec{x}_1,\vec{x}_2;\lambda)
- c^{(0)}(\vec{x}_1, \vec{x}_2; \rho_0)
\right],
\end{aligned}
\tag{A.3}
\end{equation}


where $\rho_0$ is the density of the fluid and $\Delta \rho(\vec{x}) = \rho(\vec{x}) - \rho_0$.
Substitution of Eq.~(A.3) into Eq.~(A.2) leads to

\begin{equation}
\rho(\vec{x}) = \rho_0 \exp \Bigg[
-\beta u(\vec{x}_0, \vec{x}) 
+ \int d\vec{x}_2 \, \Delta \rho(\vec{x}_2)\, c^{(0)}(\vec{x}, \vec{x}_2)
+ B(\vec{x}, \vec{x}_0)
\Bigg],
\tag{A.4}
\end{equation}

where the quantity

\begin{equation}
B(\vec{x}, \vec{x}_0) = \int d\vec{x}_2 \, \Delta \rho(\vec{x}_2)
\left[
\int_0^1 d\lambda \, c^{(0)}(\vec{x}, \vec{x}_2;\lambda)
- c^{(0)}(\vec{x}, \vec{x}_2; \rho_0)
\right],
\tag{A.5}
\end{equation}

can be recognized as the \textit{bridge function} ~\cite{Hansen2006} and

\begin{equation*}
\rho_0 = \frac{1}{\Lambda^3} \exp \left[ -\beta \mu + c^{(1)}(\rho_0) \right].
\end{equation*}

Using relations

\begin{equation}
\rho(\vec{x}) = \rho_0 \, g(\vec{x}_0, \vec{x}) \hspace{1.0cm} and 
\qquad 
\Delta \rho(\vec{x}) = \rho_0 \, h^{(0)}(\vec{x}_0, \vec{x}),
\tag{A.6}
\end{equation}

we get following expression for pair function:

\begin{equation}
g(\vec{x}_0, \vec{x}) = \exp \Bigg[
-\beta u(\vec{x}_0, \vec{x}) 
+ \rho_0 \int d\vec{x}_2 \, h^{(0)}(\vec{x}_0, \vec{x}_2)\, c^{(0)}(\vec{x}, \vec{x}_2)
+ B(\vec{x}_0, \vec{x})
\Bigg].
\tag{A.7}
\end{equation}

Though Eq.~(A.7) is exact, it involves an unknown function $B(\vec{x}_0, \vec{x})$, defined by Eq.~(A.5).

If one takes $B = 0$ and uses the Ornstein--Zernike relation (Eq.~2.40), one gets

\begin{equation}
\ln \left[ 1 + h^{(0)}(\vec{x}_0, \vec{x}) \right]
= -\beta u(\vec{x}_0, \vec{x}) 
+ h^{(0)}(\vec{x}_0, \vec{x}) 
- c^{(0)}(\vec{x}_0, \vec{x}),
\tag{A.8}
\end{equation}

which is the well-known \textit{hypernetted chain (HNC)} relation.

Rearranging terms in Eq~(A.8) gives

\begin{equation}
c^{(0)}(\vec{x}_0, \vec{x}) 
= -\beta u(\vec{x}_0, \vec{x}) 
+ h^{(0)}(\vec{x}_0, \vec{x}) 
- \ln \left[ 1 + h^{(0)}(\vec{x}_0, \vec{x}) \right],
\tag{A.9}
\end{equation}

which is the expression of the DPCF in the HNC theory.

In order to get the Percus--Yevick (PY) relation, we linearize the second term in Eq.~(A.7) and take $B = 0$. Thus,

\begin{align*}
g(\vec{x}_0, \vec{x}) 
&= e^{-\beta u(\vec{x}_0, \vec{x})}
\left[
1 + \rho_0 \int d\vec{x}_2 \, h^{(0)}(\vec{x}_0, \vec{x}_2)\, c^{(0)}(\vec{x}, \vec{x}_2)
\right] \\
&= e^{-\beta u(\vec{x}_0, \vec{x})}
\left[
1 + h^{(0)}(\vec{x}_0, \vec{x}) - c^{(0)}(\vec{x}_0, \vec{x})
\right],
\end{align*}

or

\begin{equation}
c^{(0)}(\vec{x}_0, \vec{x}) 
= g(\vec{x}_0, \vec{x}) \left[ 1 - e^{\beta u(\vec{x}_0, \vec{x})} \right].
\tag{A.10}
\end{equation}

Rearrangement of Eq.~(A.10) gives the following relation

\begin{equation}
c^{(0)}(\vec{x}_0, \vec{x}) 
= \left(e^{-\beta u(\vec{x}_0, \vec{x})}-1\right)
\left[ 1 + h^{(0)}(\vec{x}_0, \vec{x}) - c^{(0)}(\vec{x}_0, \vec{x}) \right].
\tag{A.11}
\end{equation}

which is more common in the literature than othrer form of the relation.

\vspace{0.5cm}

The PY and HNC relations have extensively been used in the IET to calculate the PCFs of simple ~\cite{Hansen2006} as well as molecular liquids ~\cite{GrayGubbins1984}. They are found to give accurate results at low densities, but fail at higher liquid densities. Values of thermodynamic quantities calculated following the virial and compressibility routes do not match; rendering both theories thermodynamically inconsistent. Over the last many years, several attempts have been made to improve upon these closure relations and to extend the applicability of the IET to a wider range of thermodynamic  state. One way to improve the HNC approximation is to include successively higher order terms in the series expansions of $B(\vec{x}_0, \vec{x})$.

To the first order in the functional Taylor expansion and using relation of Eq(A.6) one gets from Eq~(A.5),

\begin{equation}
B(\vec{x},\vec{x_0}) 
= \rho_0^2 \int d\vec{x}_2 \, d\vec{x}_3 \;
c^{(0)}(\vec{x}_1,\vec{x}_2,\vec{x}_3;\rho_0)\,
h^{(0)}(\vec{x}_0,\vec{x}_2)\,
h^{(0)}(\vec{x}_0,\vec{x}_3),
\tag{A.12}
\end{equation}

\vspace{0.4cm}

Though inclusion of Eq.~(A.12) in Eq.~(A.8) will improve the result over the HNC approximation, however the computational cost involved makes it impractical ~\cite{Hansen2006}. In another approach one may use values of $h$, $c$ and $g$ determined from computer simulations for a model potential in Eq.~(A.7) to determine values of $B(\vec{x}_0,\vec{x})$ and make it as a basis for other systems \cite{Verlet1980}. However, bridge function does not seem to have universal character, and such efforts can, at most, have limited success. In view of this, efforts have been limited to finding 
closure relations which give accurate results for a set of potential models rather than 
being universal (see for details Ref.\cite{Verlet1980,Solana2013,Choudhury2002}).

\vspace{0.4cm}

A closure relation proposed by Rogers and Young (RY) \cite{Rogers1984} is found to give accurate values of 
correlation functions of fluids interacting via short-range repulsive potentials. The RY closure relation interpolates continuously between the PY closure at short interparticle distances and the HNC closure at larger distances by imposing the condition of thermodynamic consistency. The RY relation is written as

\begin{equation}
h(r) = e^{-\beta u(r)} 
\left[
1 + \frac{\exp\{\gamma(r) f(r)\} - 1}{f(r)}
\right] - 1,
\tag{A.13}
\end{equation}
where $\gamma(r) = h^{(0)}(r) - c^{(0)}(r)$ and $f(r) = 1 - \exp(-\psi r)$ is a switching 
function which introduces an adjustable parameter $0 \le \psi \le \infty$. The value of $\psi$ is chosen to guarantee thermodynamic consistency between the virial and compressibility routes to the equation of state.

For fluids interacting via potentials which have attractive tail, the RY closure is often unable to give accurate results. For Lennard-Jones (LJ) fluid a closure relation proposed by Zerah and Hansen (ZH) \cite{Zerah1986} is    preferred. The ZH closure is expressed as

\begin{equation}
1 + h^{(0)}(r) 
= \exp[-\beta u_0(r)]
\left[
1 + \frac{e^{f(r)(\gamma^{(0)}(r)-\beta u_p(r))} - 1}{f(r)}
\right],
\tag{A.14}
\end{equation}

where $\gamma^{(0)}(r) = h^{(0)}(r) - c^{(0)}(r)$, $f(r)$ is a mixing parameter, $u_0(r)$ and $u_p(r)$ are suitably chosen short-range part and long-range part of pair potential $u(r)$.  The function $f(r) = 1 - \exp(-\psi r)$ includes an adjustable parameter $\psi$ whose value is chosen to satisfy thermodynamic self-consistency.

The following two schemes for division of the LJ potential $u(r)$ into $u_0(r)$ and $u_p(r)$ have been used \cite{Bharadwaj2015}. In the Weeks-Chandler-Andersen (WCA) scheme (WCAS), $u_0(r)$ is the RLJ potential of Eq. (10.29) and

\begin{equation}
u_p(r) =
\begin{cases}
-\epsilon, & \text{for } r < r_m \; (= 2^{1/6}\sigma) \\
u(r), & \text{for } r > r_m
\end{cases}
\tag{A.15}
\end{equation}

In another scheme referred to as optimized division scheme (ODS) \cite{bomont2001}, $u_p(r)$ is written as

\begin{equation}
u_p(r) =
\begin{cases}
-p\epsilon, & \text{for } r < r_1 \\
a_1 + a_2 r + a_3 r^2 + a_4 r^3, & \text{for } r_1 < r < r_2 \\
u(r), & \text{for } r > r_2
\end{cases}
\tag{A.16}
\end{equation}

and $u_0(r) = u(r) - u_p(r)$, values of parameters $a_i$ are given in Ref.\cite{bomont2001}. For the RLJ potential of Eq. (A.14). $u_p(r)$ is zero and the ZH closure reduces to the RY closure \cite{Rogers1984}. Both schemes, WCAS and ODS were used in the ZH closure relation to calculate values of correlation functions $h^{(0)}(r)$, $c^{(0)}(r)$ and their derivatives with respect to density \cite{Bharadwaj2015}. Both schemes give almost identical results.
\section*{\underline{Appendix B}}
\section*{Spherical harmonic expansion of pair correlation functions of isotropic phase}

It is often convenient to expand angular correlation functions in spherical harmonics ~\cite{GrayGubbins1984}. For a homogeneous isotropic fluid, a pair function $A(\mathbf{r_1}, \mathbf\Omega_1,  \mathbf{r_2},\mathbf\Omega_2)$ can be expanded in the space-fixed (SF) frame as

\begin{align}
A(\mathbf{r}, \mathbf{\Omega}_1, \mathbf{\Omega}_2)
&= \sum_{l_1,l_2,l,m} \sum_{m_1,m_2} \sum_{n_1,n_2}
A(l_1,l_2,l;n_1,n_2,r)\,
C_g(l_1,l_2,l;m_1,m_2,m) \nonumber \\
&\quad \times
D^{l_1*}_{m_1,n_1}(\mathbf{\Omega}_1)\,
D^{l_2}_{m_2,n_2}(\mathbf{\Omega}_2)\,
Y^*_{lm}(\hat{r})
\tag{B.1}
\end{align}

where $C_g(l_1,l_2,l;m_1,m_2,m)$, $D^l_{mn}$ and $Y_{lm}$ are the Clebsch-Gordan coefficient, generalized spherical harmonics and spherical harmonics respectively and $\hat{\vec r}$ is unit vector along intermolecular axis. Eq. (B.1) holds for molecules of general shape with orientation referred to an arbitrary space-fixed reference frame. For linear molecules, the coefficient $A(l_1,l_2,l;n_1,n_2,r)$ vanishes unless $n_1 = n_2 = 0$ and

\begin{equation}
D^{l*}_{m0}(\phi,\theta,x) =
\left(\frac{16\pi^2}{2l+1}\right)^{1/2}
Y_{lm}(\theta,\phi)
\tag{B.2}
\end{equation}

Thus for linear molecules

\begin{equation}
A(\mathbf{r}, \mathbf\Omega_1, \mathbf\Omega_2) =
\sum_{l_1,l_2,l}\sum_{m,m_1,m_2}
A(l_1,l_2,l;r)\,
C_g(l_1,l_2,l;m_1,m_2,m)\,
Y_{l_1 m_1}(\mathbf\Omega_1)\,
Y_{l_2 m_2}(\mathbf\Omega_2)\,
Y^*_{lm}(\hat{r})
\tag{B.3}
\end{equation}

where

\begin{equation}
A(l_1,l_2,l;r) =
\left(\frac{4\pi}{(2l_1+1)(2l_2+1)}\right)^{1/2}
A(l_1,l_2,l;0,0;r)
\tag{B.4}
\end{equation}

In a body-fixed (intermolecular) frame, in which z-axis is parallel to $\mathbf{r}$, it follows that ~\cite{GrayGubbins1984}

\begin{equation}
A(\mathbf{r}, \mathbf\Omega'_1, \mathbf\Omega'_2) =
\sum_{l_1,l_2,m}
A(l_1,l_2,m;r)\,
Y_{l_1 m}(\mathbf\Omega_1')\,
Y_{l_2 \underline{m}}(\mathbf\Omega_2')
\tag{B.5}
\end{equation}

where the prime $\Omega_i' \equiv \theta_i', \phi_i'$ are referred to the intermolecular frame and $\underline{m} = -m$.

and

\[
A(l_1, l_2, m; r) = \sum_l \left( \frac{2l + 1}{4\pi} \right)^{1/2} C_g(l_1, l_2, l; m, \underline{m}, 0) \, A(l_1, l_2, l; r)
\tag{B.6}
\]

\vspace{0.3cm}
\noindent
The above equation relates the spherical harmonic coefficients in the BF to SF for linear molecules. The inversion of the relation leads to

\[
A(l_1, l_2, l; r) = \sum_m \left( \frac{4\pi}{2l + 1} \right)^{1/2} C_g(l_1, l_2, l; m, \underline{m}, 0) \, A(l_1, l_2, m; r)
\tag{B.7}
\]

\subsection*{B.1 Expansion of OZ equation}

If we use Fourier transform for $h(r,\Omega_1,\Omega_2)$ and $c(r,\Omega_1,\Omega_2)$ to change to $k$ space according to

\begin{equation}
h(r,\Omega_1,\Omega_2)=\frac{1}{(2\pi)^3}\int d\vec k\, e^{-i\mathbf{k}\cdot\mathbf{r}}h(\vec k,\mathbf\Omega_1,\mathbf\Omega_2)
\tag{B.8}
\end{equation}

where

\begin{equation}
h(\mathbf k,\mathbf\Omega_1,\mathbf\Omega_2)=\int_V d\mathbf{r}\,e^{i\mathbf{k}\cdot\mathbf{r}}h(\mathbf r,\mathbf \Omega_1,\mathbf \Omega_2)
\end{equation}

the OZ Equation (2.40) transforms to

\begin{equation}
h(\mathbf k,\mathbf \Omega_1,\mathbf \Omega_2)=c(\mathbf k,\mathbf\Omega_1,\mathbf\Omega_2)+\rho_0\int c(\mathbf k,\mathbf\Omega_1,\mathbf\Omega_3)h(\mathbf k,\mathbf\Omega_3,\mathbf\Omega_2)d\mathbf\Omega_3
\tag{B.9}
\end{equation}

The convolution in $r$-space on the right hand side of the OZ equation reduces to a product in $k$-space. This simplification of the equation is one of the advantages of working in $k$-space. The $k$-space functions $h(k,\Omega_1,\Omega_2)$ and $c(k,\Omega_1,\Omega_2)$ can also be expanded in same way as the $r$-space function in harmonics in the SF frame

\begin{equation}
A(k,\Omega_1,\Omega_2)=\sum_{l_1 l_2 l}A_{l_1 l_2 l}(k)\sum_{m_1 m_2 m}
C_g(l_1,l_2,l;m_1,m_2,m)
Y_{l_1 m_1}(\Omega_1)Y_{l_2 m_2}(\Omega_2)Y^*_{l m}(\Omega_k)
\tag{B.10}
\end{equation}

where $\mathbf\Omega_k$ denote the direction of $\mathbf{k}$ in the SF frame and $A$ represents any invariant pairwise function like $h$ or $c$. The $r$-space and $k$-space harmonic expansion coefficients are mutual Hankel transforms.

\begin{equation*}
A_{l_1 l_2 l}(k)=4\pi i^l \int_0^\infty dr\,r^2 j_l(kr)A_{l_1 l_2 l}(r)
\end{equation*}

\begin{equation*}
A_{l_1 l_2 l}(r)=\frac{1}{2\pi^2} (-1)^l \int_0^\infty dk\,k^2 j_l(kr)A_{l_1 l_2 l}(k)
\tag{B.11}
\end{equation*}

where $j_l$ is the spherical Bessel function. Substituting (B.10) into (B.9) and after some algebra we get

\begin{align*}
h_{l_1 l_2 l}(k) &= c_{l_1 l_2 l}(k) + (4\pi)^{-3/2}\rho_0 \sum_{l_3,l',l''}
c_{l_1 l_3 l'}(k) h_{l_3 l_2 l''}(k)
(-1)^{l_1+l_2+l_3}(2l'+1)(2l''+1) \\
&\quad \hspace{5cm}
\left(
\begin{array}{ccc}
l' & l'' & l \\
0 & 0 & 0
\end{array}
\right)
\left(
\begin{array}{ccc}
l_1 & l_2 & l \\
l'' & l' & l_3
\end{array}
\right)
\tag{B.12}
\end{align*}
where $
\left(
\begin{array}{ccc}
l' & l'' & l \\
0 & 0 & 0
\end{array}
\right)$ and  $\left(
\begin{array}{ccc}
l_1 & l_2 & l \\
l'' & l' & l_3
\end{array}
\right)$ are $3j$ and $6j$ symbols respectively. For definitions and properties of these symbols interested readers may consult Ref. ~\cite{GrayGubbins1984}.

Eq.(B.12) has the same structure as the original OZ equation and couples together the harmonics of $h$ and $c$. The OZ equation can also be written in $k-$frame ( a body--fixed frame with $z$-axis along $\mathbf{k}$) harmonics. In this frame, we take the polar axis along $\mathbf{k}$. The $k$-frame harmonics are related to the space-fixed harmonics by ~\cite{GrayGubbins1984}

\begin{equation*}
A_{l_1 l_2 m}(k)=\sum_{l} \left( \frac{(2l+1)}{4\pi} \right)^{1/2} C_g(l_1,l_2,l;m,m',0)A_{l_1 l_2 l}(k)
\end{equation*}

\begin{equation}
A_{l_1 l_2 l}(k)=\sum_m \left(\frac{4\pi}{2l+1}\right)^{1/2}
C_g(l_1,l_2,l;m,m',0)A_{l_1 l_2 m}(k)
\tag{B.13}
\end{equation}

Using (B.13) to transform (B.12) from $A_{l_1 l_2 l}(k)$ to $A_{l_1 l_2 m}(k)$ we get

\begin{equation}
h_{l_1 l_2 lm}(k)=c_{l_1 l_2 l m}(k)+(4\pi)^{-1}\rho_0(-1)^m \sum_{l_3}
\left[h_{l_1 l_2 m}(k)c_{l_3 l_2 m}(k)\right]
\tag{B.14}
\end{equation}

This form of the OZ equation is simple both in structure and appearance and easy to use in practice.

\subsection*{B.2 Expansion of closure relation}

The PY relation given by Eq.(A.11) can be rewritten as

\begin{equation}
c(r,\Omega_1,\Omega_2)=m(r,\Omega_1,\Omega_2)\left[1+\gamma(r,\Omega_1,\Omega_2)\right]
\tag{B.15}
\end{equation}

where

\begin{equation*}
m(r,\Omega_1,\Omega_2)=\exp(-\beta u(r,\Omega_1,\Omega_2))-1
\tag{B.16}
\end{equation*}

and

\begin{equation*}
\gamma(r,\Omega_1,\Omega_2)=h(r,\Omega_1,\Omega_2)-c(r,\Omega_1,\Omega_2)
\tag{B.17}
\end{equation*}

Since all functions appearing in Eq.(B.17) are invarient pair wise function, they an be expanded in spherical harmonics either in SF-frame or in BF-frame. In the BF-frame one finds ~\cite{GrayGubbins1984}, 

\[
c_{l_1 l_2 m}(r_{12}) 
= \sum_{l_1' l_1'' l_2' l_2''} 
\frac{1}{4\pi}
\left[
\frac{(2l_1' + 1)(2l_2' + 1)(2l_1'' + 1)(2l_2'' + 1)}
{(2l_1 + 1)(2l_2 + 1)}
\right]^{1/2}
C_g(l_1' l_1'' l_1; 000)
\]
\[
\times C_g(l_2' l_2'' l_2; 000)
\sum_{m' m''}
C_g(l_1' l_1'' l_1; m' m'' m)\,
C_g(l_2' l_2'' l_2; \underline{m}'\underline{m}'' \underline{m})
\]
\[
\times M_{l_1' l_2' m'}(r_{12})
\left[
4\pi \delta_{000}^{\,l_1'' l_2'' m''}
+ \gamma_{l_1'' l_2'' m''}(r_{12})
\right]
\]
where \underline{m} = -m and the summation is over allowed values of $l_1'$, $l_1''$, $l_2'$, $l_2''$, $m'$ and $m''$. A similar expression can be written in the SF frame. However it is easier  numerically to calculate BF harmonic coefficients than SF harmonic coefficients.

\section*{\underline{Appendix C}}

\section*{Structure factor of a crystalline solid}

The structure factor S(q) of a crystal which is measured by scattering experiments, is defined as
\[
S(q) = \frac{1}{N} \int d\vec r_1 d\vec r_2 e^{-iq\cdot(\vec r_2-\vec r_1)} \langle \hat{\rho}(\vec r_1)\hat{\rho}(\vec r_2) \rangle 
\tag{C.1}
\]
\[
\hspace{5cm} = \frac{1}{N} \int d\vec r_1 d\vec r_2 e^{-iq\cdot(\vec r_2-\vec r_1)} \left( \rho(\vec r_1)\rho(\vec r_2) + H^{(2)}(\vec r_1,\vec r_2) \right) 
\tag{C.2}
\]
In writing Eq.(C.1) use of Eq.(2.20) has been made.

Using the relation
\[
\rho(\vec r) = \rho \sum_G \mu_G e^{iG\cdot r}
\tag{C.3}
\]
one gets from Eq. (C.2),
\[
S(q) = (2\pi)^3 \rho \sum_q |\mu_q|^2 \delta(q-G) + \frac{1}{N} H^{(2)}(q).
\tag{C.4}
\]
where
\[
\frac{1}{N} H^{(2)}(\vec q) = \frac{1}{N} \int \vec dr_1 \int \vec dr_2 e^{-iq\cdot(\vec r_2-\vec r_1)} \rho(\vec r_1)\rho(\vec r_2)
\left[ h^{(0)}(|\vec r_2-\vec r_1|) + h^{(b)}(\vec r_1,\vec r_2) \right]
\]
\[
= 1 + \rho \sum_\vec G |\mu_G|^2 h^{(0)}(|q-\vec G|) + \rho \sum_\vec G \sum_{\vec G_1} \mu_{G_1}\mu_{G-G_1} h^{(b)}\left(\vec q+\frac{1}{2}\vec G+\vec G_1\right)
\tag{C.5}
\]
In Eq. (C.4), the first term represents the Bragg peaks at reciprocal lattice vectors G, the
second term, which contains information about correlations between density fluctuations, contributes to the background corresponding to the non-Bragg angle region of the scattering function. For G = 0, one gets
\[
S(q) = (2\pi)^3 \rho \delta(q) + 1 + \rho h^{(0)}(q),
\tag{C.6}
\]
which is the structure factor of a fluid ~\cite{Hansen2006}. The $\delta-$function term corresponds to the radiation that
passes through the sample unscattered and is neglected. PCFs of fluids are determined from S(q) given by Eq. (C.6); however, to find PCFs of a crystals from $H^{(2)}(q)$ given by Eq. (C.5) is impractical.

\section*{\underline{Acknowledgments}}
\addcontentsline{toc}{section}{Acknowledgments}
I am thankful to my collaborators, J. Ram, S. Kumar, P. Mishra, S. L. Singh, A. S. Bharadwaj and A. Jaiswal for their help. My most sincere thanks are due to Mayank Raj Gupta and Savesh Kumar Upadhyay for careful reading and typing the manuscript. Without the help of Mayank this project could not have been completed.
\addcontentsline{toc}{section}{Disclosure statement}
\section*{\underline{Disclosure statement}}

The author declares that he has no known competing financial interests or personal relationships that could have appeared to influence the work reported in this article.

\section*{\underline{Funding}}

No fund from any source was received.

\addcontentsline{toc}{section}{References}
\bibliographystyle{tfq}
\bibliography{paper}

\end{document}